\documentclass[amsmath,amssymb,10pt,aps,prb,twocolumn,longbibliography]{revtex4-2} 
 \usepackage[colorlinks=true,linkcolor=blue,citecolor=blue,urlcolor=blue]{hyperref}
 \usepackage{lipsum}
 \usepackage{graphicx}
\usepackage{latexsym}
\usepackage{amsmath}
\usepackage{amsthm}
\usepackage{amssymb}
\usepackage{enumitem}
\usepackage{setspace}
\usepackage{dcolumn}
\usepackage{bm}
\usepackage{setspace} 
\usepackage{slashed}
\usepackage{color}
\usepackage{youngtab}
\usepackage{tikz}
\usepackage{tikz-3dplot}
\usepackage{tikz-3dplot}
\usepackage{braket}
\usepackage{subfigure}
\usepackage{makecell}

\DeclareMathAlphabet{\mathpzc}{OT1}{pzc}{m}{it}

\begin{document}

\title{Vison condensation and spinon confinement in a kagome-lattice $\mathbb{Z}_2$ spin liquid:\\A numerical study of a quantum dimer model}

\author{Kyusung Hwang}
\email{khwang@kias.re.kr}
\affiliation{School of Physics, Korea Institute for Advanced Study, Seoul 02455, Korea}

\date{\today}
\begin{abstract}
Quantum spin liquids are exotic many-body states featured with long-range entanglement and fractional anyon quasiparticles. Quantum phase transitions of spin liquids are particularly interesting problems related with novel phenomena of anyon condensation and anyon confinement. Here we study a quantum dimer model which implements a transition between a $\mathbb{Z}_2$ spin liquid ($\mathbb{Z}_2$SL) and a valence bond solid (VBS) on the kagome lattice. The transition is driven by the condensation of vison excitation of the $\mathbb{Z}_2$ spin liquid, which impacts on other anyon excitations especially leading to the confinement of spinon excitations. By numerical exact diagonalization of the dimer model, we directly measure the vison condensation using vison string operators, and explicitly check a confining potential acting on spinon excitations in the VBS state. It is observed that topological degeneracy of the spin-liquid state is lifted concomitantly with the vison condensation. The dimer ordering pattern of the VBS state is identified by investigating dimer structure factor. Furthermore, we find an interesting state that exhibits features of spin liquid and VBS simultaneously. We discuss the origin of the mixed behaviors and possible scenarios expected in thermodynamic limit. This work complements the previous analytical studies on the dimer model [\href{https://doi.org/10.1103/PhysRevB.87.104408}{Phys.~Rev.~B~87,~104408~(2013)} and \href{https://journals.aps.org/prb/abstract/10.1103/PhysRevB.92.205131} {Phys.~Rev.~B~92,~205131~(2015)}] by providing numerical evidences on the vison condensation and the spinon confinement in the $\mathbb{Z}_2$SL-to-VBS transition.
\end{abstract}
\maketitle 


\section{Introduction}

Quantum spin liquids are highly entangled quantum states of localized spins featured with fractional anyonic quasiparticles~\cite{Anderson1973,WenBook,FradkinBook,SachdevBook,Kitaev2003,Kitaev2006,LevinWen2005,Savary2016,Norman2016,Ng2017,Knolle2019,Takagi2019,Broholm2020,Motome2020}. Depending on the entanglement structure, quantum spin liquids support different types of anyon quasiparticles.
$\mathbb{Z}_2$ spin liquids, which are one of the most extensively studied states in frustrated magnetism, realize the $\mathbb{Z}_2$ lattice gauge theory with four different anyons: trivial boson ($1$), bosonic spinon ($e$), bosonic vison ($m$), and fermionic spinon ($\psi=e \times m$)~\cite{ReadSachdev1991,Wen1991,Sachdev1992,Read1989,Senthil2000,Moessner2001,Fradkin2001,Misguich2002,Wen2002,Balents2002,Kitaev2003,Wen2003,Kitaev2006,Wang2006,White2011,Depenbrock2012,Motrunich2011,Lu2011,Iqbal2011,Zhang2012,Jiang2012,Messio2012,Hermele2013,Huh2011,Wan2013,Hwang2015,Zaletel2015,Lu2017,Poilblanc2010,Hao2014,Rousochatzakis2014,Ralko2018,Iqbal2020,Sachdev2010,Messio2010,Dodds2013,Sachdev2014,Messio2017,Halimeh2016,Sheng2019,Halimeh2019,Meng2018BFG,Wessel2018BFG,White2015,Sheng2015,Mila2007,Misguich2008,Slagle2014,Meng2021QDM,Meng2022QDM}.
Although $e$-anyon and $m$-anyon are self-bosons, they can sense each other via the Aharonov-Bohm effect (or nontrivial mutual statistics). Namely, they see each other as a $\pi$-flux when $e$-anyon moves around $m$-anyon and vice versa~\cite{Read1989}.
The mutual statistics renders the bound state of $e$-anyon and $m$-anyon ($\psi=e \times m$) to be a self-fermion.
Such nontrivial anyons are created in pairs from the vacuum by physical processes as indicated by the fusion rules, $e \times e = m \times m = \psi \times \psi =1$.
All these anyon properties of the $\mathbb{Z}_2$ spin liquid are collectively called $\mathbb{Z}_2$ topological order~\cite{WenBook,Kitaev2003,Kitaev2006}.
Resonating valence bond (RVB) state on the kagome lattice is a good example of the $\mathbb{Z}_2$ spin liquid~\cite{Sachdev1992,Lu2011,Lu2017}.
Specifically, quantum superposition of all possible dimer states (nearest-neighbor spin-singlets) on the kagome lattice implements the $\mathbb{Z}_2$ topological order~\cite{Misguich2002}.
Recently, such $\mathbb{Z}_2$ spin liquids have been experimentally realized in quantum simulators of superconducting qubits and Rydberg atom arrays~\cite{Satzinger2021,Semeghini2021,Samajdar2020,Verresen2021,Verresen2022,Samajdar2023}.

\begin{figure*}[tb]
\centering
\includegraphics[width=\linewidth]{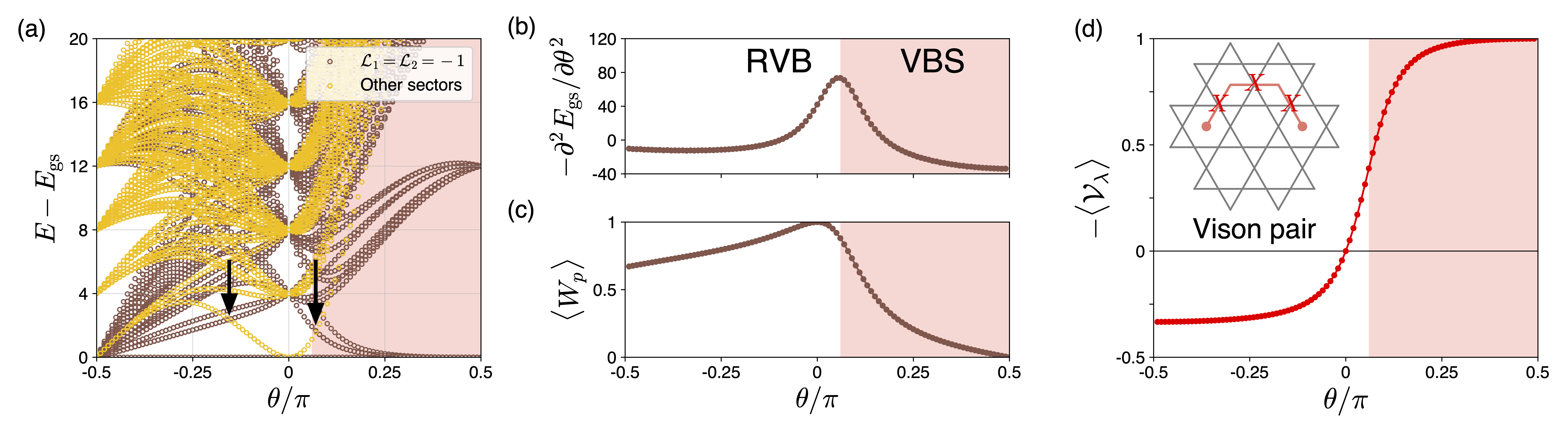}
\caption{ED results of the dimer model on the 36-site cluster.
(a) Energy spectrum as a function of the parameter $\theta$. Different colors distinguish the four distinct topological sectors. Brown: $(\mathcal{L}_{1},\mathcal{L}_{2})=(-1,-1)$. Yellow: $(\mathcal{L}_{1},\mathcal{L}_{2})=(1,1),(-1,1),(1,-1)$. The latter three sectors have an identical energy spectrum.
The two arrows mark the topological degeneracy lift where the threefold degenerate states (yellow) move upward crossing the lowest vison excitation energy level (brown).
(b) Second derivative of the ground-state energy, $-\partial^2 E_{\rm gs}/\partial \theta^2$. The peak indicates the phase boundary between the RVB state and the VBS state ($\theta_{\rm c}\simeq0.06\pi$).
(c) Expectation value of the $\mathbb{Z}_2$-flux operator, $\langle W_p \rangle$.
(d) Expectation value of the vison string operator, $-\langle \mathcal{V}_{\lambda} \rangle$. The inset depicts the vison string operator ($\mathcal{V}_{\lambda}$) used in the calculations.
}
\label{fig:ED-results}
\end{figure*}

Anyon quasiparticles not only define the underlying topological orders of quantum spin liquids, but also determine possible continuous transitions to other phases.
To be specific, condensing vison excitations in the $\mathbb{Z}_2$ spin liquid ($\mathbb{Z}_2$SL) results in a transition to a valence bond solid (VBS) state~\cite{Mila2007,Misguich2008,Huh2011,Slagle2014,Hwang2015,Meng2021QDM}.
\begin{equation}
\left\{
\begin{array}{c}
{\mathbb{\bf Z}_2{\rm \bf SL}}
\\
\\
1,e,m,\psi
\end{array}
\right\}
\xrightarrow[]{\langle m \rangle \ne 0} 
\left\{
\begin{array}{c}
{{\rm \bf VBS}}
\\
\\
1
\end{array}
\right\}
:~~\makecell{\rm Spinons~({\it e}~and~\psi)\\ \rm confined~and \\ \rm symmetries~broken}
\nonumber
\end{equation}
In the $\mathbb{Z}_2$SL-to-VBS transition, spinon excitations are confined due to their nontrivial braiding with the condensed anyon, the vison.
In fact, this kind of anyon condensation transition is not limited to symmetry-breaking transitions to long-range orders, but can be further generalized to cover topological transitions between distinct quantum spin liquids~\cite{Bais2009,Burnell2018,Hwang2024}.
The $\mathbb{Z}_2$SL-to-VBS transition provides not only a simplest setup of anyon condensation transition but also a condensed matter analog of the quark confinement in high-energy physics~\cite{Wilson1974,Kogut1979}.

In this work, we study the $\mathbb{Z}_2$SL-to-VBS transition focusing on numerical detection of vison condensation and spinon confinement. As a concrete model for the transition, we consider the kagome-lattice quantum dimer model (QDM) studied in Refs.~\cite{Misguich2002,Wan2013,Hwang2015}.
Interestingly, the model has three equivalent descriptions: (i) quantum dimer model on the kagome lattice, (ii) $\mathbb{Z}_2$ gauge theory on a honeycomb lattice, and (iii) transverse field Ising model on a triangular lattice.
Here we focus on the dimer model description. The model is defined in the dimer Hilbert space with each dimer state satisfying the so-called hardcore dimer constraint, i.e., every site of the lattice is covered by only a single dimer.
The Hamiltonian is given by
\begin{equation}
H_{\rm QDM} = - h \sum_p {W}_p + K \sum_p {V}_p,
\label{eq:QDM}
\end{equation}
where the $h$ term generates kinetic motions of dimers and the $K$ term represents interactions of dimers.
Specifically, the operator $W_p$ moves dimers along closed loops around a local hexagon plaquette $p$:
\begin{eqnarray}
{W}_p 
= 
\sum_{D} | \bar{D} \rangle \langle {D} | 
&=&
| \parbox{0.6cm}{\includegraphics[width=\linewidth]{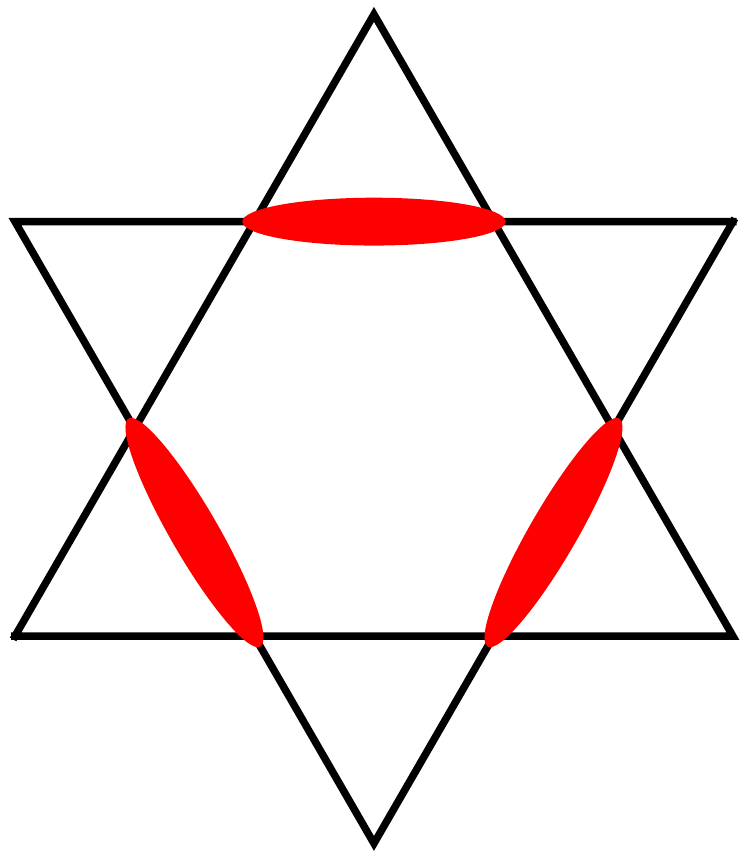}} \rangle
\langle \parbox{0.6cm}{\includegraphics[width=\linewidth]{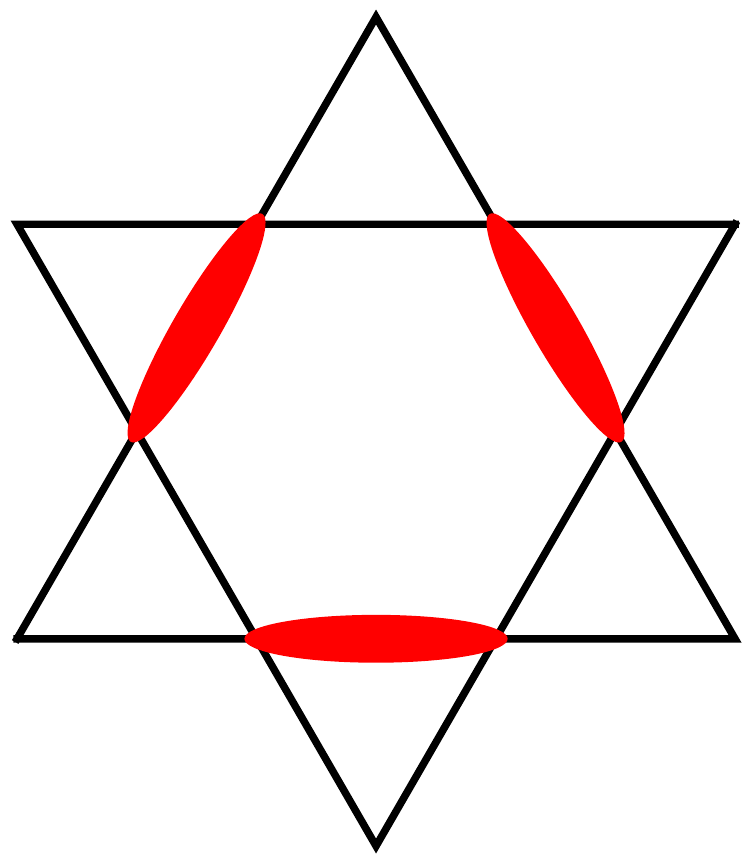}} |
+
| \parbox{0.6cm}{\includegraphics[width=\linewidth]{./hexagon2.pdf}} \rangle
\langle \parbox{0.6cm}{\includegraphics[width=\linewidth]{./hexagon.pdf}} |
+
\cdots 
\nonumber\\
&+&
| \parbox{0.6cm}{\includegraphics[width=\linewidth]{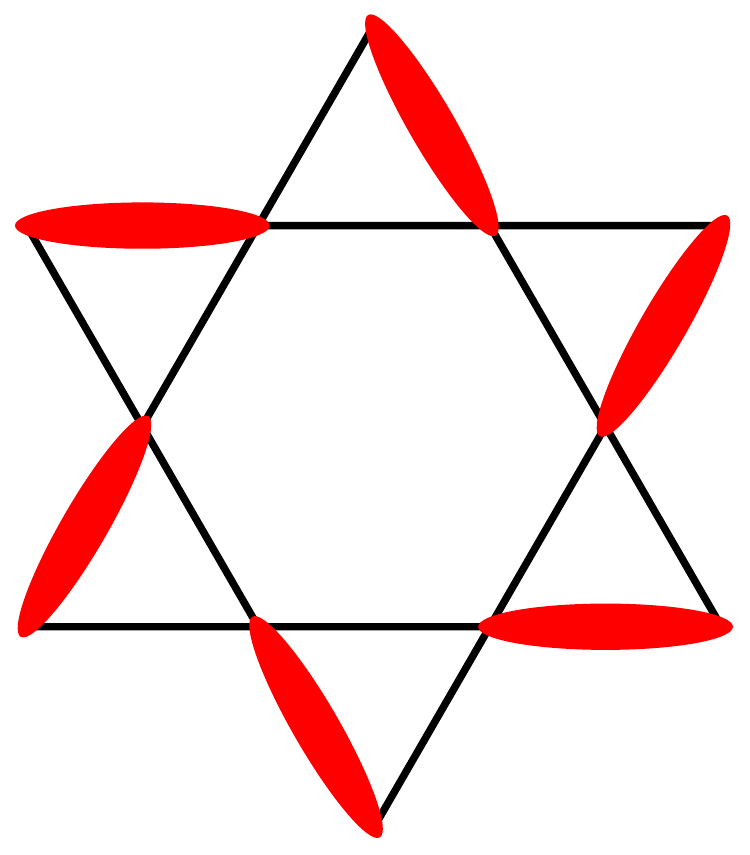}} \rangle
\langle \parbox{0.6cm}{\includegraphics[width=\linewidth]{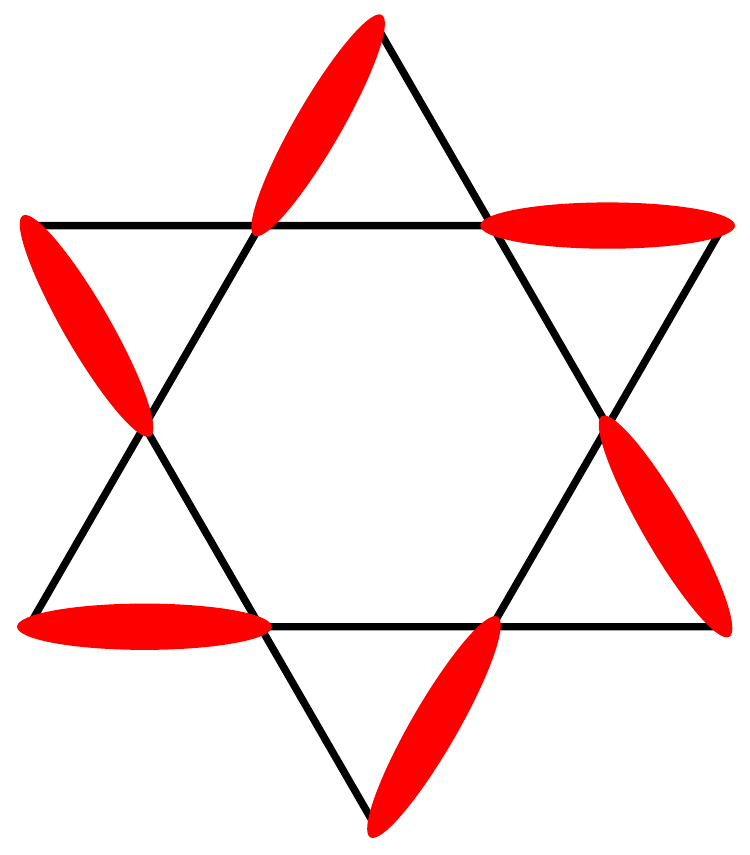}} |
+
| \parbox{0.6cm}{\includegraphics[width=\linewidth]{./pinwheel2.pdf}} \rangle
\langle \parbox{0.6cm}{\includegraphics[width=\linewidth]{./pinwheel.pdf}} | ,
\end{eqnarray}
where $| D \rangle$ denotes a dimer covering on the 12-site David star enclosing the hexagon $p$, and $|\bar{D} \rangle$ means the conjugate dimer covering that is obtained by shifting dimers of $| D \rangle$ along a closed loop by one lattice spacing.
Acting on each dimer covering $| D \rangle$, the operator $V_p$ assigns an interaction energy $E_D$:
\begin{eqnarray}
{V}_p = \sum_{D} E_D | D \rangle \langle D |
&=&
3| \parbox{0.6cm}{\includegraphics[width=\linewidth]{./hexagon.pdf}} \rangle
\langle \parbox{0.6cm}{\includegraphics[width=\linewidth]{./hexagon.pdf}} |
+
3| \parbox{0.6cm}{\includegraphics[width=\linewidth]{./hexagon2.pdf}} \rangle
\langle \parbox{0.6cm}{\includegraphics[width=\linewidth]{./hexagon2.pdf}} |
+
\cdots 
\nonumber\\
&-&3
| \parbox{0.6cm}{\includegraphics[width=\linewidth]{./pinwheel.pdf}} \rangle
\langle \parbox{0.6cm}{\includegraphics[width=\linewidth]{./pinwheel.pdf}} |
-3
| \parbox{0.6cm}{\includegraphics[width=\linewidth]{./pinwheel2.pdf}} \rangle
\langle \parbox{0.6cm}{\includegraphics[width=\linewidth]{./pinwheel2.pdf}} |
. 
\end{eqnarray}
The whole list of dimer coverings, dimer motion graphs, and interaction energies is provided in Table~\ref{tab:1}.
The QDM with no dimer interactions ($K=0$) was initially introduced by Misguich, Serban, and Pasquier as an exactly solved model for a short-ranged RVB state~\cite{Misguich2002}. 
Afterwards, Wan and Tchernyshyov investigated the extended model in Eq.~(\ref{eq:QDM}) as a low-energy effective theory of the spin-1/2 kagome-lattice Heisenberg antiferromagnet~\cite{Wan2013}. It was shown that the effective QDM well captures the low-energy spin-singlet dimer fluctuations observed in the density matrix renormalization group (DMRG) study by Yan, Huse, and White~\cite{White2011}.
In a previous work, we also have studied the QDM and possible VBS orders that may arise from the RVB state by using projective symmetry group (PSG) analyses on vison excitations and Ginzburg-Landau theories~\cite{Hwang2015}.
Other forms of quantum dimer models have been derived microscopically from the spin-1/2 kagome-lattice Heisenberg model~\cite{Poilblanc2010,Rousochatzakis2014,Ralko2018}. Such models would be good choices if one wishes to more accurately capture the low-energy sector of the spin model. Among various QDMs, we consider the model in Eq.~(\ref{eq:QDM}) due its simplicity. In this setup, we can efficiently demonstrate the anyon physics of vison condensation and spinon confinement.

Here we investigate the QDM in Eq.~(\ref{eq:QDM}) by numerical exact diagonalization (ED).
Using the parametrization,
\begin{equation}
h = \cos \theta ~~~{\rm and}~~~ K=\sin \theta ,
\end{equation}
we find that the system has (i) the RVB state over a fairly extended region containing $\theta=0$ and (ii) a 12-site pinwheel VBS state around $\theta=\pi/2$, separated by a continuous transition at $\theta_{\rm c}\simeq0.06\pi$ (see Fig.~\ref{fig:ED-results}).
The RVB state and the VBS state exhibit distinguished behaviors in our ED calculations on finite-size clusters.
We identify the nature of the RVB-to-VBS transition by calculating topological degeneracy, dimer structure factor, vison condensation, and spinon confinement.
We measure the vison condensation and the spinon confinement by employing two types of string operators that create a pair of visons and a pair of spinons, respectively.
For an intuitive understanding of the underlying physics, the string operators and the associated excitations are interpreted in dimer picture.

This work provides a complete understanding on the RVB-to-VBS transition of the dimer model by demonstrating the anyon physics of vison condensation and spinon confinement that have been missing in previous studies.
On the other hand, we find that on negative $\theta$ the system simultaneously shows behaviors of spin liquid and VBS.
We discuss the origin of the mixed behaviors and possible scenarios expected in thermodynamic limit.

The rest of this paper is structured in the following way.
In Sec.~\ref{sec:II}, the quantum dimer model is constructed in a slightly different fashion from previous studies but more efficiently.
We discuss expected phases in the dimer model. The RVB spin liquid and anyon excitations are described along with anyon string operators.
Also, VBS orders proposed by the previous Ginzburg-Landau theories are introduced together with the expected dimer structure factor.
In Sec.~\ref{sec:III}, the results of numerical exact diagonalization are presented, where we confirm the vison condensation in the RVB-to-VBS transition by using vison string operators and also identify the dimer ordering pattern of the VBS state via the calculated dimer structure factor.
In Sec.~\ref{sec:IV}, we discuss numerical evidences on spinon confinement.
In Sec.~\ref{sec:V}, we discuss the negative-$\theta$ region where the system shows behaviors of spin liquid and VBS simultaneously.
We point out the origin of the mixed behaviors and suggest possible scenarios in thermodynamic limit.
Lastly, we summarize this study and conclude in Sec.~\ref{sec:VI}.

\begin{table*}
\caption{Dimer motions and interactions in the quantum dimer model [Eq.~(\ref{eq:QDM})]. The first row illustrates the dimer motion graph for each dimer covering. By the action of a $W_p$ operator, dimers (red) move by one lattice spacing along the closed loop (light blue). The second row indicates the dimer interaction energy ($E_D$) for each dimer covering. Other cases related by symmetries are dropped for simplicity.
\label{tab:1}}
\begin{ruledtabular}
\begin{tabular}{ccccccccc}
\makecell{Dimer covering and \\ dimer motion graph} &
\parbox{5em}{\includegraphics[width=0.6\linewidth]{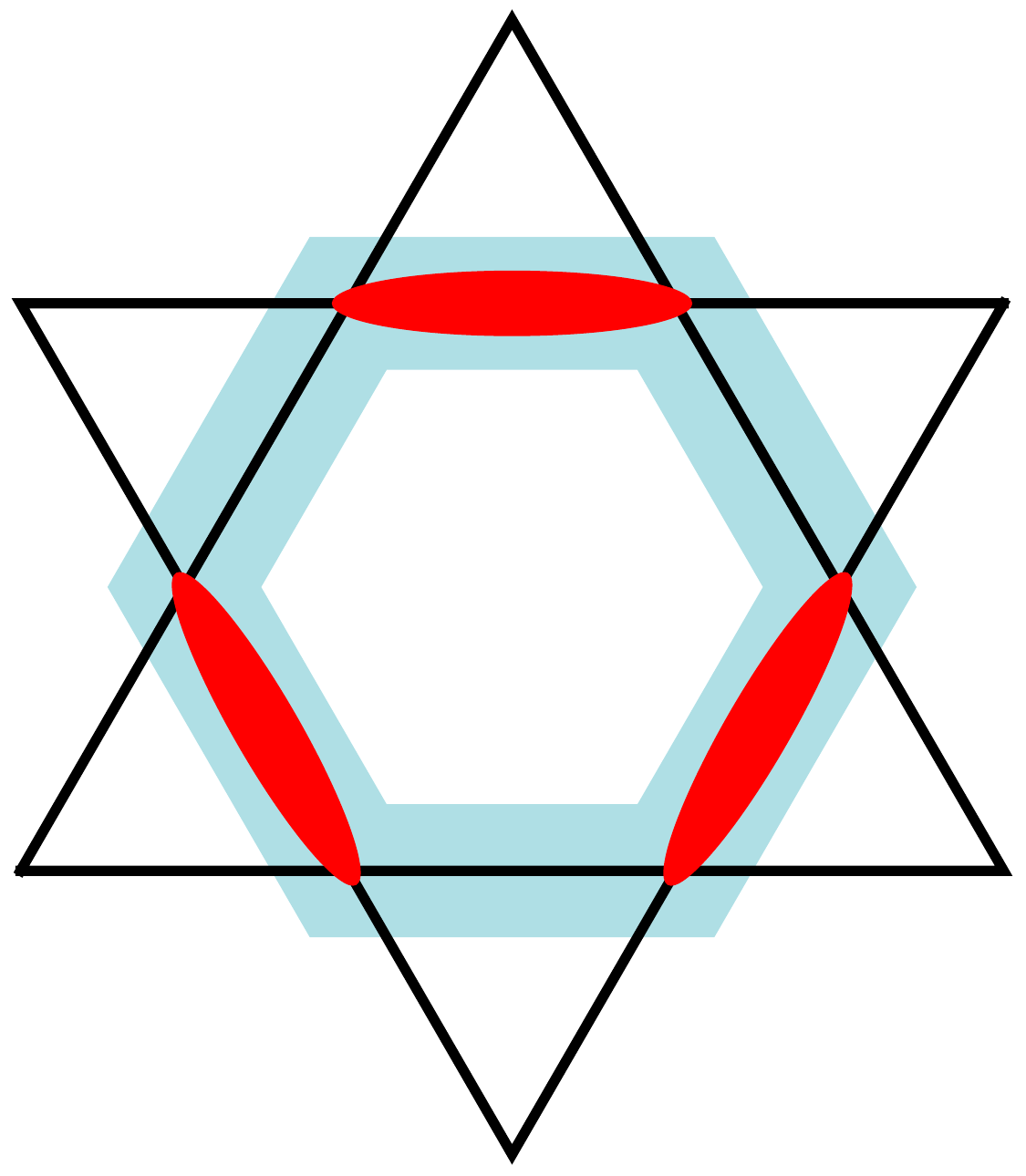}} & 
\parbox{5em}{\includegraphics[width=0.6\linewidth]{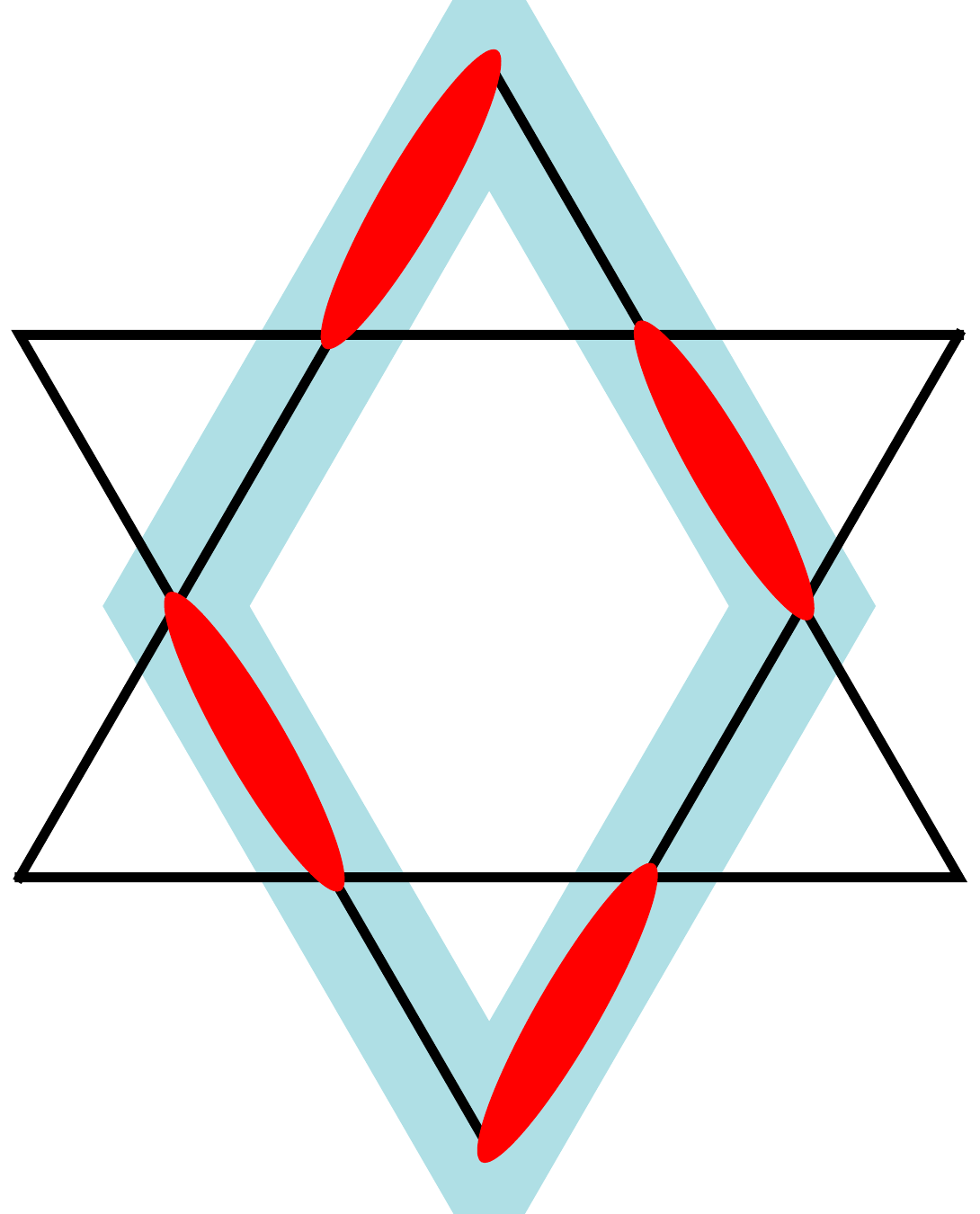}} & 
\parbox{5em}{\includegraphics[width=0.6\linewidth]{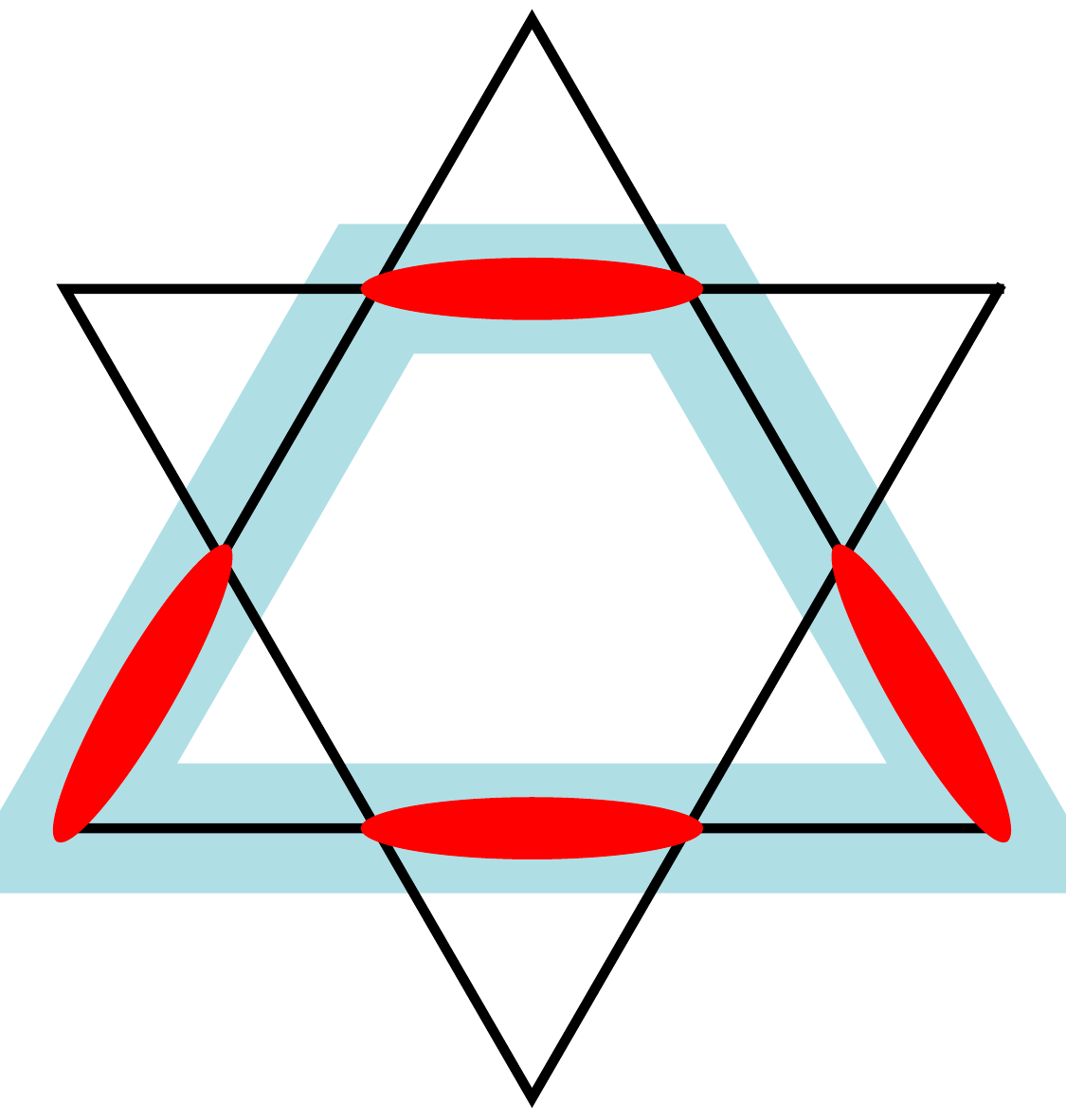}} & 
\parbox{5em}{\includegraphics[width=0.6\linewidth]{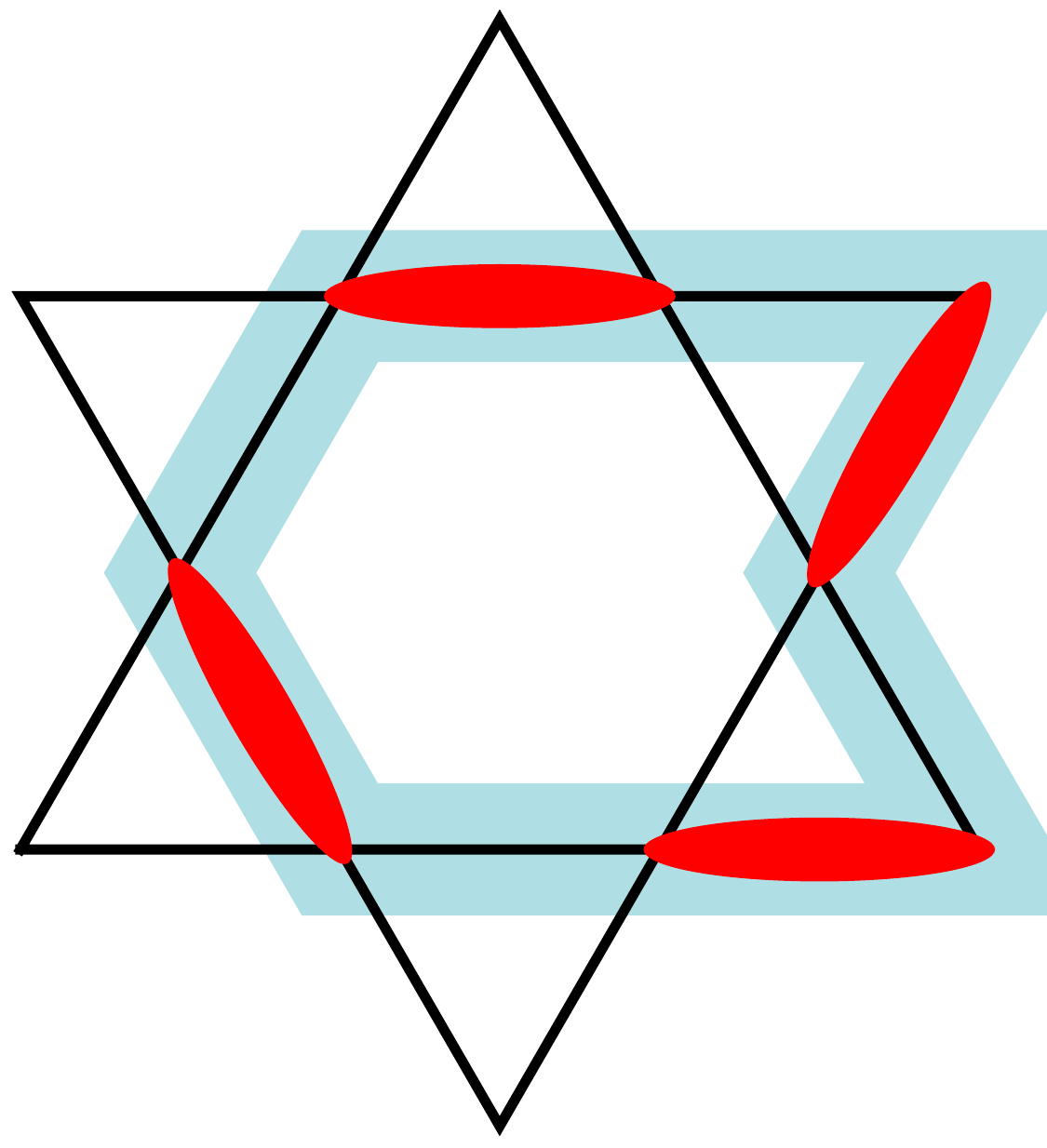}} & 
\parbox{5em}{\includegraphics[width=0.6\linewidth]{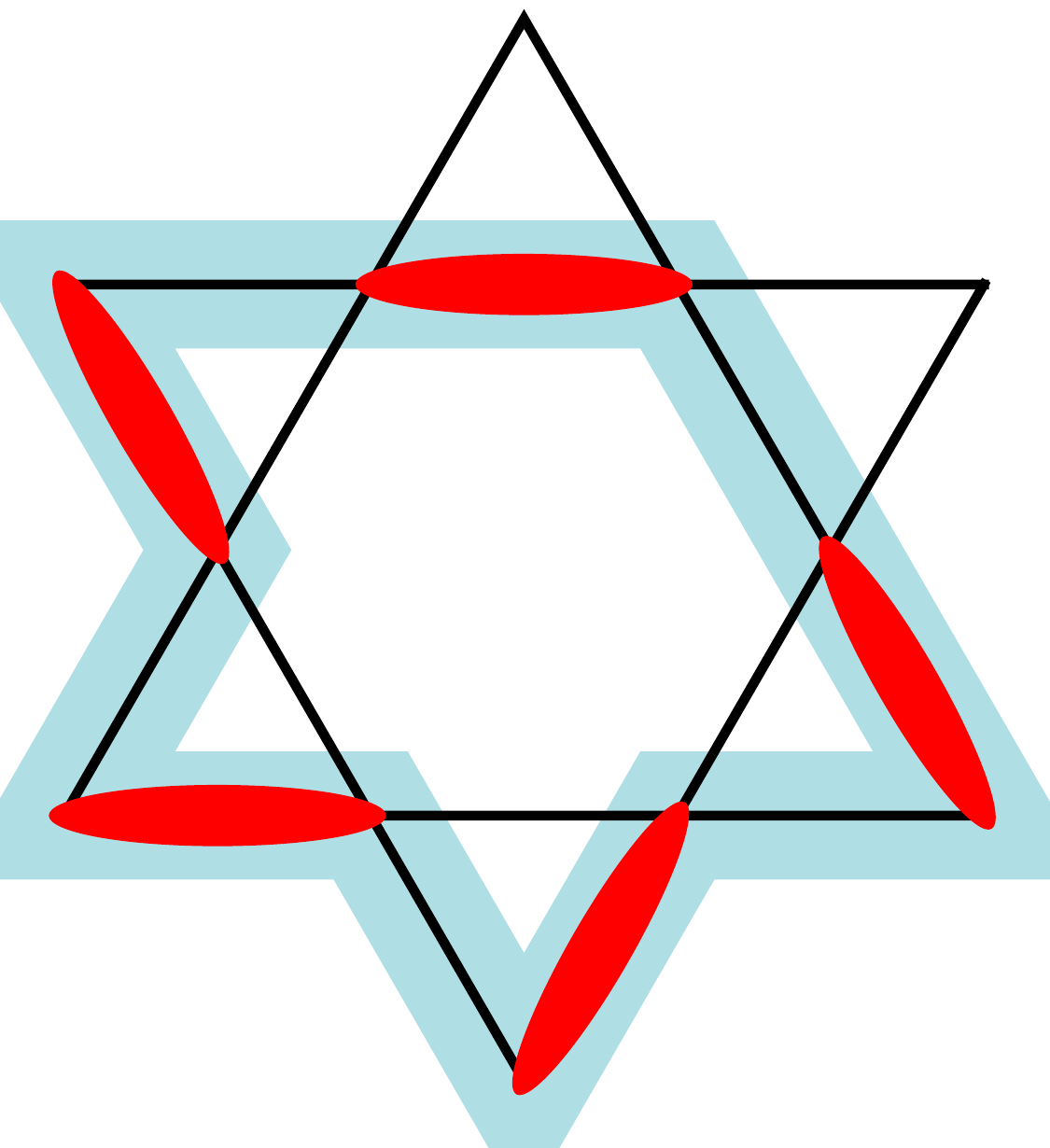}} & 
\parbox{5em}{\includegraphics[width=0.6\linewidth]{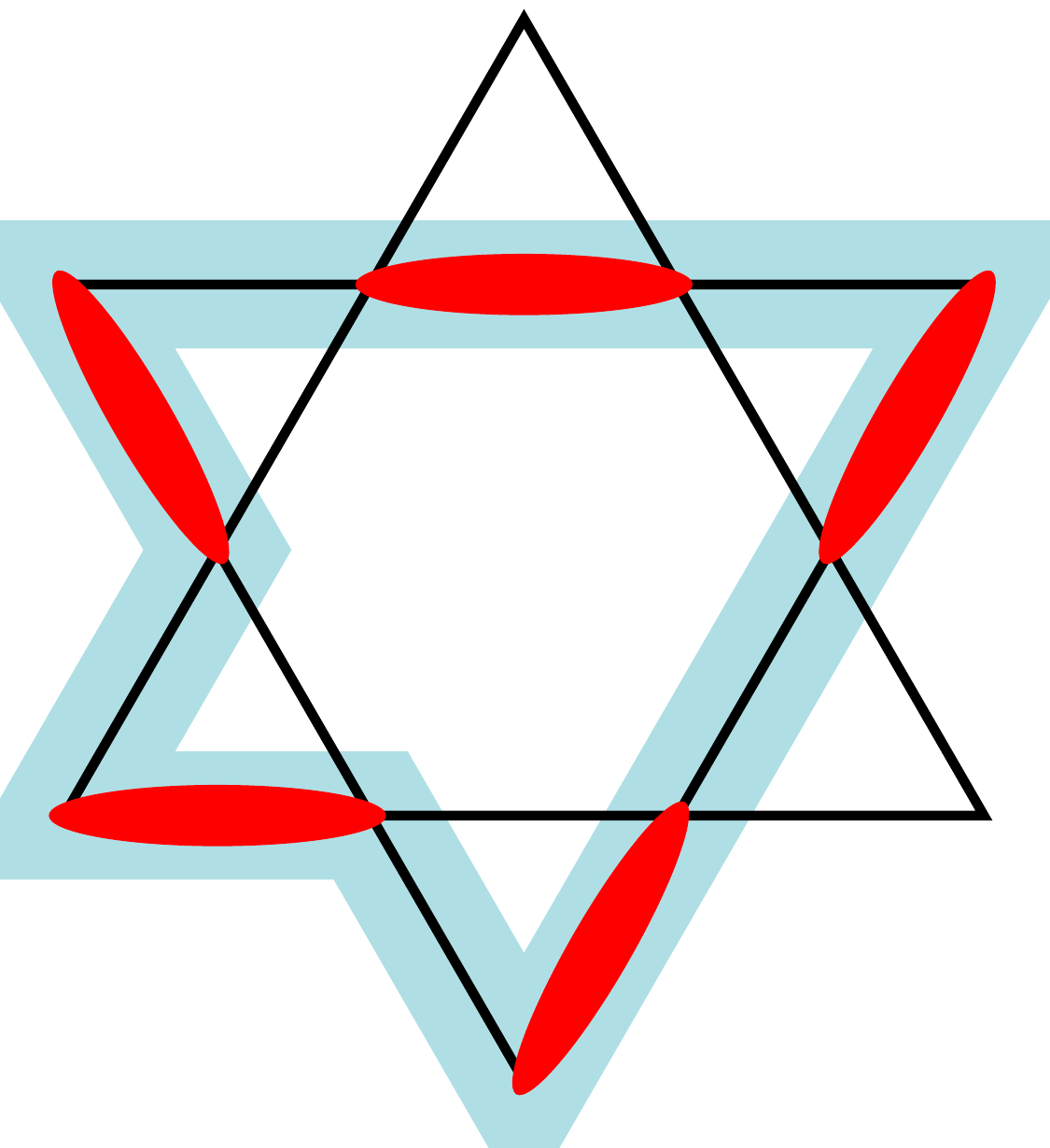}} & 
\parbox{5em}{\includegraphics[width=0.6\linewidth]{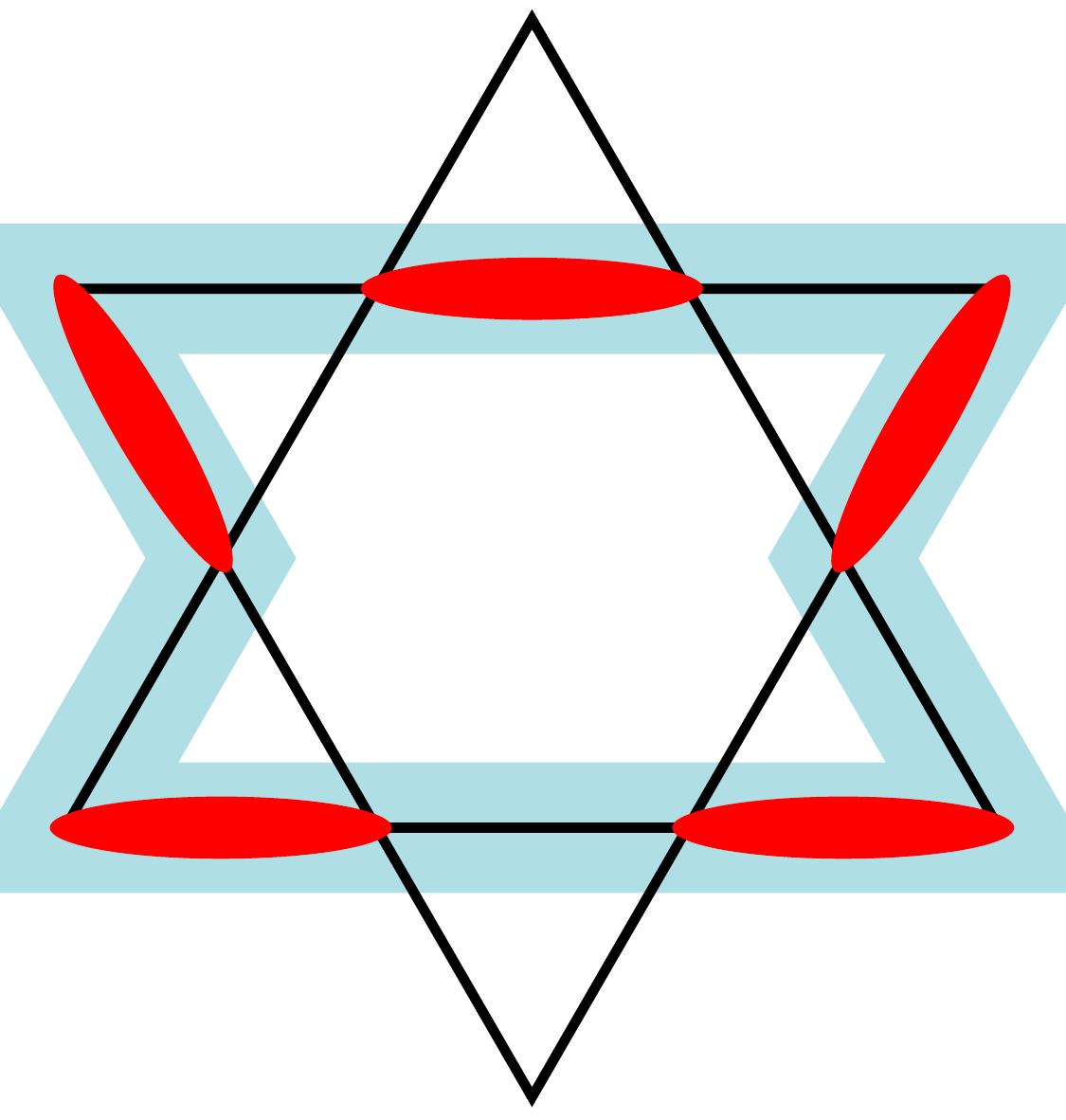}} & 
\parbox{5em}{\includegraphics[width=0.6\linewidth]{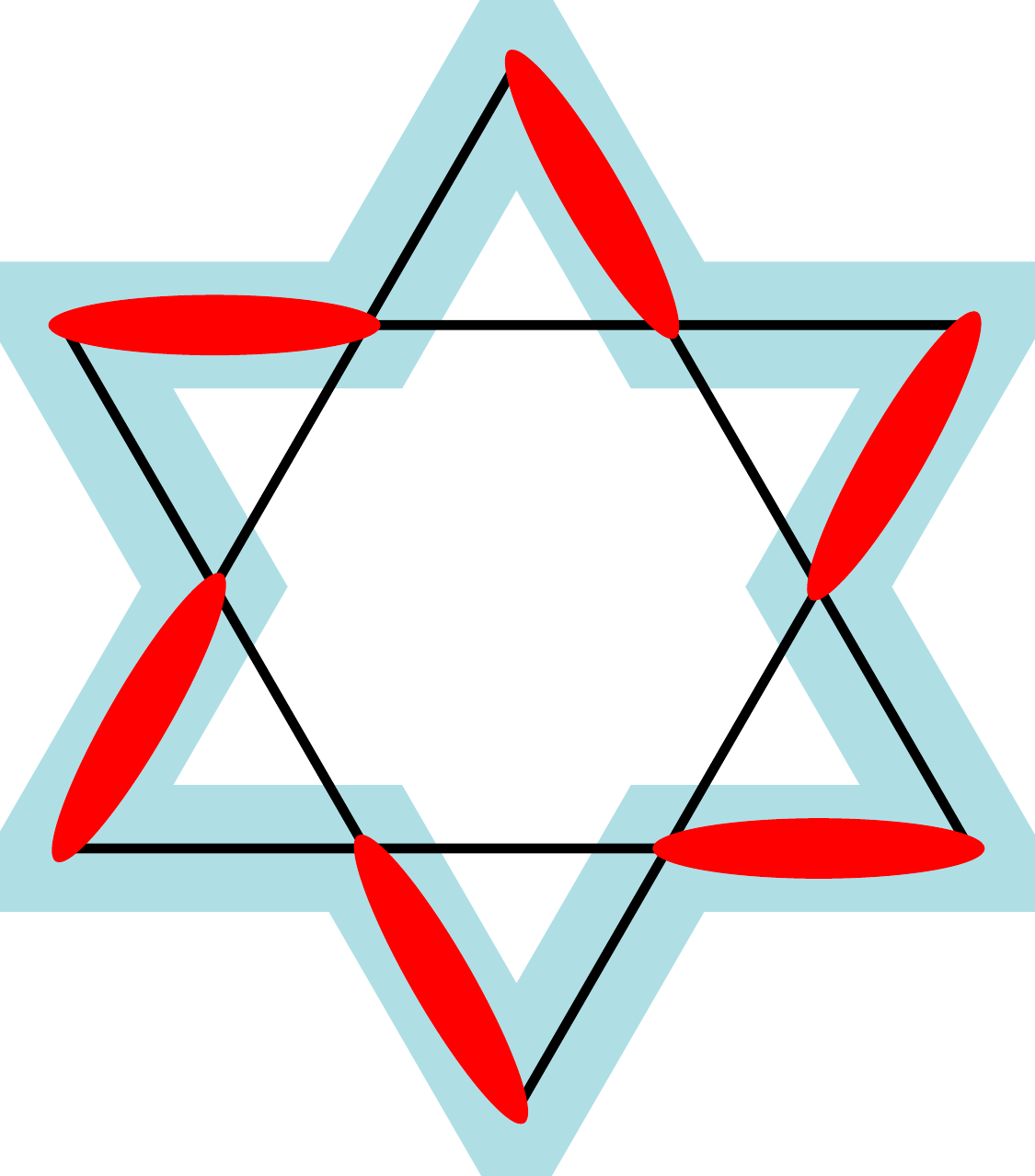}}
\\
\\
$E_D$ &
$3$ &
$-3$ &
$-1$ &
$1$ &
$-1$ &
$1$ &
$3$ &
$-3$
\end{tabular}
\end{ruledtabular}
\end{table*}

\begin{figure}[tb]
\centering
\includegraphics[width=0.8\linewidth]{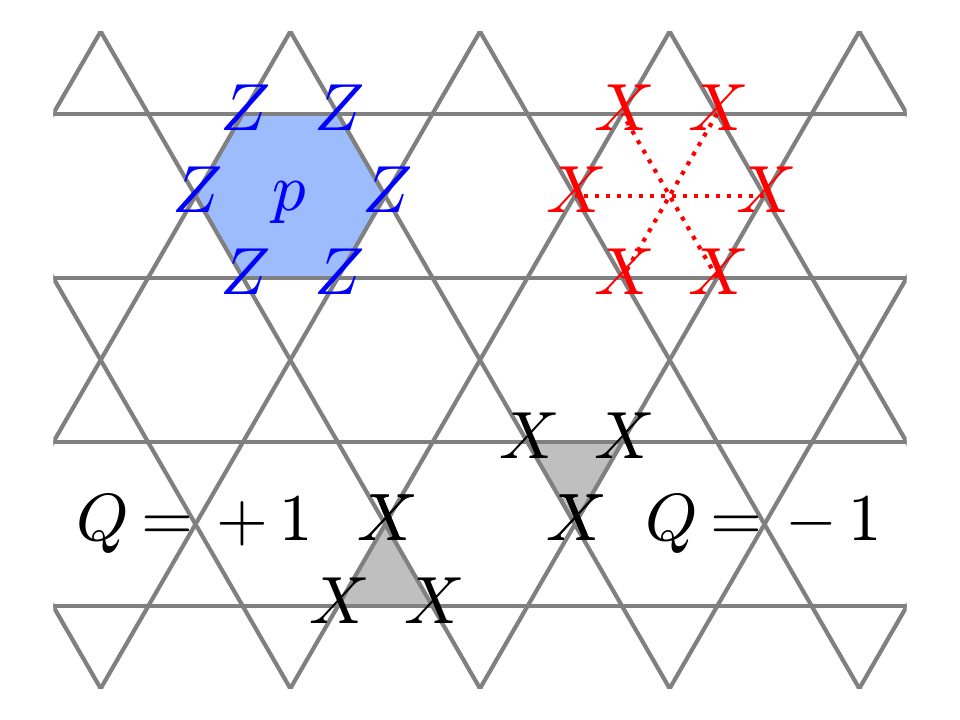}
\caption{Qubit representation of the dimer model. The two terms of Eq.~(\ref{eq:qubit-rep}) and the Gauss law constraint in Eq.~(\ref{eq:Gauss-law}) are visualized in the figure.}
\label{fig:qubit-model}
\end{figure}

\begin{figure*}
\centering
\includegraphics[width=\linewidth]{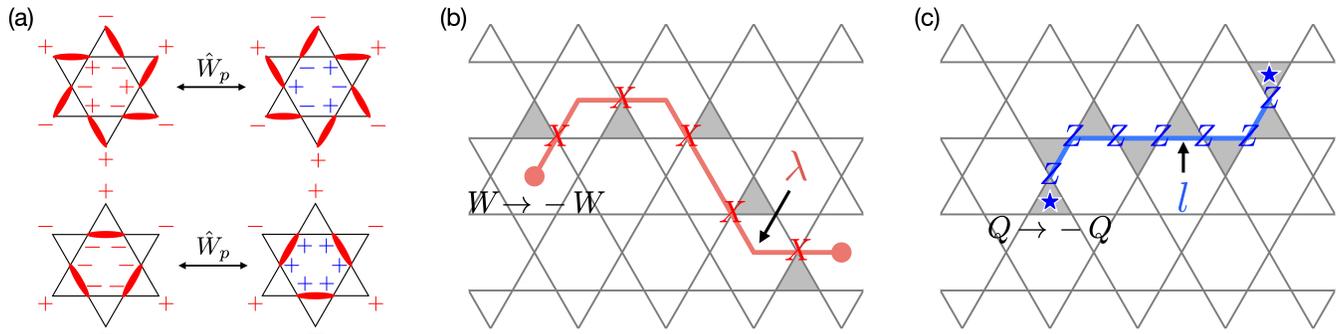}
\caption{Illustrations of $\mathbb{Z}_2$-flux operator and anyon string operators.
(a) $\mathbb{Z}_2$-flux operator ($W_p$) acting on dimer/qubit states. The operator switches the qubit state of the hexagon plaquette $p$ ($X_i=+1 \rightleftarrows X_i=-1$ for $i\in p$), reproducing the dimer resonance motions of the original QDM.
(b) Vison string operator ($\mathcal{V}_\lambda$). The string operator measures the dimer parity at up triangles (gray) touched by the string $\lambda$. The string operator changes the quantum number of ${W}_p$ and the energy only at the ends of the string.
(c) Spinon string operator ($\mathcal{S}_l$). The string operator moves dimers along the triangles (gray) touched by the string $l$, resulting in the violation of the hardcore dimer constraint in the two triangles located at the ends of the string.
}
\label{fig:vison-string}
\end{figure*}

\section{Dimer model\label{sec:II}}

We recast the quantum dimer model in terms of site variables on the kagome lattice. Specifically, a qubit is assigned to each site of the kagome lattice. 
We note that each site is shared by up-pointing and down-pointing triangles on the kagome lattice. In other words, each site can be covered by either a dimer in the up-pointing triangle or a dimer in the down-pointing triangle.
We represent the dimer covering of each local site by the qubit states $|X=\pm1\rangle$.
For instance, if the site is covered by a dimer in the ``down'' triangle, this dimer covering is represented by the qubit state $|X=+1\rangle$.
If the site is covered by a dimer in the ``up'' triangle, this dimer covering is represented by the qubit state $|X=-1\rangle$.
The dimer-qubit mapping is illustrated in the following figure.
\begin{equation}
\parbox{0.7cm}{\includegraphics[width=\linewidth]{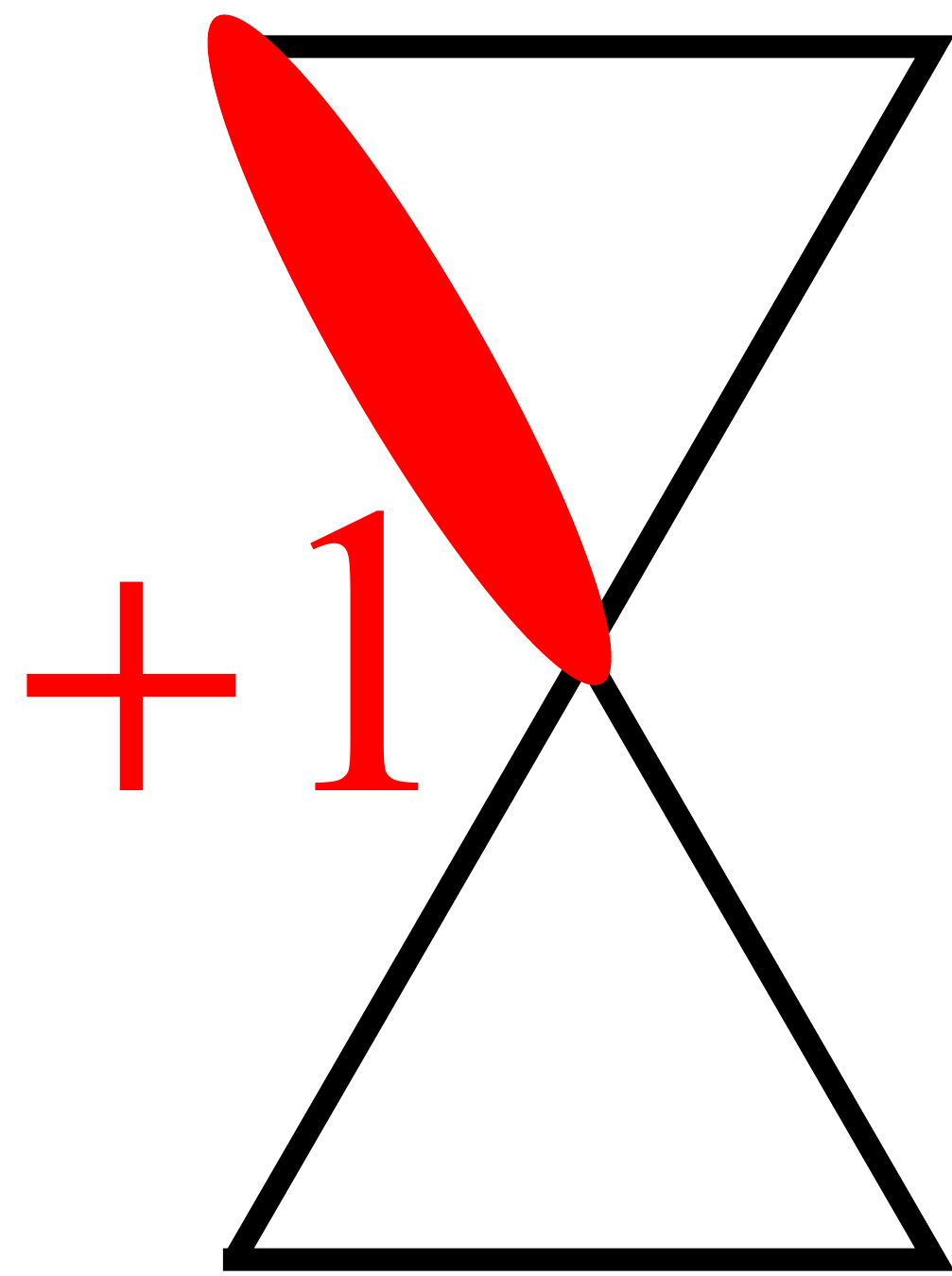}}
~~~
\parbox{0.7cm}{\includegraphics[width=\linewidth]{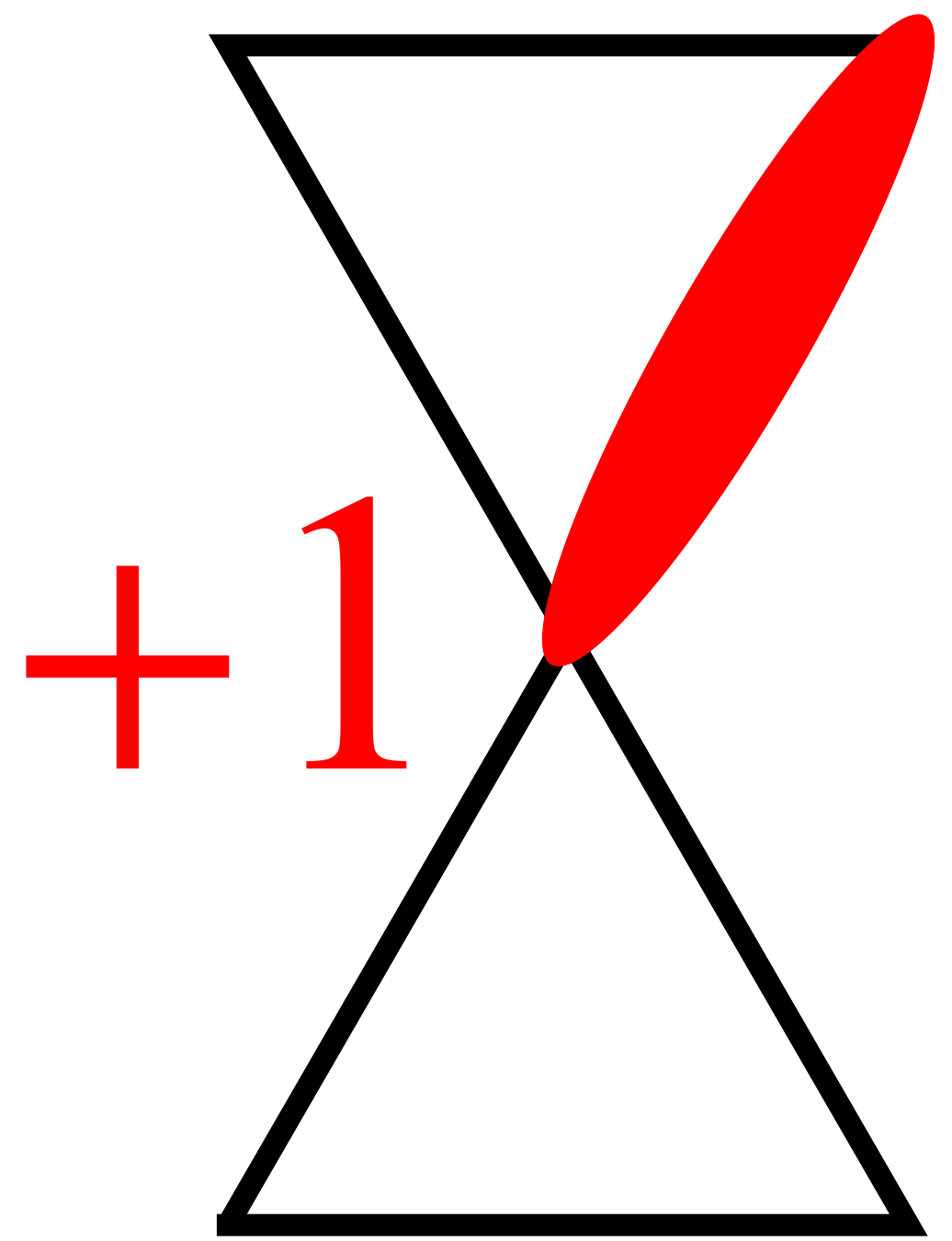}}
~~~
\parbox{0.7cm}{\includegraphics[width=\linewidth]{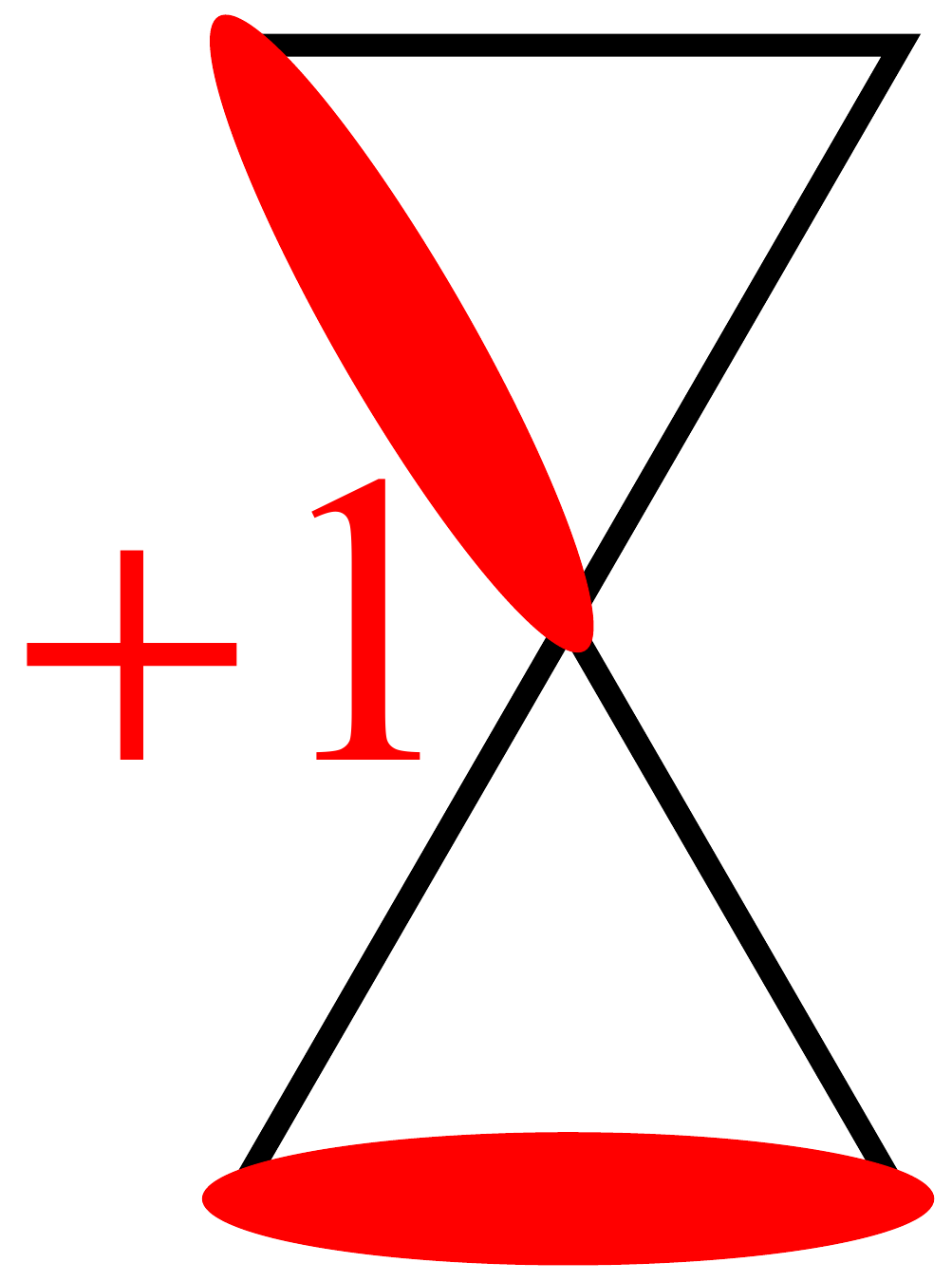}}
~~~
\parbox{0.7cm}{\includegraphics[width=\linewidth]{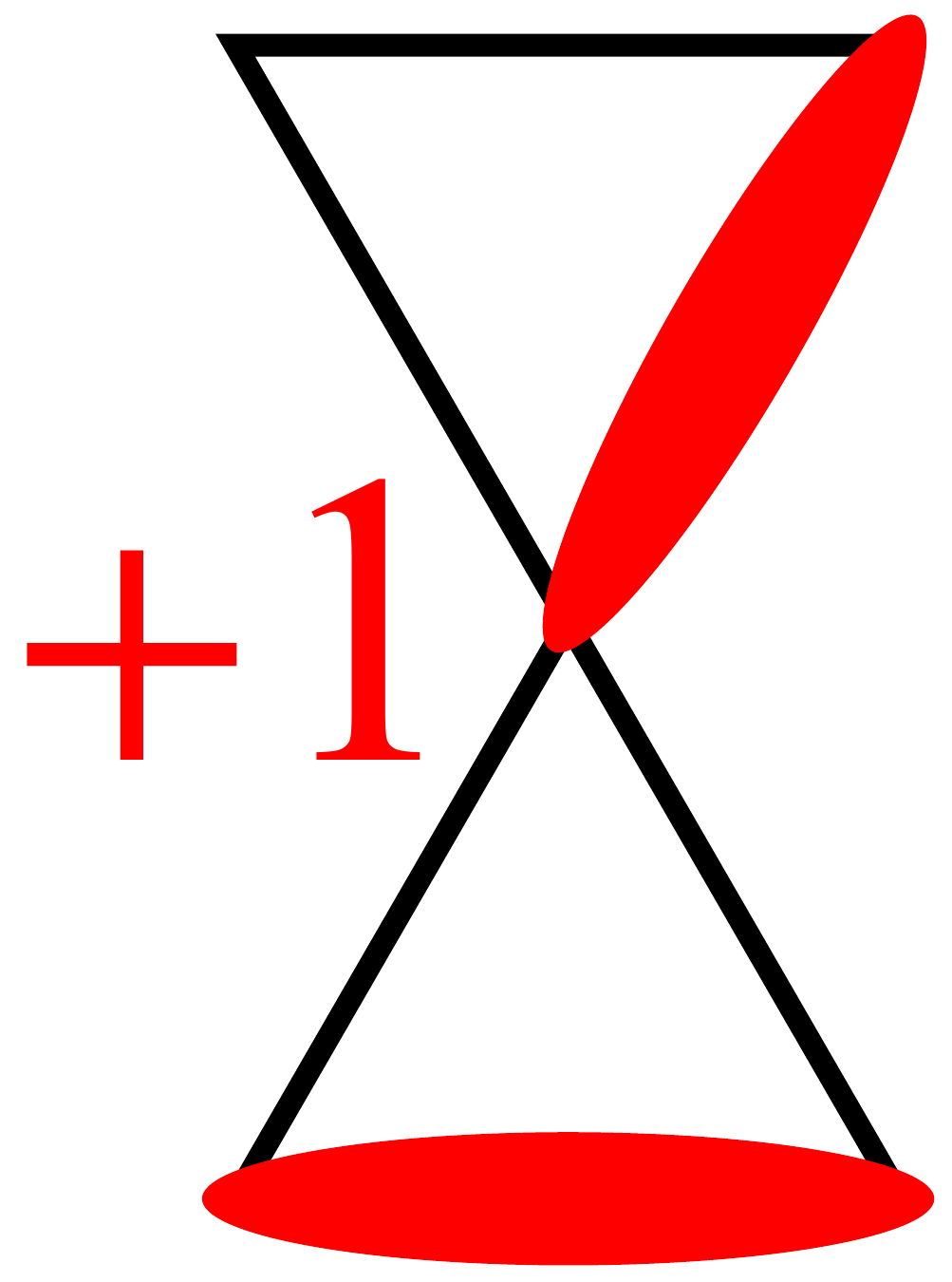}}
~~~
\parbox{0.7cm}{\includegraphics[width=\linewidth]{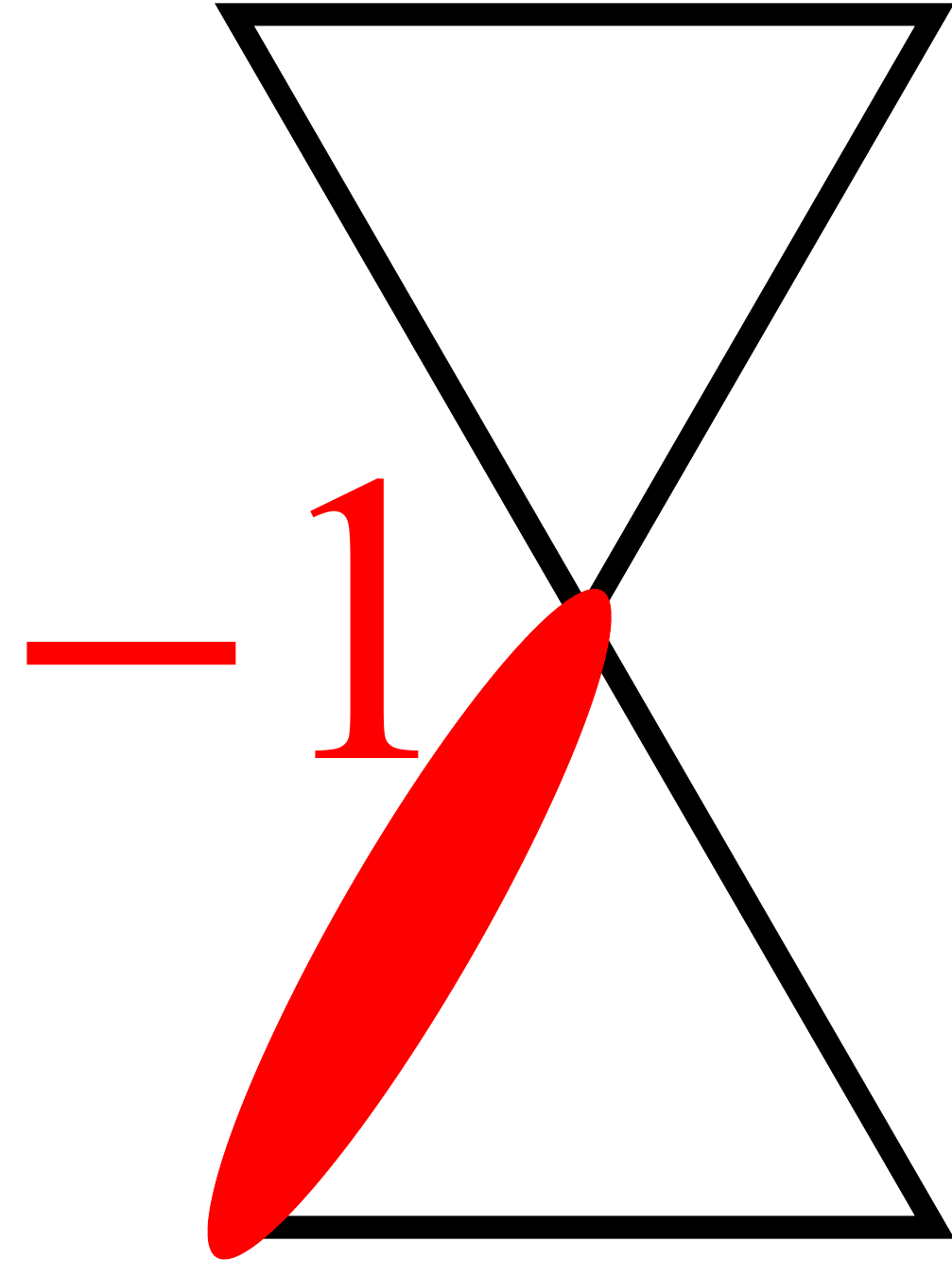}}
~~~
\parbox{0.7cm}{\includegraphics[width=\linewidth]{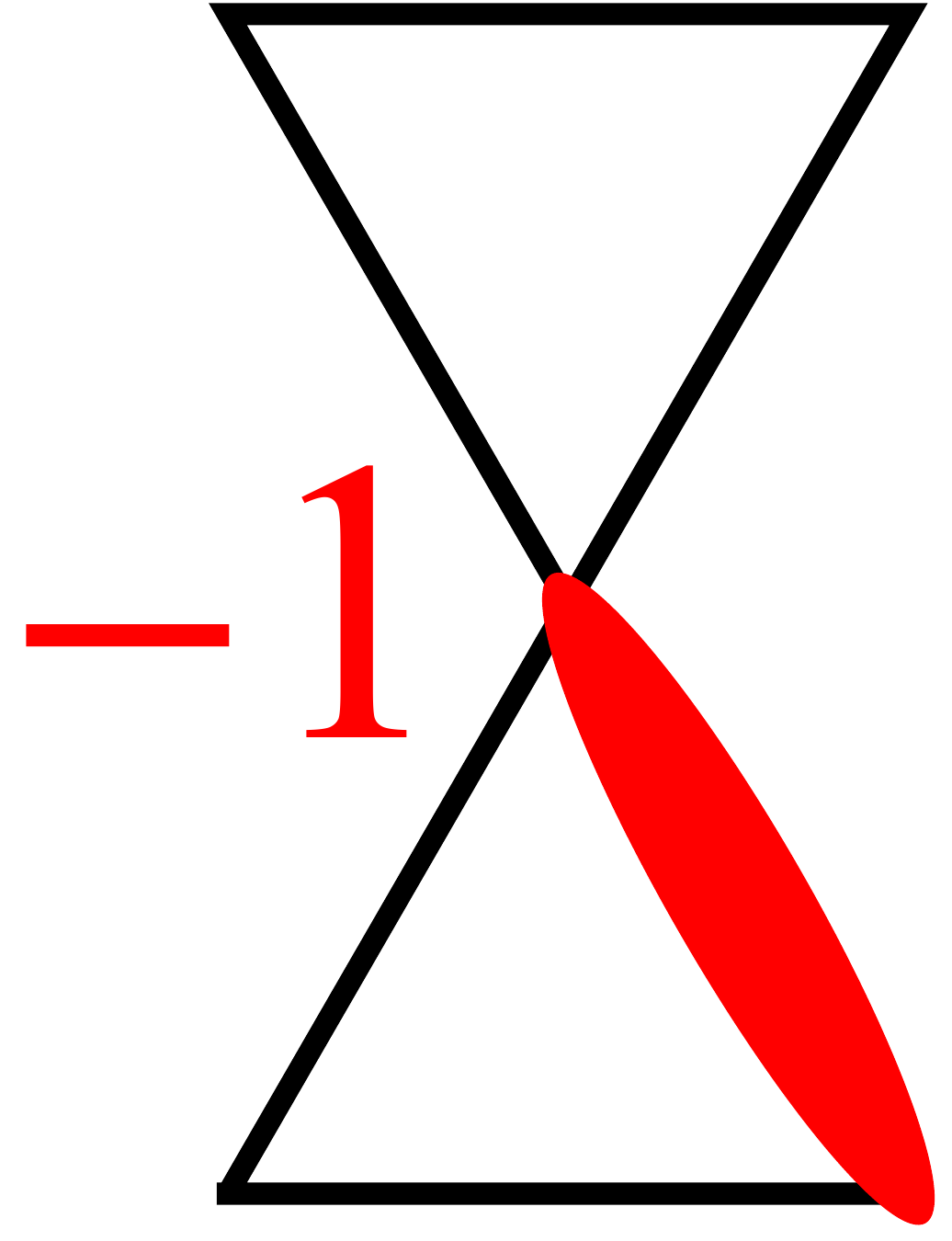}}
~~~
\parbox{0.7cm}{\includegraphics[width=\linewidth]{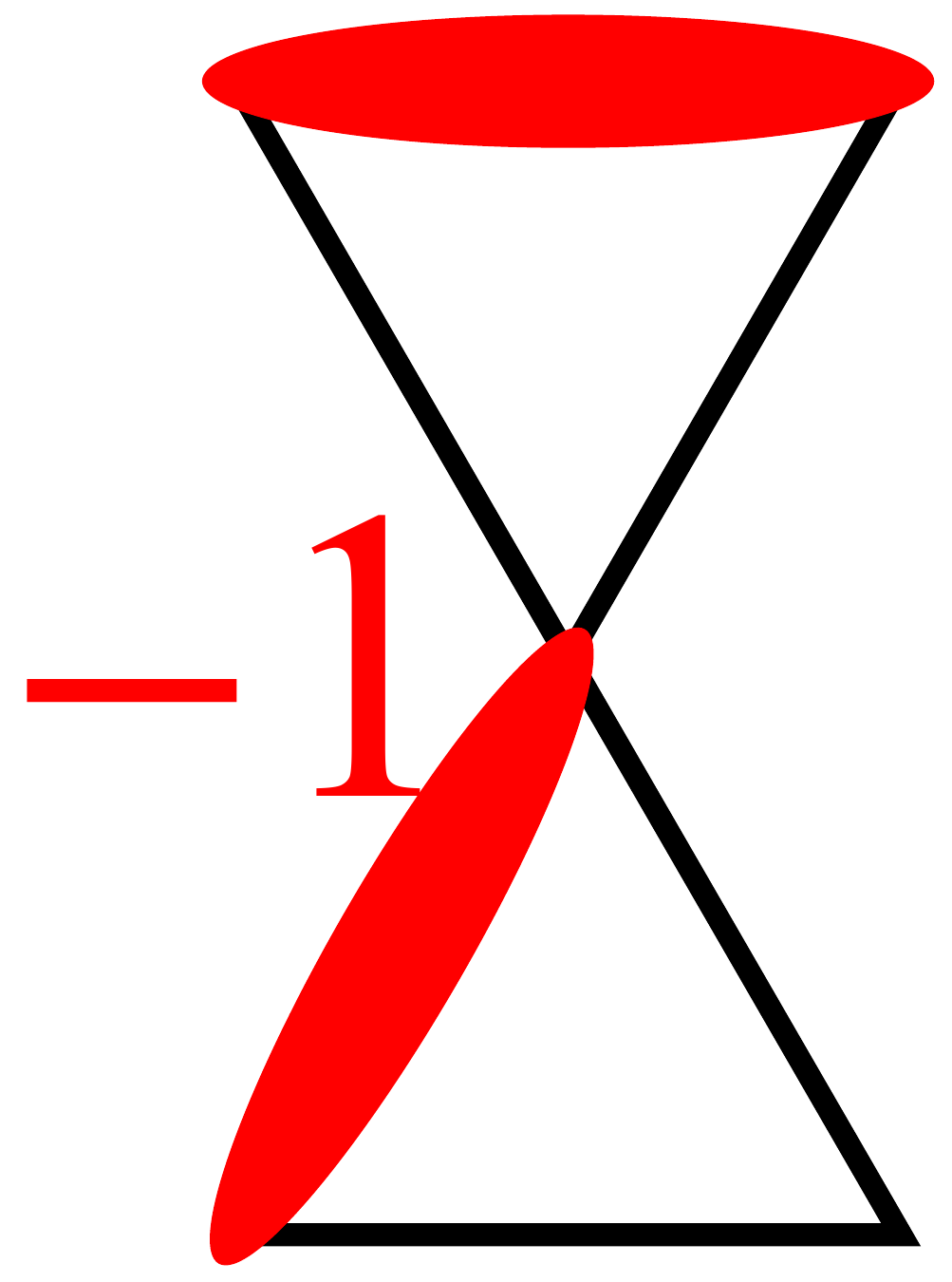}}
~~~
\parbox{0.7cm}{\includegraphics[width=\linewidth]{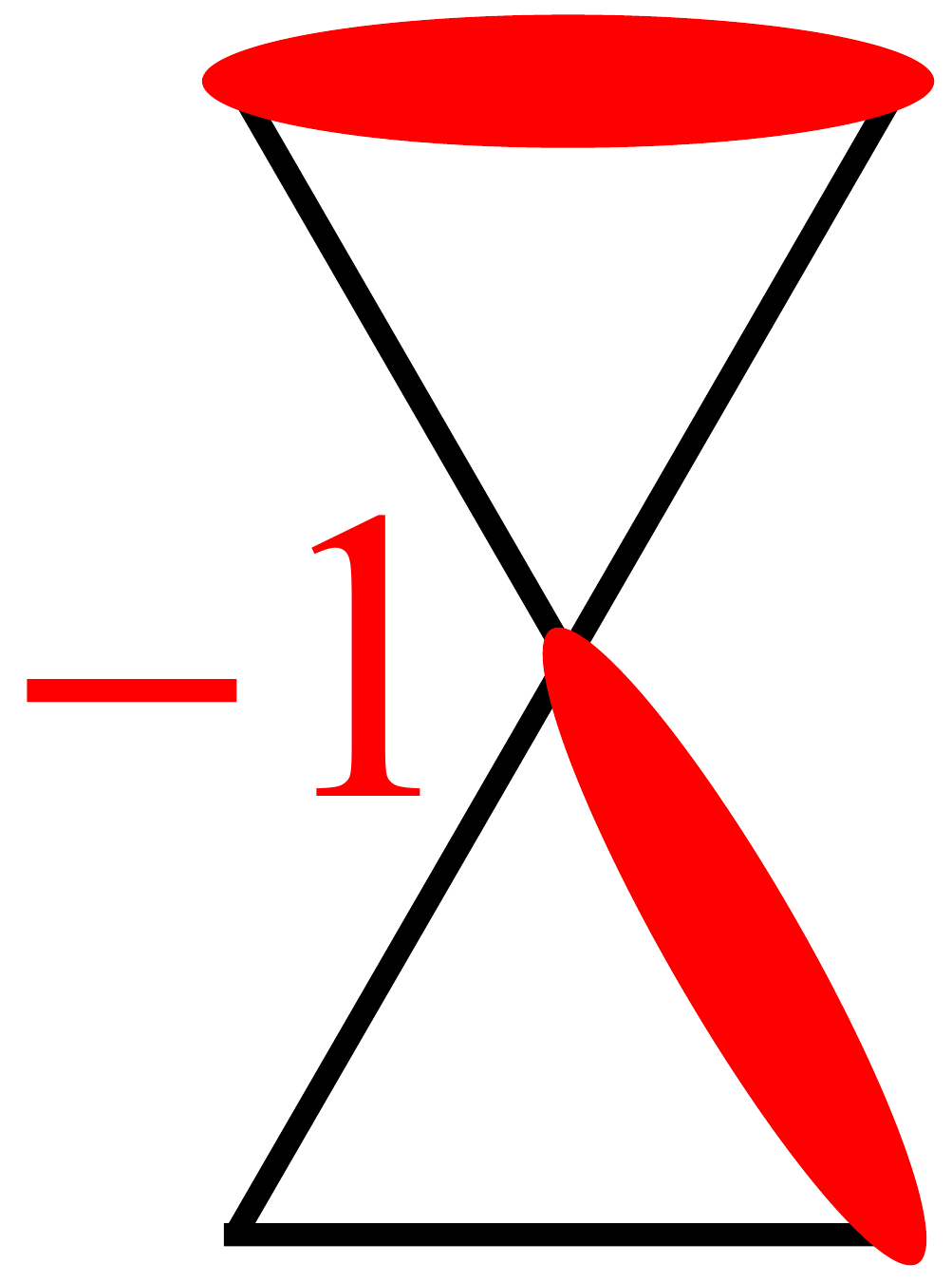}}
\end{equation}
By applying the mapping to hardcore dimer states, one can find that each local triangle has only four distinct qubit states as follows.
\begin{equation}
\parbox{7.5cm}{\includegraphics[width=\linewidth]{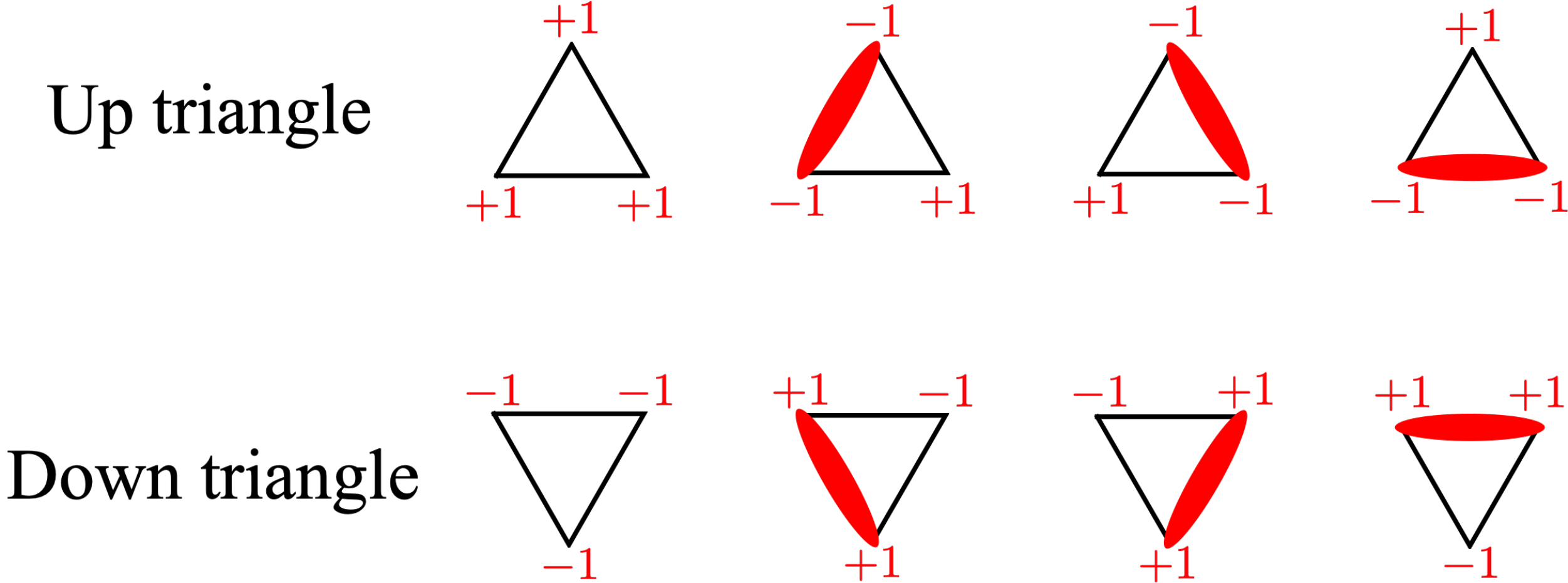}}
\label{fig:dimer-qubit-mapping}
\end{equation}
It is important to note that the qubit states of up and down triangles satisfy the condition,
\begin{equation}
Q_{ijk}
=
X_i X_j X_k
=
\left\{
\begin{array}{cc}
+1 & (\textup{up~triangle})
\\
-1 & (\textup{down~triangle})
\end{array}
\right\} ,
\label{eq:Gauss-law}
\end{equation}
where $i,j,k$ denote the three sites of a given triangle.
The above condition defines the qubit Hilbert space which corresponds to the dimer Hilbert space, and is called {\it hardcore dimer constraint} in the QDM and {\it Gauss law constraint} in the $\mathbb{Z}_2$ gauge theory description~\cite{Wan2013,Hwang2015}.
Hence, we shall call the $Q$ operator ``$\mathbb{Z}_2$-charge'' operator.
If the hardcore dimer constraint in Eq.~(\ref{eq:Gauss-law}) is violated at some local triangles, some sites of the triangles become doubly occupied or unoccupied by dimers, resulting in spinon excitations as will be discussed later.


In the qubit representation~\cite{Note1}, the QDM takes the Hamiltonian in the following form.
\begin{eqnarray}
H = 
-h \sum_p {W}_p +K \sum_{\langle \langle \langle ij \rangle \rangle \rangle} X_i X_j.
\label{eq:qubit-rep}
\end{eqnarray}
The $h$ term is now represented by the hexagon plaquette operator,
\begin{equation}
{W}_p = \prod_{i \in p} Z_i,
\end{equation}
i.e., the product of Pauli-$Z$ operators belonging to the hexagon plaquette $p$.
Following the $\mathbb{Z}_2$ gauge theory description~\cite{Wan2013,Hwang2015}, we shall often call the $W$ operator ``$\mathbb{Z}_2$-flux'' operator.
The $K$ term is simply given by third-nearest-neighbor interactions of Pauli-$X$ operators. 
See Fig.~\ref{fig:qubit-model} for a visualization of the model.
Notice that $\mathbb{Z}_2$-charge operators commute with $\mathbb{Z}_2$-flux operators and the Hamiltonian:
\begin{equation}
[Q_{ijk},{W}_p]=[Q_{ijk},H]=0.
\end{equation}
In Fig.~\ref{fig:vison-string}(a), we illustrate effects of a $\mathbb{Z}_2$-flux operator on dimer/qubit states. The operator ${W}_p$ switches the qubit state of the hexagon plaquette $p$ ($X_i=+1 \rightleftarrows X_i=-1$ for $i\in p$), reproducing the dimer resonance motions of the original QDM.
One can also check the interaction energy of the $K$ term from the examples given in the figure.

Dimer occupation can be quantified by using qubit variables.
We denote the dimer occupation at a link $ij$ by (i) $d_{ij}=0$ for the absence of a dimer and (ii) $d_{ij}=1$ for the presence of a dimer. We find that the dimer occupation can be represented by qubit variables in the following fashion.
\begin{equation}
d_{ij} = \frac{1}{4}
\left(
1
+
X_i X_j
-
X_i X_k
-
X_j X_k
\right)
\label{eq:dimer-qubit-relationship}
\end{equation}
Note that $k$ is the nearest-neighbor site of $i,j$ (i.e., $i,j,k$ form a local triangle).
One may check this relationship using the examples in Eq.~(\ref{fig:dimer-qubit-mapping}).

On the other hand, qubit operators ($X$ and $Z$) can be also understood in the dimer basis.
By using Eq.~(\ref{eq:dimer-qubit-relationship}), we obtain the relationship,
\begin{equation}
X_k Q_{ijk}
=
X_i X_j
=
(-1)^{d_{ik}+d_{jk}} ,
\label{eq:dimer-parity}
\end{equation}
which reveals that $X_k$ operator measures the dimer parity over the two links, $ik$ and $jk$, in a local triangle $ijk$.
By contrast, $Z$ operator switches the qubit state between $|X=+1\rangle$ and $|X=-1\rangle$, which is equivalent to moving a dimer from down triangle to up triangle or vice versa [Fig.~\ref{fig:vison-string}(a)].

The dimer model is exactly solved in two special cases, (i) when $K=0$ and (ii) when $h=0$. In the former case ($K=0$), a resonating valence bond spin liquid state appears as the exact ground state of the system.
In the latter case ($h=0$), different types of valence bond solid states emerge depending on the sign of $K$.

\begin{figure*}
\centering
\includegraphics[width=\linewidth]{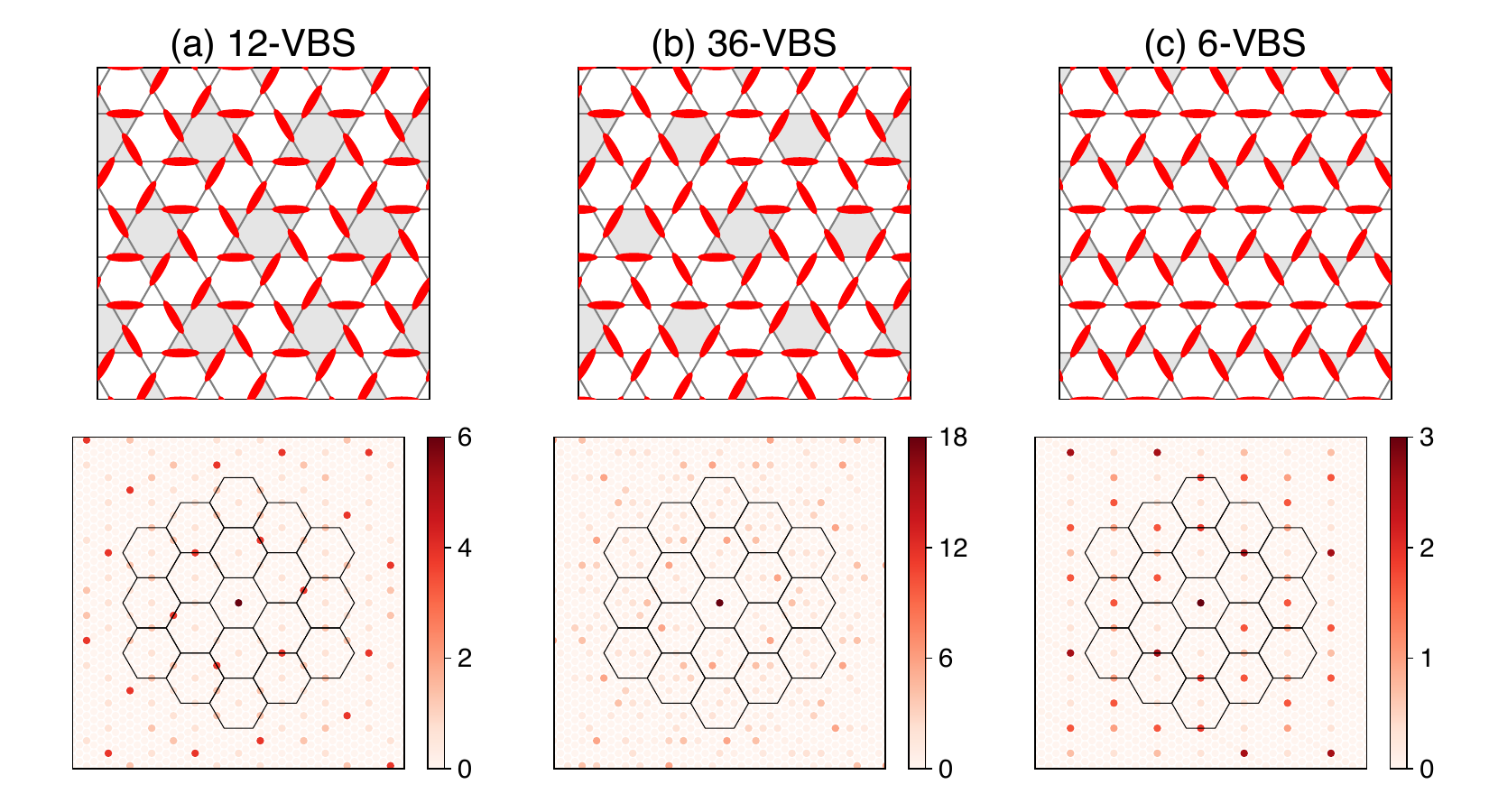}
\caption{Three types of VBS orders and the dimer structure factors.
(a) 12-site VBS state. 
(b) 36-site VBS state.
(c) 6-site VBS state.
In each case, characteristic dimer structures are highlighted by gray shade.
The bottom plots display the dimer structure factors obtained by Eq.~(\ref{eq:DSF}) only with the dimer-dimer correlator $\langle d_{ij} d_{kl} \rangle$ (we drop the other term since $\langle d_{ij} \rangle \langle d_{kl} \rangle=\langle d_{ij} d_{kl} \rangle$ in pure dimer states). The hexagons denote repeated Brillouin zones in momentum space. The central one corresponds to the first Brillouin zone.
}
\label{fig:VBS}
\end{figure*}

\subsection{RVB spin liquid and anyon excitations}

When $K=0$, the ground-state wave function is given by 
\begin{equation}
| {\rm RVB} \rangle = \mathcal{N} \prod_p \frac{1+{W}_p}{2} | \Phi \rangle ,
\end{equation}
where $| \Phi \rangle$ is an arbitrary state satisfying Eq.~(\ref{eq:Gauss-law}) and $\mathcal{N}$ is a constant for normalization. In the dimer language, the state is an equal-weight superposition of all possible hardcore dimer states, i.e., a short-ranged resonating valence bond spin liquid~\cite{Misguich2002}.
One can check that this state is equivalent to the $\mathbb{Z}_2$ toric code state on the dual honeycomb lattice~\cite{Iqbal2020}. We also remark that robustness of the kagome-lattice RVB state under perturbations generating spinon/vison excitations has been investigated using the numerical methods of tensor network states~\cite{Iqbal2020}.

The spin-liquid state is characterized by the quantum number, $W_p=+1$, at every plaquette. Visons are elementary excitations of the state carrying the quantum number, $W_p=-1$, at some local plaquettes, and they are created in pairs by the string operator,
\begin{equation}
\mathcal{V}_{\lambda} = \prod_{i \in \lambda} X_i = (-1)^{D_\lambda} .
\label{eq:vison-string-op}
\end{equation}
Here $\lambda$ is an open string passing through kagome sites and ending at hexagon plaquettes as shown in Fig.~\ref{fig:vison-string}(b).
The string operator measures the dimer parity at up triangles touched by the string $\lambda$:
\begin{equation}
D_\lambda=\sum_{i \in \lambda~{\rm or}~j \in \lambda}^{\rm up~triangles} d_{ij}.
\end{equation}
The string operator anticommutes with the $\mathbb{Z}_2$-flux operators at the ends of the string:
\begin{equation}
\mathcal{V}_\lambda {W}_p =
\left\{
\begin{array}{cc}
- {W}_p \mathcal{V}_\lambda & (p\in\partial \lambda )
\\
 {W}_p \mathcal{V}_\lambda & (\textup{otherwise})
\end{array} 
\right\} .
\end{equation}
The string operator changes the quantum number of ${W}_p$ and the energy only at the ends of the string.
Each end point carries the excitation energy, $\Delta E = 2 h$, and the $\mathbb{Z}_2$ flux, $W_p=-1$. Such point-like particles appearing at the ends of string operators are the vison excitations ($m$-anyons).
Although the visons are immobile (completely static) at $K=0$,
in general they become mobile by nonzero $K$.

Spinons are another elementary excitations in the RVB spin liquid, excited by violating the hardcore dimer constraint or the Gauss law constraint in Eq.~(\ref{eq:Gauss-law}).
They are created in pairs by the string operator,
\begin{equation}
{\mathcal S}_l = \prod_{i \in l} Z_i,
\label{eq:spinon-string}
\end{equation}
where $l$ is an open string lying on links of the kagome lattice as shown in Fig.~\ref{fig:vison-string}(c).
The string operator moves dimers along the triangles touched by the string, resulting in the violation of the hardcore dimer constraint in the two triangles located at the ends of the string.
The string operator anticommutes with the $\mathbb{Z}_2$-charge operators at the ends of the string:
\begin{equation}
\mathcal{S}_l Q_{ijk} =
\left\{
\begin{array}{cc}
- Q_{ijk} \mathcal{S}_l & (ijk:~{\rm at~the~ends~of}~l)
\\
 Q_{ijk} \mathcal{S}_l & (\textup{otherwise})
\end{array} 
\right\} .
\end{equation}
The $\mathbb{Z}_2$-charge excitations created at the ends of the string correspond to spinons ($e$-anyons).

Notice that the vison operator ($\mathcal{V}_\lambda$) and the spinon operator ($\mathcal{S}_l$) anticommute if there is an intersection between the two string operators.
This implies that there is mutual statistics between vison and spinon; specifically, the wave function undergoes an overall sign change ($|\Psi\rangle \rightarrow e^{i\pi} |\Psi\rangle$) if a spinon moves around a vison or vice versa.

\subsection{VBS orders}

When $h=0$, the Hamiltonian is diagonal in the basis of $X$ operators or in the dimer basis.
By examining the energetics of dimer states, the ground state can be identified for the two cases, $K>0$ and $K<0$~\cite{Hwang2015}.
When $K>0$, the 12-site VBS state with the characteristic pinwheel dimer structure appears as the ground state with the energy $E_{\rm gs}=-N|K|$ ($N$: number of kagome sites); see Fig.~\ref{fig:VBS}(a).
In the opposite case of $K<0$, the ground-state manifold has the energy $E_{\rm gs}=-N|K|/3$ with substantial degeneracy consisting of various states such as 36-site VBS and 6-site VBS.
The 36-site VBS is composed of the characteristic pinwheel and hexagon dimer structures [Fig.~\ref{fig:VBS}(b)].
The 6-site VBS is featured with the arrangement of parallel dimers and zigzag dimers [Fig.~\ref{fig:VBS}(c)].
For the three types of VBS orders, the predicted dimer structure factors are presented together in Fig.~\ref{fig:VBS}.

The three types of VBS orders have been studied by the author in a previous work using Ginzburg-Landau theories~\cite{Hwang2015}.
In this paper, we investigate the RVB spin liquid, the VBS orders, and their transitions by numerical exact diagonalization.

\section{Exact diagonalization\label{sec:III}}

\begin{figure}[tb]
\centering
\includegraphics[width=\linewidth]{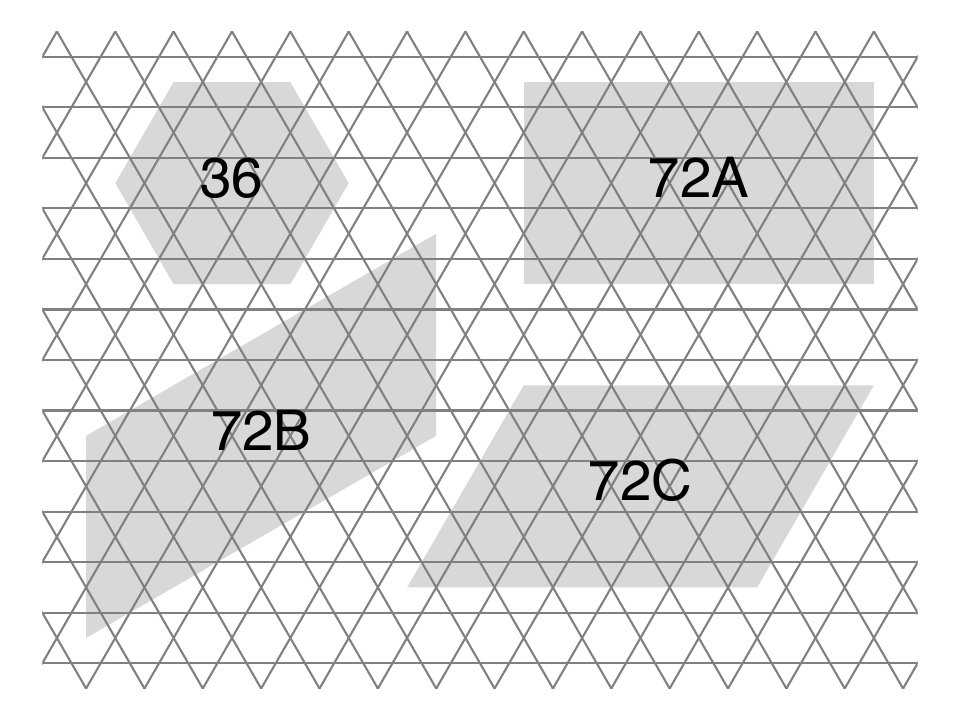}
\caption{Clusters for the exact diagonalization.}
\label{fig:ED-clusters}
\end{figure}

\begin{figure}
\centering
\includegraphics[width=0.8\linewidth]{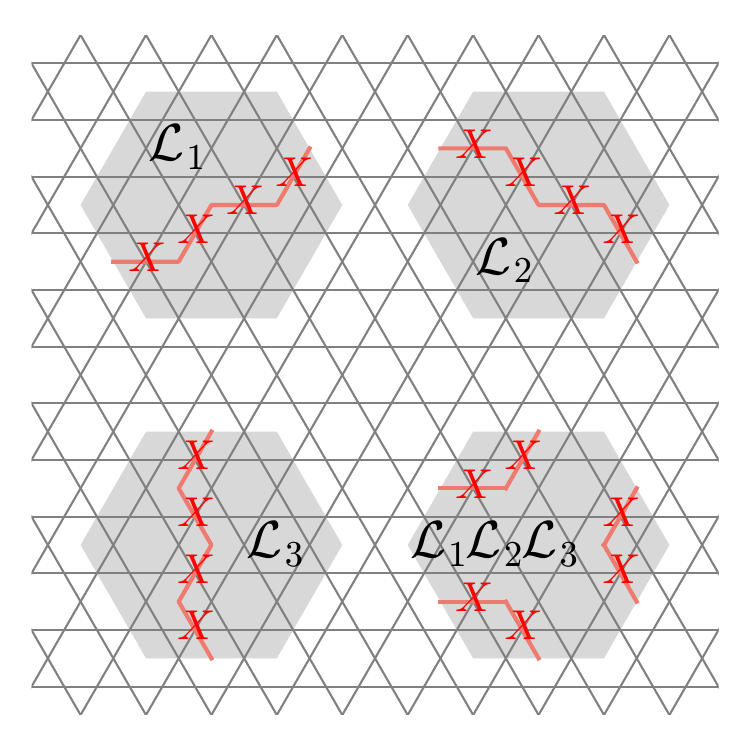}
\caption{Non-contractable loop operators ($\mathcal{L}_{1,2,3}$) of the 36-site cluster.}
\label{fig:loop-op}
\end{figure}

\begin{figure*}[tb]
\centering
\includegraphics[width=\linewidth]{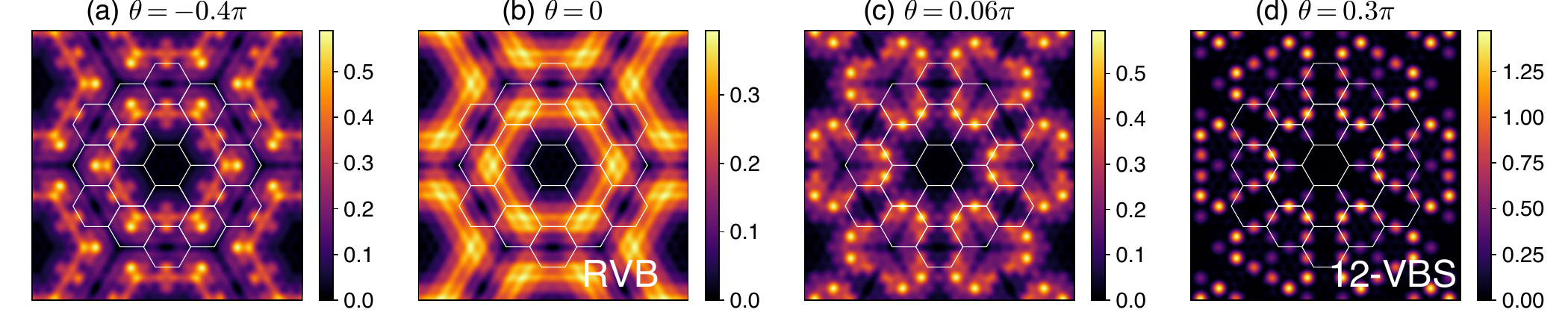}
\caption{Dimer structure factor $\mathcal{D}({\bf k})$ obtained with the 36-site cluster at (a) $\theta=-0.4\pi$, (b) $\theta=0$, (c) $\theta=0.06\pi$, and (d) $\theta=0.3\pi$. In each plot, $\mathcal{D}({\bf k})$ is presented as a color map on momentum space. The white hexagons denote repeated Brillouin zones with the central one being the first Brillouin zone.
See Fig.~\ref{fig:DSF-peak} for the strength of $\mathcal{D}({\bf k})$ at several peak positions.}
\label{fig:DSF}
\end{figure*}

The dimer model is solved by exact diagonalization (ED) on finite-size clusters. We consider a 36-site cluster and three different 72-site clusters with periodic boundary conditions shown in Fig.~\ref{fig:ED-clusters}.
For efficient ED calculations, we reduce the size of the Hilbert space by utilizing conserved quantities of the system. With periodic boundary conditions, we may define non-contractible loop operators that commute with the Hamiltonian. As illustrated in Fig.~\ref{fig:loop-op}, for the 36-site cluster there are three different non-contractible loop operators:
\begin{equation}
\mathcal{L}_1=\prod_{i \in \Lambda_1} X_i,
~~~
\mathcal{L}_2=\prod_{i \in \Lambda_2} X_i,
~~~
\mathcal{L}_3=\prod_{i \in \Lambda_3} X_i.
\end{equation}
In fact, among $\mathcal{L}_{1,2,3}$ only two are independent due to the identity,
\begin{equation}
\mathcal{L}_1\mathcal{L}_2\mathcal{L}_3
=
\prod_{ijk\in\mathcal{C}} Q_{ijk},
\end{equation}
i.e., the product of the loop operators is identical to the product of $Q$-operators inside the closed path ($\mathcal{C}$) formed by the three non-contractible loops ($\Lambda_{1,2,3}$).
Since the loop operators commute with the Hamiltonian ($[\mathcal{L}_n,{W}_p]=[\mathcal{L}_n,H]=0$), the Hilbert space is partitionized into four distinct topological sectors labeled by the eigenvalues of the loop operators, $\mathcal{L}_{1}=\pm1$ and $\mathcal{L}_{2}=\pm1$.
In the case of the 36-site cluster, each topological sector has the dimension,
\begin{equation}
\frac{2^{36}}{2^{23} \times 2^{2}}=2^{11},
\end{equation}
where $2^{23}$ comes from the hardcore dimer constraints [Eq.~(\ref{eq:Gauss-law})] with periodic boundary conditions and $2^2$ counts the four topological sectors.
For 72-site clusters, each topological sector has the dimension,
\begin{equation}
\frac{2^{72}}{2^{47} \times 2^{2}}=2^{23}.
\end{equation}
We run our ED calculations in each of the four topological sectors for a given cluster.

Figure~\ref{fig:ED-results} displays the ED results on the 36-site cluster.
We identify two phases, the RVB spin liquid and the 12-site VBS, separated by a continuous transition at $\theta_{\rm c}\simeq0.06\pi$ as indicated by the second derivative of the ground-state energy, $-\partial^2 E_{\rm gs}/\partial \theta^2$ [Fig.~\ref{fig:ED-results}(b)].
It is observed that the $\mathbb{Z}_2$-flux expectation value $\langle {W}_p \rangle$ remains pretty large in the RVB spin liquid whereas it is substantially suppressed in the 12-site VBS [Fig.~\ref{fig:ED-results}(c)].
The transition is also verified through the energy spectrum.
Around the point $\theta=0$, the system shows fourfold ground-state degeneracy below the two-vison excitation gap, $\Delta E = 4h$, which is exactly the topological degeneracy expected for the $\mathbb{Z}_2$ spin liquid.
Across the transition point $\theta_{\rm c}\simeq0.06\pi$, the ground-state degeneracy is fully lifted and the lowest energy levels of vison excitations come down close to the ground state; see the right arrow in Fig.~\ref{fig:ED-results}(a). 

The nature of the two phases and their transition are further clarified via the investigation of vison condensation and dimer structure factor.

\subsection{Vison condensation}

We measure vison condensation by calculating the expectation value of vison string operator, $\langle \mathcal{V}_{\lambda} \rangle$, as shown in Fig.~\ref{fig:ED-results}(d).
The RVB spin liquid phase has relatively small values of $\langle \mathcal{V}_{\lambda} \rangle$ whereas the VBS phase is characterized by substantially large values ($\langle \mathcal{V}_{\lambda} \rangle\approx -1$), which confirms that the VBS phase is indeed a vison-condensed state.

\begin{figure}[b]
\centering
\includegraphics[width=\linewidth]{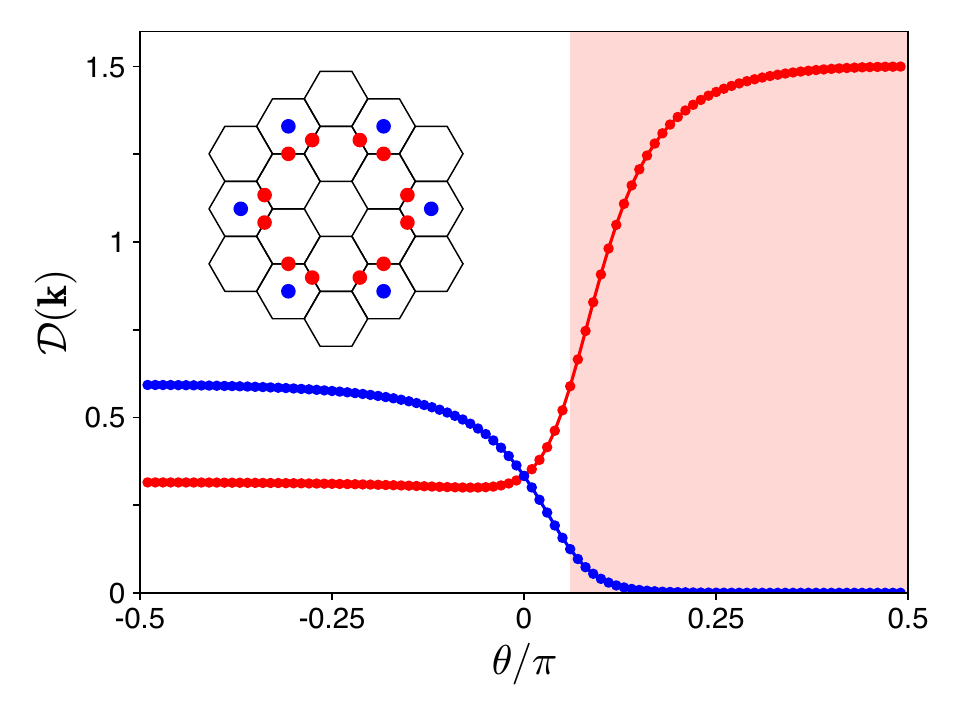}
\caption{Strength of the dimer structure factor $\mathcal{D}({\bf k})$ at ${\bf k}$ points where peak structures appear. The inset depicts the peak positions over the repeated Brillouin zones. Red (blue) indicates the largest peak of DSF appearing on positive (negative) $\theta$.
}
\label{fig:DSF-peak}
\end{figure}

\subsection{Dimer structure factor}

Structure factors are useful quantities in identifying symmetry-broken long-range orders and also quantum spin liquids~\cite{Messio2010,Dodds2013,Sachdev2014,Messio2017,Halimeh2016,Sheng2019,Halimeh2019,Meng2018BFG,Wessel2018BFG}.
We determine the dimer ordering pattern of the VBS phase by investigating the dimer structure factor (DSF),
\begin{equation}
\mathcal{D}({\bf k}) = \frac{1}{N_{\rm d}} \sum_{ij,kl} e^{i{\bf k}\cdot({\bf r}_{ij}-{\bf r}_{kl})} \left( \langle d_{ij} d_{kl} \rangle -  \langle d_{ij} \rangle \langle d_{kl} \rangle \right) ,
\label{eq:DSF}
\end{equation}
where $ij~\&~kl$ denote nearest-neighbor bonds (dimers), ${\bf r}_{ij}~\&~{\bf r}_{kl}$ are their position vectors, and ${\bf k}$ is momentum ($N_{\rm d}$: number of dimers).
The calculated dimer structure factor is displayed in Fig.~\ref{fig:DSF}.
The RVB spin liquid exhibits broad features in the DSF due to the absence of dimer ordering [Fig.~\ref{fig:DSF}(b)].
As the system enters the VBS phase across the transition point ($\theta_{\rm c}\simeq 0.06 \pi$), sharp peak structures are developed in the DSF [Figs.~\ref{fig:DSF}(c)~and~\ref{fig:DSF}(d)].
By comparing Fig.~\ref{fig:DSF}(d) with Fig.~\ref{fig:VBS}, we identify that the 12-site VBS emerges via the vison condensation transition~\cite{Note2}.


\section{Spinon confinement\label{sec:IV}}

In general, anyon condensation gives rise to novel phenomena on other anyons called confinement effects. If an anyon has nontrivial braiding with the condensed anyon, then the (uncondensed) anyon becomes confined, i.e., the anyon cannot be isolated/observed as a single particle in the low-energy physics of the condensed phase~\cite{Bais2009,Burnell2018,Hwang2024}. In our system, the vison condensation generates such confinement effects on the two spinons ($e$- and $\psi$-anyons).

The spinon confinement can be understood in several ways.
In the perspective of anyon theory, bosonic spinon and fermionic spinon become identical particles ($e=e\times m=\psi$) under the vison condensation ($\langle m \rangle \ne 0$) since we can freely take visons out of the vacuum.
However, $e$- and $\psi$-anyons have different topological spins due to the nontrivial braiding between $e$- and $m$-anyons; the $e$-anyon is a self-boson but the $\psi$-anyon is a self-fermion.
For the condensed phase to have a self-consistent anyon theory, $e$- and $\psi$-anyons should not appear as deconfined anyons in the low-energy physics of the condensed phase, i.e., $e$- and $\psi$-anyons are confined by the condensation of the $m$-anyon.

At the level of a microscopic model, the confinement effects can be understood by using vison and spinon string operators.
Nontrivial braiding between vison and spinon is encoded in the string operators, $\mathcal{V}_\lambda$ and $\mathcal{S}_l$.
When there is a crossing point between the two string operators, the two operators have the anticommutation relation, $\mathcal{V}_\lambda \mathcal{S}_l = - \mathcal{S}_l \mathcal{V}_\lambda$ (since the two operators are defined by the conjugate variables, $X$ and $Z$), which implies the existence of nontrivial braiding between vison and spinon.
Dimer interpretations of the string operators offer an intuitive picture about spinon confinement.
First, we note that vison condensation sets a specific dimer pattern in the system. The spinon string operator creates a pair of spinons by moving dimers between adjacent up and down triangles along the string, obviously leading to an energy cost proportional to the length of the string~[Eq.~(\ref{eq:spinon-string}) and Fig.~\ref{fig:vison-string}(c)].
Therefore, spinons are confined due to the linearly increasing energy cost (confining potential).

\begin{figure}[tb]
\centering
\includegraphics[width=\linewidth]{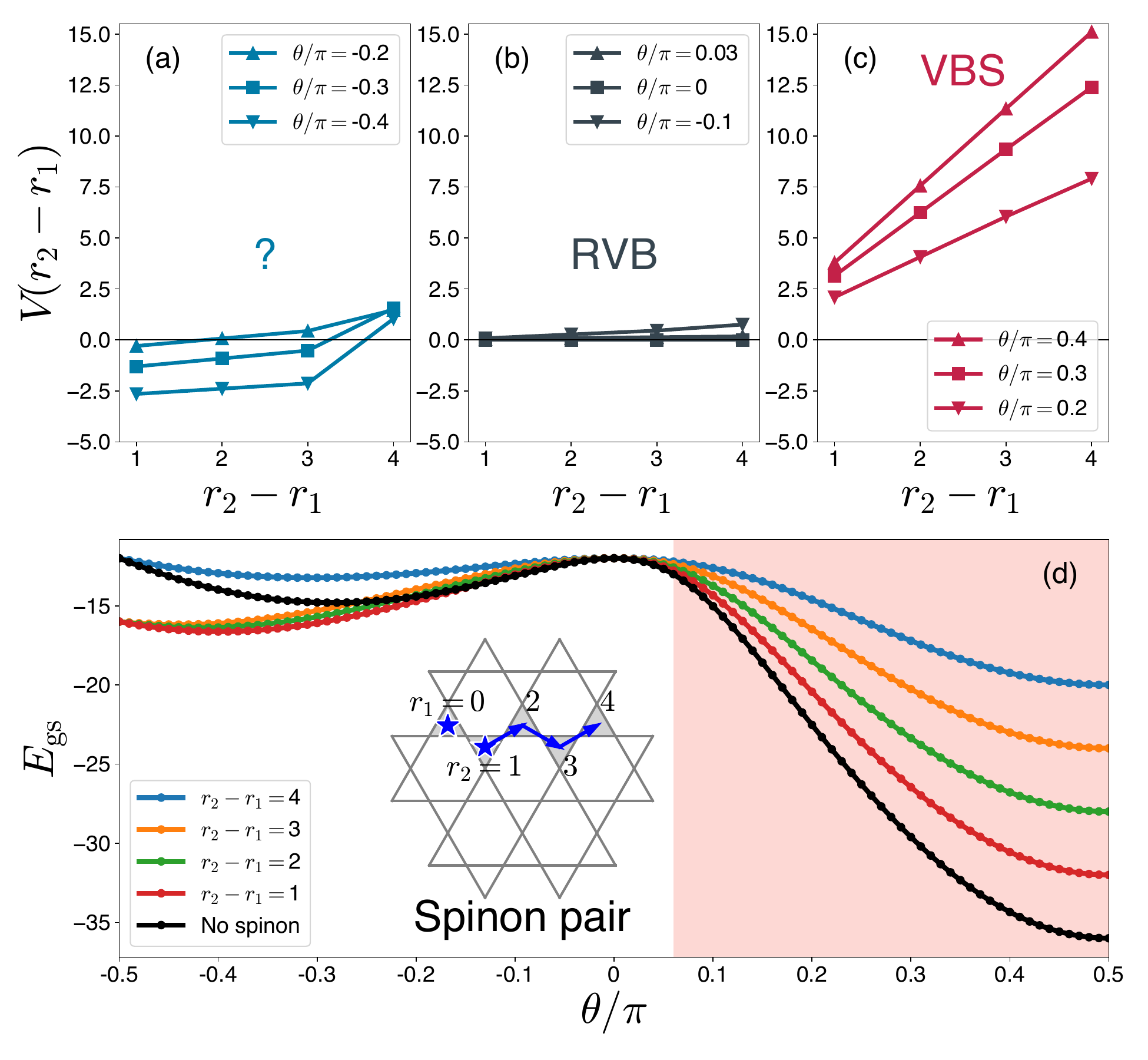}
\caption{Interaction energy of the inserted two spinons.
(a),(b),(c) The minimum excitation energy, $V(r_2-r_1)$, of the two-spinon state as a function of the separation length, $r_2-r_1$.
(d) Energy profile of the lowest two-spinon state for various separations ($r_2-r_1=1,2,3,4$). For comparison, the ground-state energy (with no spinon) is plotted together (black). The inset depicts the locations of the spinon pair.
One spinon is fixed at $r_1=0$ and the other spinon is moved over different locations ($r_2=1,2,3,4$).
The results are obtained with the 36-site cluster.
}
\label{fig:spinon-confinement}
\end{figure}

We confirm the spinon confinement in our numerical calculations.
Figure~\ref{fig:spinon-confinement} displays the energy cost of creating a spinon pair as a function of distance.
In these calculations, a pair of spinons are inserted to the system by switching two $\mathbb{Z}_2$-charge eigenvalues in the original Gauss law constraint [Eq.~(\ref{eq:Gauss-law})]: 
\begin{equation}
Q_{ijk}=+1 \rightleftarrows Q_{ijk}=-1~~~{\rm at}~~~r_1~{\rm and}~r_2.
\end{equation}
We fix a $\mathbb{Z}_2$ charge at $r_1=0$ and change the location of the other $\mathbb{Z}_2$ charge from $r_2=1$ to $r_2=4$ [see the inset of Fig.~\ref{fig:spinon-confinement}(d)].
For each spinon-pair configuration, we obtain the lowest energy level, $E_{\rm gs}({r_2-r_1})$, as shown in Fig.~\ref{fig:spinon-confinement}(d).
Then, we calculate the spinon-pair excitation energy with respect to the ground state with no spinon, 
\begin{equation}
V(r_2-r_1)=E_{\rm gs}({r_2-r_1})-E_{\rm gs}({\rm no~spinon}). 
\end{equation}
In the RVB phase, the excitation energy is almost constant regardless of the separation [Fig.~\ref{fig:spinon-confinement}(b)].
In stark contrast, in the VBS phase, the excitation energy increases proportional to the separation length [Fig.~\ref{fig:spinon-confinement}(c)].
As the system enters deep inside of the VBS phase, the excitation energy more rapidly increases upon separating two spinons further.
This result clearly demonstrates the phenomena of spinon confinement in the VBS phase.

\begin{figure*}[tb]
\centering
\includegraphics[width=\linewidth]{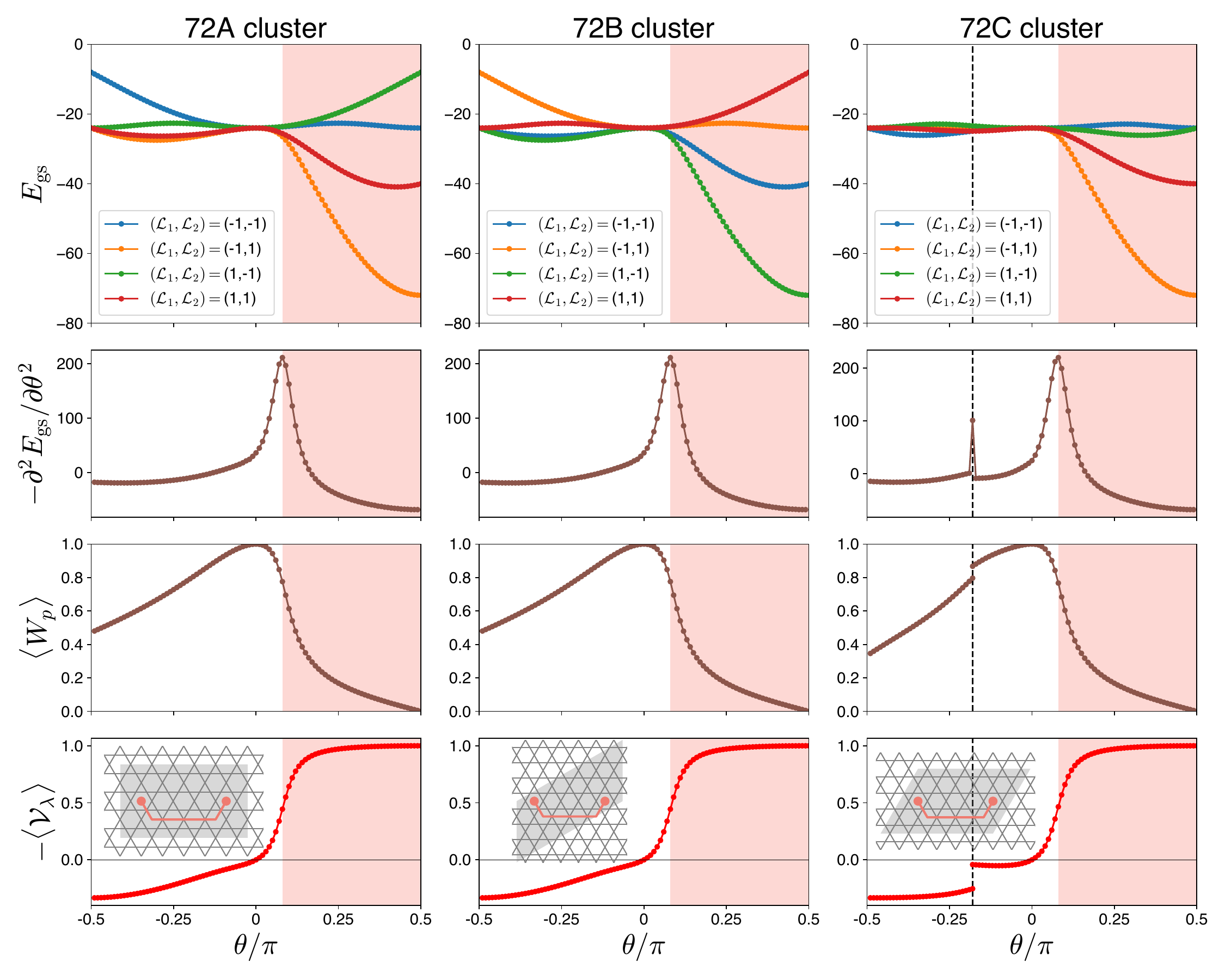}
\caption{ED results on the 72-site clusters. 
Left: 72A cluster. Middle: 72B cluster. Right: 72C cluster.
Each column shows the lowest energy levels ($E_{\rm gs}$) of the four topological sectors $(\mathcal{L}_1=\pm1,\mathcal{L}_2=\pm1)$, the ground-state energy second derivative ($-\partial^2 E_{\rm gs}/ \partial \theta^2$), the $\mathbb{Z}_2$-flux expectation value ($\langle W_p \rangle$), and the vison condensation ($-\langle \mathcal{V}_\lambda \rangle$) calculated with the vison string operator depicted in the inset. 
In the case of the 72C cluster, there is a level crossing transition at $\theta\simeq-0.18\pi$ (marked by dashed line).
}
\label{fig:ED-72}
\end{figure*}

\section{Discussion\label{sec:V}}

\begin{figure*}[tb]
\centering
\includegraphics[width=\linewidth]{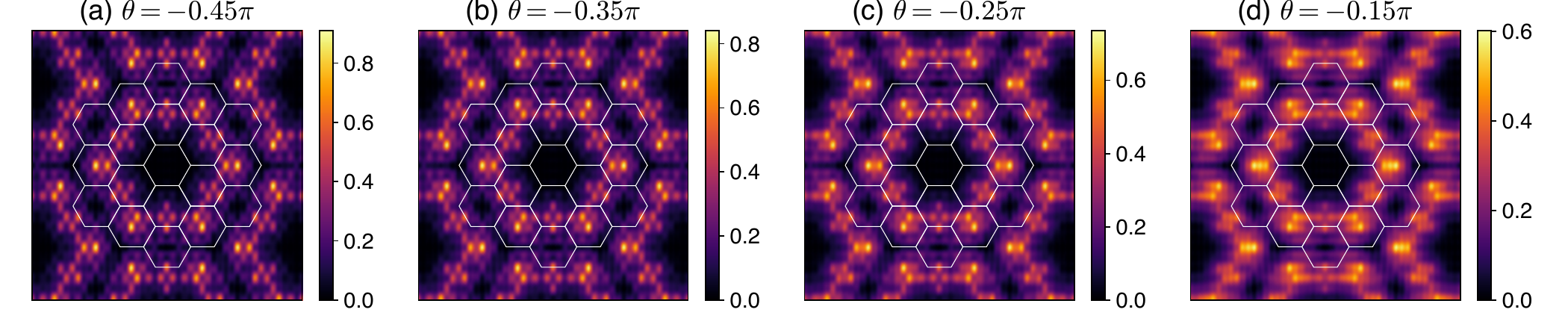}
\caption{Dimer structure factor obtained with the 72A cluster at (a) $\theta=-0.45\pi$, (b) $\theta=-0.35\pi$, (c) $\theta=-0.25\pi$, and (d) $\theta=-0.15\pi$. In each plot, $\mathcal{D}({\bf k})$ is presented as a color map on momentum space. The white hexagons denote repeated Brillouin zones with the central one being the first Brillouin zone.
}
\label{fig:DSF-72A}
\end{figure*}

We identified the anyon physics of vison condensation and spinon confinement occurring in the RVB-to-VBS transition of the dimer model.
Until this point, we have focused on positive $\theta$ in the parameter space.
We now turn our attention to negative $\theta$ and discuss some unexpected behaviors observed in this dimer model.

\subsection{Mixed behaviors in negative $\theta$}

When the $K$ term (dimer interaction) becomes dominating over the $h$ term (dimer motion), we generally expect a transition from the RVB state to a VBS state.
This is what happens on the positive side of $\theta$ as already discussed in previous sections.
However, the expectation seems betrayed on the negative side of $\theta$. 
In the parameter region of $\theta \lesssim -0.15\pi$, features of RVB and VBS states are simultaneously observed as we summarize below.
\begin{itemize}
\item RVB features: We do not see any clear sign of transition on negative $\theta$ in Fig.~\ref{fig:ED-results}(b). Moreover, the $\mathbb{Z}_2$ flux is substantially large over the entire region of negative $\theta$ as shown in Fig.~\ref{fig:ED-results}(c). Broad features are observed in the dimer structure factor [Fig.~\ref{fig:DSF}(a)].
\item VBS features: Topological degeneracy of the RVB state is lifted near $\theta\simeq-0.15\pi$; marked by the left arrow in Fig.~\ref{fig:ED-results}(a). The vison condensation has non-negligible strengths over the entire region of negative $\theta$ as shown in Fig.~\ref{fig:ED-results}(d). Also, peak structures are observed in the dimer structure factor [Figs.~\ref{fig:DSF}(a) and \ref{fig:DSF-peak}].
\end{itemize}
In addition to the above features, the spinon excitation energy becomes negative when $\theta \lesssim -0.15\pi$ [Figs.~\ref{fig:spinon-confinement}(a)~and~\ref{fig:spinon-confinement}(d)], suggesting that the nature of the RVB state is changing across $\theta\simeq-0.15\pi$.

\subsection{Results on 72-site clusters}

\begin{figure}[tb]
\centering
\includegraphics[width=\linewidth]{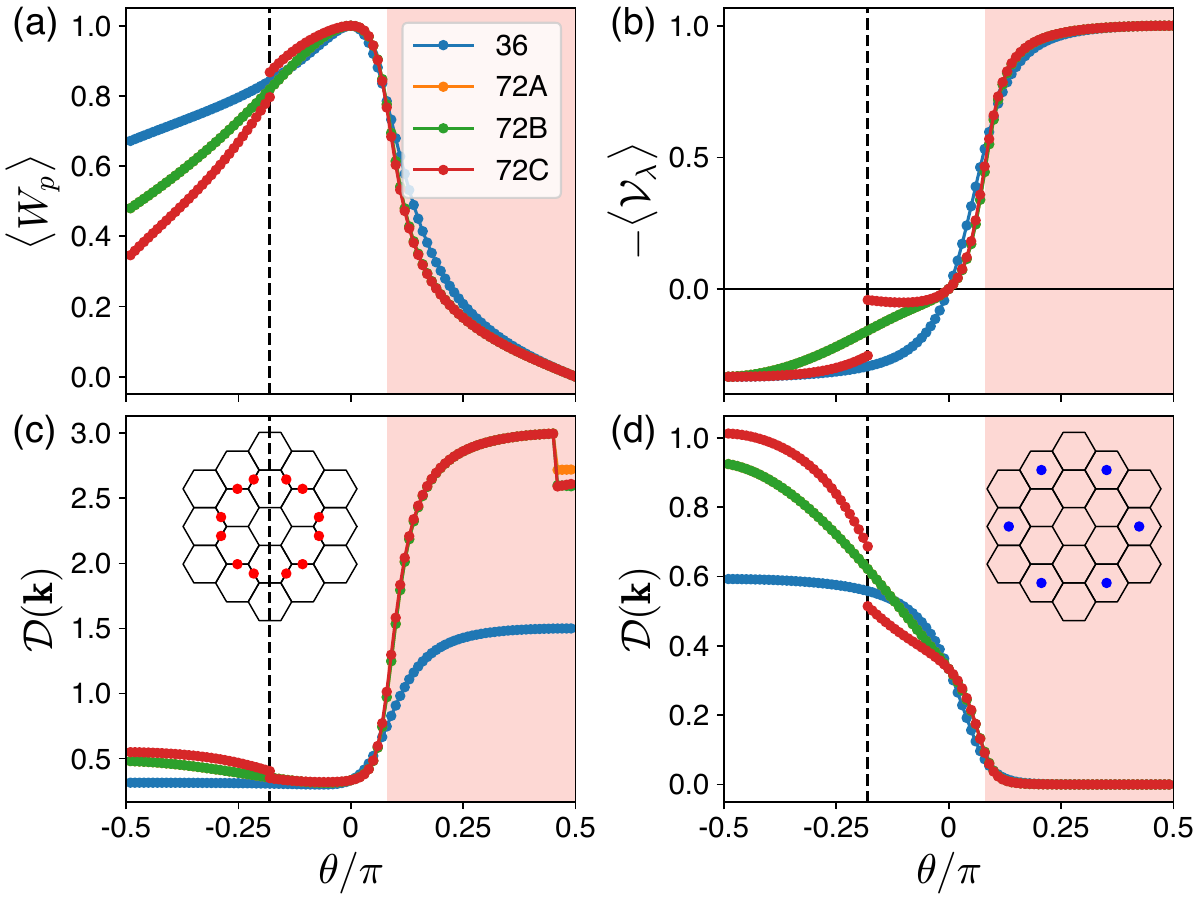}
\caption{Comparison of all the ED results obtained with the 36, 72A, 72B, and 72C clusters.
(a) The $\mathbb{Z}_2$-flux expectation value ($\langle W_p \rangle$). 
(b) The vison condensation ($-\langle \mathcal{V}_\lambda \rangle$).
(c),(d) Strength of the dimer structure factor $\mathcal{D}({\bf k})$ at several peak positions. 
Identical results are obtained for the 72A and 72B clusters, hence the 72A cluster results are overlayed by the 72B cluster results in each plot. 
In the panel (c), 72-site clusters show abrupt changes in $\mathcal{D}({\bf k})$ around $\theta=0.46\pi$, which are caused by ground state degeneracies.
The dashed line marks the level crossing point ($\theta\simeq-0.18\pi$)
observed in the 72C cluster.
}
\label{fig:ED-all-clusters}
\end{figure}

To better understand the mixed behaviors shown in the 36-site cluster, we perform ED calculations on larger clusters of 72 sites.
Overall, we find similar behaviors in the 72-site clusters as summarized in Fig.~\ref{fig:ED-72}.
In the 72C cluster, a level crossing transition occurs at $\theta\simeq-0.18\pi$ (third column of Fig.~\ref{fig:ED-72}).
We find that VBS features become slightly more prominent than RVB features in negative $\theta$ of 72-site clusters (Figs.~\ref{fig:DSF-72A}~and~\ref{fig:ED-all-clusters}).
To be specific, suppressed $\mathbb{Z}_2$-flux values [Fig.~\ref{fig:ED-all-clusters}(a)] and enhanced DSF peak strengths [Figs.~\ref{fig:DSF-72A} and~\ref{fig:ED-all-clusters}(d)] are observed in 72-site clusters compared to the 36-site cluster.
Nonetheless, the $\mathbb{Z}_2$-flux values are still non-negligible ($\langle W_p \rangle \gtrsim 0.4$) in negative $\theta$ and there is no clear sign of transition in the 72A and 72B clusters. 
It is unclear even on 72-site clusters what is the true nature of the state in negative $\theta$.

\subsection{Three possibilities}

The mixed behaviors in negative $\theta$ are attributed to the dimer interactions which allow extensive ground-state degeneracy when $K<0$.
As illustrated in Fig.~\ref{fig:ED-results}(a), lots of states come down to the ground state as we approach the $\theta=-0.5\pi$ point (in sharp contrast to the energy spectrum at $\theta=0.5\pi$).
Considering the extensive degeneracy at $\theta=-0.5\pi$, we may think of three possibilities about the phase appearing on negative $\theta$ in thermodynamic limit.
\begin{itemize}
\item VBS order: Out of the extensively degenerate ground-state manifold at $\theta=-0.5\pi$, a VBS state can be selected by the so-called quantum order-by-disorder effect. Perturbative quantum dimer resonance motions lift the degeneracy by generating a kind of zero-point energy in the ground-state manifold.
\item Spin liquid: Dimer resonance motions can mix the degenerate dimer states, yielding some kind of RVB state. But this state is expected to be somewhat different from the RVB state near $\theta=0$ due to the difference in the participating dimer states.
\item Mixture of VBS order and spin liquid: Based on our numerical results, another plausible state is a spin liquid with broken lattice symmetries. This state can be viewed as some mixture of VBS and spin-liquid states, which may exhibit both features of VBS and spin liquid as we have seen in our results.
\end{itemize}
The last possibility is particularly interesting since such a spin-liquid state with partial symmetry breaking is quite rare and usually not expected to occur in dimer models.
Clarifying the true nature of the system requires additional large-scale numerical simulations.
We note that large-scale simulations of quantum dimer models have been performed in several studies by using quantum Monte Carlo (QMC) techniques~\cite{Mila2007,Hao2014,Ralko2018,Meng2022QDM}. We hope to elucidate the mixed behaviors shown in negative $\theta$ via QMC or DMRG simulations in future.

\section{Conclusions\label{sec:VI}}

In this work, we studied a quantum dimer model that implements a transition from a RVB-type $\mathbb{Z}_2$ spin liquid to a valence bond solid on the kagome lattice.
The RVB state supports two types of anyon excitations, vison and spinon, whereas the VBS state has a crystalline order of dimers but no anyon excitation.
By using numerical exact diagonalization methods, we identified the RVB-to-VBS transition mechanism, which is represented by the two anyon phenomena, vison condensation and spinon confinement.
We further clarified the dimer ordering pattern of the VBS state by investigating dimer structure factor.
On the other hand, we found mixed behaviors of spin liquid and VBS states in a certain parameter region of the dimer model, suggesting a possibility of a new phase different from conventional phases of dimer models.
For the clarification of this matter, it is necessary to perform large-scale numerical computations.
We leave this nontrivial problem for future study.

\acknowledgements
This work was supported by a KIAS Individual Grant (No. PG071403) at Korea Institute for Advanced Study (KIAS).
Computations were performed on clusters at the Center for Advanced Computation (CAC) of KIAS.


\begin{thebibliography}{75}%
\makeatletter
\providecommand \@ifxundefined [1]{%
 \@ifx{#1\undefined}
}%
\providecommand \@ifnum [1]{%
 \ifnum #1\expandafter \@firstoftwo
 \else \expandafter \@secondoftwo
 \fi
}%
\providecommand \@ifx [1]{%
 \ifx #1\expandafter \@firstoftwo
 \else \expandafter \@secondoftwo
 \fi
}%
\providecommand \natexlab [1]{#1}%
\providecommand \enquote  [1]{``#1''}%
\providecommand \bibnamefont  [1]{#1}%
\providecommand \bibfnamefont [1]{#1}%
\providecommand \citenamefont [1]{#1}%
\providecommand \href@noop [0]{\@secondoftwo}%
\providecommand \href [0]{\begingroup \@sanitize@url \@href}%
\providecommand \@href[1]{\@@startlink{#1}\@@href}%
\providecommand \@@href[1]{\endgroup#1\@@endlink}%
\providecommand \@sanitize@url [0]{\catcode `\\12\catcode `\$12\catcode
  `\&12\catcode `\#12\catcode `\^12\catcode `\_12\catcode `\%12\relax}%
\providecommand \@@startlink[1]{}%
\providecommand \@@endlink[0]{}%
\providecommand \url  [0]{\begingroup\@sanitize@url \@url }%
\providecommand \@url [1]{\endgroup\@href {#1}{\urlprefix }}%
\providecommand \urlprefix  [0]{URL }%
\providecommand \Eprint [0]{\href }%
\providecommand \doibase [0]{https://doi.org/}%
\providecommand \selectlanguage [0]{\@gobble}%
\providecommand \bibinfo  [0]{\@secondoftwo}%
\providecommand \bibfield  [0]{\@secondoftwo}%
\providecommand \translation [1]{[#1]}%
\providecommand \BibitemOpen [0]{}%
\providecommand \bibitemStop [0]{}%
\providecommand \bibitemNoStop [0]{.\EOS\space}%
\providecommand \EOS [0]{\spacefactor3000\relax}%
\providecommand \BibitemShut  [1]{\csname bibitem#1\endcsname}%
\let\auto@bib@innerbib\@empty
\bibitem [{\citenamefont {Anderson}(1973)}]{Anderson1973}%
  \BibitemOpen
  \bibfield  {author} {\bibinfo {author} {\bibfnamefont {P.}~\bibnamefont
  {Anderson}},\ }\bibfield  {title} {\bibinfo {title} {{Resonating valence
  bonds: A new kind of insulator?}},\ }\href
  {https://doi.org/https://doi.org/10.1016/0025-5408(73)90167-0} {\bibfield
  {journal} {\bibinfo  {journal} {Materials Research Bulletin}\ }\textbf
  {\bibinfo {volume} {8}},\ \bibinfo {pages} {153} (\bibinfo {year}
  {1973})}\BibitemShut {NoStop}%
\bibitem [{\citenamefont {Wen}(2004)}]{WenBook}%
  \BibitemOpen
  \bibfield  {author} {\bibinfo {author} {\bibfnamefont {X.~G.}\ \bibnamefont
  {Wen}},\ }\href@noop {} {\emph {\bibinfo {title} {{Quantum Field Theory of
  Many-Body Systems}}}}\ (\bibinfo  {publisher} {Oxford University Press}, {Oxford},\
  \bibinfo {year} {2004})\BibitemShut {NoStop}%
\bibitem [{\citenamefont {Fradkin}(2013)}]{FradkinBook}%
  \BibitemOpen
  \bibfield  {author} {\bibinfo {author} {\bibfnamefont {E.}~\bibnamefont
  {Fradkin}},\ }\href@noop {} {\emph {\bibinfo {title} {{Field Theories of
  Condensed Matter Physics}}}}\ (\bibinfo  {publisher} {Cambridge University
  Press}, {Cambridge}, \ \bibinfo {year} {2013})\BibitemShut {NoStop}%
\bibitem [{\citenamefont {Sachdev}(2023)}]{SachdevBook}%
  \BibitemOpen
  \bibfield  {author} {\bibinfo {author} {\bibfnamefont {S.}~\bibnamefont
  {Sachdev}},\ }\href@noop {} {\emph {\bibinfo {title} {{Quantum Phases of
  Matter}}}}\ (\bibinfo  {publisher} {Cambridge University Press}, {Cambridge}, \ \bibinfo
  {year} {2023})\BibitemShut {NoStop}%
\bibitem [{\citenamefont {Kitaev}(2003)}]{Kitaev2003}%
  \BibitemOpen
  \bibfield  {author} {\bibinfo {author} {\bibfnamefont {A.}~\bibnamefont
  {Kitaev}},\ }\bibfield  {title} {\bibinfo {title} {{Fault-tolerant quantum
  computation by anyons}},\ }\href
  {https://doi.org/https://doi.org/10.1016/S0003-4916(02)00018-0} {\bibfield
  {journal} {\bibinfo  {journal} {Ann. Phys. (NY)}\ }\textbf {\bibinfo {volume}
  {303}},\ \bibinfo {pages} {2} (\bibinfo {year} {2003})}\BibitemShut {NoStop}%
\bibitem [{\citenamefont {Kitaev}(2006)}]{Kitaev2006}%
  \BibitemOpen
  \bibfield  {author} {\bibinfo {author} {\bibfnamefont {A.}~\bibnamefont
  {Kitaev}},\ }\bibfield  {title} {\bibinfo {title} {{Anyons in an exactly
  solved model and beyond}},\ }\href
  {https://doi.org/https://doi.org/10.1016/j.aop.2005.10.005} {\bibfield
  {journal} {\bibinfo  {journal} {Ann. Phys. (NY)}\ }\textbf {\bibinfo {volume}
  {321}},\ \bibinfo {pages} {2} (\bibinfo {year} {2006})}\BibitemShut {NoStop}%
\bibitem [{\citenamefont {Levin}\ and\ \citenamefont
  {Wen}(2005)}]{LevinWen2005}%
  \BibitemOpen
  \bibfield  {author} {\bibinfo {author} {\bibfnamefont {M.~A.}\ \bibnamefont
  {Levin}}\ and\ \bibinfo {author} {\bibfnamefont {X.-G.}\ \bibnamefont
  {Wen}},\ }\bibfield  {title} {\bibinfo {title} {{String-net condensation: A
  physical mechanism for topological phases}},\ }\href
  {https://doi.org/10.1103/PhysRevB.71.045110} {\bibfield  {journal} {\bibinfo
  {journal} {Phys. Rev. B}\ }\textbf {\bibinfo {volume} {71}},\ \bibinfo
  {pages} {045110} (\bibinfo {year} {2005})}\BibitemShut {NoStop}%
\bibitem [{\citenamefont {Savary}\ and\ \citenamefont
  {Balents}(2016)}]{Savary2016}%
  \BibitemOpen
  \bibfield  {author} {\bibinfo {author} {\bibfnamefont {L.}~\bibnamefont
  {Savary}}\ and\ \bibinfo {author} {\bibfnamefont {L.}~\bibnamefont
  {Balents}},\ }\bibfield  {title} {\bibinfo {title} {{Quantum spin liquids: a
  review}},\ }\href {https://doi.org/10.1088/0034-4885/80/1/016502} {\bibfield
  {journal} {\bibinfo  {journal} {Rep. Prog. Phys.}\ }\textbf {\bibinfo
  {volume} {80}},\ \bibinfo {pages} {016502} (\bibinfo {year}
  {2016})}\BibitemShut {NoStop}%
\bibitem [{\citenamefont {Norman}(2016)}]{Norman2016}%
  \BibitemOpen
  \bibfield  {author} {\bibinfo {author} {\bibfnamefont {M.~R.}\ \bibnamefont
  {Norman}},\ }\bibfield  {title} {\bibinfo {title} {{Colloquium:
  Herbertsmithite and the search for the quantum spin liquid}},\ }\href
  {https://doi.org/10.1103/RevModPhys.88.041002} {\bibfield  {journal}
  {\bibinfo  {journal} {Rev. Mod. Phys.}\ }\textbf {\bibinfo {volume} {88}},\
  \bibinfo {pages} {041002} (\bibinfo {year} {2016})}\BibitemShut {NoStop}%
\bibitem [{\citenamefont {Zhou}\ \emph {et~al.}(2017)\citenamefont {Zhou},
  \citenamefont {Kanoda},\ and\ \citenamefont {Ng}}]{Ng2017}%
  \BibitemOpen
  \bibfield  {author} {\bibinfo {author} {\bibfnamefont {Y.}~\bibnamefont
  {Zhou}}, \bibinfo {author} {\bibfnamefont {K.}~\bibnamefont {Kanoda}},\ and\
  \bibinfo {author} {\bibfnamefont {T.-K.}\ \bibnamefont {Ng}},\ }\bibfield
  {title} {\bibinfo {title} {{Quantum spin liquid states}},\ }\href
  {https://doi.org/10.1103/RevModPhys.89.025003} {\bibfield  {journal}
  {\bibinfo  {journal} {Rev. Mod. Phys.}\ }\textbf {\bibinfo {volume} {89}},\
  \bibinfo {pages} {025003} (\bibinfo {year} {2017})}\BibitemShut {NoStop}%
\bibitem [{\citenamefont {Knolle}\ and\ \citenamefont
  {Moessner}(2019)}]{Knolle2019}%
  \BibitemOpen
  \bibfield  {author} {\bibinfo {author} {\bibfnamefont {J.}~\bibnamefont
  {Knolle}}\ and\ \bibinfo {author} {\bibfnamefont {R.}~\bibnamefont
  {Moessner}},\ }\bibfield  {title} {\bibinfo {title} {{A Field Guide to Spin
  Liquids}},\ }\href {https://doi.org/10.1146/annurev-conmatphys-031218-013401}
  {\bibfield  {journal} {\bibinfo  {journal} {Annu. Rev. Condens. Matter
  Phys.}\ }\textbf {\bibinfo {volume} {10}},\ \bibinfo {pages} {451} (\bibinfo
  {year} {2019})}\BibitemShut {NoStop}%
\bibitem [{\citenamefont {Takagi}\ \emph {et~al.}(2019)\citenamefont {Takagi},
  \citenamefont {Takayama}, \citenamefont {Jackeli}, \citenamefont
  {Khaliullin},\ and\ \citenamefont {Nagler}}]{Takagi2019}%
  \BibitemOpen
  \bibfield  {author} {\bibinfo {author} {\bibfnamefont {H.}~\bibnamefont
  {Takagi}}, \bibinfo {author} {\bibfnamefont {T.}~\bibnamefont {Takayama}},
  \bibinfo {author} {\bibfnamefont {G.}~\bibnamefont {Jackeli}}, \bibinfo
  {author} {\bibfnamefont {G.}~\bibnamefont {Khaliullin}},\ and\ \bibinfo
  {author} {\bibfnamefont {S.~E.}\ \bibnamefont {Nagler}},\ }\bibfield  {title}
  {\bibinfo {title} {{Concept and realization of Kitaev quantum spin
  liquids}},\ }\href {https://doi.org/10.1038/s42254-019-0038-2} {\bibfield
  {journal} {\bibinfo  {journal} {Nat. Rev. Phys.}\ }\textbf {\bibinfo {volume}
  {1}},\ \bibinfo {pages} {264} (\bibinfo {year} {2019})}\BibitemShut {NoStop}%
\bibitem [{\citenamefont {Broholm}\ \emph {et~al.}(2020)\citenamefont
  {Broholm}, \citenamefont {Cava}, \citenamefont {Kivelson}, \citenamefont
  {Nocera}, \citenamefont {Norman},\ and\ \citenamefont
  {Senthil}}]{Broholm2020}%
  \BibitemOpen
  \bibfield  {author} {\bibinfo {author} {\bibfnamefont {C.}~\bibnamefont
  {Broholm}}, \bibinfo {author} {\bibfnamefont {R.~J.}\ \bibnamefont {Cava}},
  \bibinfo {author} {\bibfnamefont {S.~A.}\ \bibnamefont {Kivelson}}, \bibinfo
  {author} {\bibfnamefont {D.~G.}\ \bibnamefont {Nocera}}, \bibinfo {author}
  {\bibfnamefont {M.~R.}\ \bibnamefont {Norman}},\ and\ \bibinfo {author}
  {\bibfnamefont {T.}~\bibnamefont {Senthil}},\ }\bibfield  {title} {\bibinfo
  {title} {{Quantum spin liquids}},\ }\href
  {https://www.science.org/doi/full/10.1126/science.aay0668} {\bibfield
  {journal} {\bibinfo  {journal} {Science}\ }\textbf {\bibinfo {volume}
  {367}},\ \bibinfo {pages} {eaay0668} (\bibinfo {year} {2020})}\BibitemShut
  {NoStop}%
\bibitem [{\citenamefont {Motome}\ and\ \citenamefont
  {Nasu}(2020)}]{Motome2020}%
  \BibitemOpen
  \bibfield  {author} {\bibinfo {author} {\bibfnamefont {Y.}~\bibnamefont
  {Motome}}\ and\ \bibinfo {author} {\bibfnamefont {J.}~\bibnamefont {Nasu}},\
  }\bibfield  {title} {\bibinfo {title} {{Hunting Majorana Fermions in Kitaev
  Magnets}},\ }\href {https://doi.org/10.7566/JPSJ.89.012002} {\bibfield
  {journal} {\bibinfo  {journal} {J. Phys. Soc. Jpn.}\ }\textbf {\bibinfo
  {volume} {89}},\ \bibinfo {pages} {012002} (\bibinfo {year}
  {2020})}\BibitemShut {NoStop}%
\bibitem [{\citenamefont {Read}\ and\ \citenamefont
  {Sachdev}(1991)}]{ReadSachdev1991}%
  \BibitemOpen
  \bibfield  {author} {\bibinfo {author} {\bibfnamefont {N.}~\bibnamefont
  {Read}}\ and\ \bibinfo {author} {\bibfnamefont {S.}~\bibnamefont {Sachdev}},\
  }\bibfield  {title} {\bibinfo {title} {{Large-N expansion for frustrated
  quantum antiferromagnets}},\ }\href
  {https://doi.org/10.1103/PhysRevLett.66.1773} {\bibfield  {journal} {\bibinfo
   {journal} {Phys. Rev. Lett.}\ }\textbf {\bibinfo {volume} {66}},\ \bibinfo
  {pages} {1773} (\bibinfo {year} {1991})}\BibitemShut {NoStop}%
\bibitem [{\citenamefont {Wen}(1991)}]{Wen1991}%
  \BibitemOpen
  \bibfield  {author} {\bibinfo {author} {\bibfnamefont {X.~G.}\ \bibnamefont
  {Wen}},\ }\bibfield  {title} {\bibinfo {title} {Mean-field theory of
  spin-liquid states with finite energy gap and topological orders},\ }\href
  {https://doi.org/10.1103/PhysRevB.44.2664} {\bibfield  {journal} {\bibinfo
  {journal} {Phys. Rev. B}\ }\textbf {\bibinfo {volume} {44}},\ \bibinfo
  {pages} {2664} (\bibinfo {year} {1991})}\BibitemShut {NoStop}%
\bibitem [{\citenamefont {Sachdev}(1992)}]{Sachdev1992}%
  \BibitemOpen
  \bibfield  {author} {\bibinfo {author} {\bibfnamefont {S.}~\bibnamefont
  {Sachdev}},\ }\bibfield  {title} {\bibinfo {title} {{Kagome- and
  triangular-lattice Heisenberg antiferromagnets: Ordering from quantum
  fluctuations and quantum-disordered ground states with unconfined bosonic
  spinons}},\ }\href {https://doi.org/10.1103/PhysRevB.45.12377} {\bibfield
  {journal} {\bibinfo  {journal} {Phys. Rev. B}\ }\textbf {\bibinfo {volume}
  {45}},\ \bibinfo {pages} {12377} (\bibinfo {year} {1992})}\BibitemShut
  {NoStop}%
\bibitem [{\citenamefont {Read}\ and\ \citenamefont
  {Chakraborty}(1989)}]{Read1989}%
  \BibitemOpen
  \bibfield  {author} {\bibinfo {author} {\bibfnamefont {N.}~\bibnamefont
  {Read}}\ and\ \bibinfo {author} {\bibfnamefont {B.}~\bibnamefont
  {Chakraborty}},\ }\bibfield  {title} {\bibinfo {title} {Statistics of the
  excitations of the resonating-valence-bond state},\ }\href
  {https://doi.org/10.1103/PhysRevB.40.7133} {\bibfield  {journal} {\bibinfo
  {journal} {Phys. Rev. B}\ }\textbf {\bibinfo {volume} {40}},\ \bibinfo
  {pages} {7133} (\bibinfo {year} {1989})}\BibitemShut {NoStop}%
\bibitem [{\citenamefont {Senthil}\ and\ \citenamefont
  {Fisher}(2000)}]{Senthil2000}%
  \BibitemOpen
  \bibfield  {author} {\bibinfo {author} {\bibfnamefont {T.}~\bibnamefont
  {Senthil}}\ and\ \bibinfo {author} {\bibfnamefont {M.~P.~A.}\ \bibnamefont
  {Fisher}},\ }\bibfield  {title} {\bibinfo {title} {${Z}_{2}$ gauge theory of
  electron fractionalization in strongly correlated systems},\ }\href
  {https://doi.org/10.1103/PhysRevB.62.7850} {\bibfield  {journal} {\bibinfo
  {journal} {Phys. Rev. B}\ }\textbf {\bibinfo {volume} {62}},\ \bibinfo
  {pages} {7850} (\bibinfo {year} {2000})}\BibitemShut {NoStop}%
\bibitem [{\citenamefont {Moessner}\ and\ \citenamefont
  {Sondhi}(2001)}]{Moessner2001}%
  \BibitemOpen
  \bibfield  {author} {\bibinfo {author} {\bibfnamefont {R.}~\bibnamefont
  {Moessner}}\ and\ \bibinfo {author} {\bibfnamefont {S.~L.}\ \bibnamefont
  {Sondhi}},\ }\bibfield  {title} {\bibinfo {title} {{Resonating Valence Bond
  Phase in the Triangular Lattice Quantum Dimer Model}},\ }\href
  {https://doi.org/10.1103/PhysRevLett.86.1881} {\bibfield  {journal} {\bibinfo
   {journal} {Phys. Rev. Lett.}\ }\textbf {\bibinfo {volume} {86}},\ \bibinfo
  {pages} {1881} (\bibinfo {year} {2001})}\BibitemShut {NoStop}%
\bibitem [{\citenamefont {Moessner}\ \emph {et~al.}(2001)\citenamefont
  {Moessner}, \citenamefont {Sondhi},\ and\ \citenamefont
  {Fradkin}}]{Fradkin2001}%
  \BibitemOpen
  \bibfield  {author} {\bibinfo {author} {\bibfnamefont {R.}~\bibnamefont
  {Moessner}}, \bibinfo {author} {\bibfnamefont {S.~L.}\ \bibnamefont
  {Sondhi}},\ and\ \bibinfo {author} {\bibfnamefont {E.}~\bibnamefont
  {Fradkin}},\ }\bibfield  {title} {\bibinfo {title} {{Short-ranged resonating
  valence bond physics, quantum dimer models, and Ising gauge theories}},\
  }\href {https://doi.org/10.1103/PhysRevB.65.024504} {\bibfield  {journal}
  {\bibinfo  {journal} {Phys. Rev. B}\ }\textbf {\bibinfo {volume} {65}},\
  \bibinfo {pages} {024504} (\bibinfo {year} {2001})}\BibitemShut {NoStop}%
\bibitem [{\citenamefont {Misguich}\ \emph {et~al.}(2002)\citenamefont
  {Misguich}, \citenamefont {Serban},\ and\ \citenamefont
  {Pasquier}}]{Misguich2002}%
  \BibitemOpen
  \bibfield  {author} {\bibinfo {author} {\bibfnamefont {G.}~\bibnamefont
  {Misguich}}, \bibinfo {author} {\bibfnamefont {D.}~\bibnamefont {Serban}},\
  and\ \bibinfo {author} {\bibfnamefont {V.}~\bibnamefont {Pasquier}},\
  }\bibfield  {title} {\bibinfo {title} {{Quantum Dimer Model on the Kagome
  Lattice: Solvable Dimer-Liquid and Ising Gauge Theory}},\ }\href
  {https://doi.org/10.1103/PhysRevLett.89.137202} {\bibfield  {journal}
  {\bibinfo  {journal} {Phys. Rev. Lett.}\ }\textbf {\bibinfo {volume} {89}},\
  \bibinfo {pages} {137202} (\bibinfo {year} {2002})}\BibitemShut {NoStop}%
\bibitem [{\citenamefont {Wen}(2002)}]{Wen2002}%
  \BibitemOpen
  \bibfield  {author} {\bibinfo {author} {\bibfnamefont {X.-G.}\ \bibnamefont
  {Wen}},\ }\bibfield  {title} {\bibinfo {title} {Quantum orders and symmetric
  spin liquids},\ }\href {https://doi.org/10.1103/PhysRevB.65.165113}
  {\bibfield  {journal} {\bibinfo  {journal} {Phys. Rev. B}\ }\textbf {\bibinfo
  {volume} {65}},\ \bibinfo {pages} {165113} (\bibinfo {year}
  {2002})}\BibitemShut {NoStop}%
\bibitem [{\citenamefont {Balents}\ \emph {et~al.}(2002)\citenamefont
  {Balents}, \citenamefont {Fisher},\ and\ \citenamefont
  {Girvin}}]{Balents2002}%
  \BibitemOpen
  \bibfield  {author} {\bibinfo {author} {\bibfnamefont {L.}~\bibnamefont
  {Balents}}, \bibinfo {author} {\bibfnamefont {M.~P.~A.}\ \bibnamefont
  {Fisher}},\ and\ \bibinfo {author} {\bibfnamefont {S.~M.}\ \bibnamefont
  {Girvin}},\ }\bibfield  {title} {\bibinfo {title} {Fractionalization in an
  easy-axis kagome antiferromagnet},\ }\href
  {https://doi.org/10.1103/PhysRevB.65.224412} {\bibfield  {journal} {\bibinfo
  {journal} {Phys. Rev. B}\ }\textbf {\bibinfo {volume} {65}},\ \bibinfo
  {pages} {224412} (\bibinfo {year} {2002})}\BibitemShut {NoStop}%
\bibitem [{\citenamefont {Wen}(2003)}]{Wen2003}%
  \BibitemOpen
  \bibfield  {author} {\bibinfo {author} {\bibfnamefont {X.-G.}\ \bibnamefont
  {Wen}},\ }\bibfield  {title} {\bibinfo {title} {Quantum orders in an exact
  soluble model},\ }\href {https://doi.org/10.1103/PhysRevLett.90.016803}
  {\bibfield  {journal} {\bibinfo  {journal} {Phys. Rev. Lett.}\ }\textbf
  {\bibinfo {volume} {90}},\ \bibinfo {pages} {016803} (\bibinfo {year}
  {2003})}\BibitemShut {NoStop}%
\bibitem [{\citenamefont {Wang}\ and\ \citenamefont
  {Vishwanath}(2006)}]{Wang2006}%
  \BibitemOpen
  \bibfield  {author} {\bibinfo {author} {\bibfnamefont {F.}~\bibnamefont
  {Wang}}\ and\ \bibinfo {author} {\bibfnamefont {A.}~\bibnamefont
  {Vishwanath}},\ }\bibfield  {title} {\bibinfo {title} {{Spin-liquid states on
  the triangular and Kagom\'e lattices: A projective-symmetry-group analysis of
  Schwinger boson states}},\ }\href
  {https://doi.org/10.1103/PhysRevB.74.174423} {\bibfield  {journal} {\bibinfo
  {journal} {Phys. Rev. B}\ }\textbf {\bibinfo {volume} {74}},\ \bibinfo
  {pages} {174423} (\bibinfo {year} {2006})}\BibitemShut {NoStop}%
\bibitem [{\citenamefont {Yan}\ \emph {et~al.}(2011)\citenamefont {Yan},
  \citenamefont {Huse},\ and\ \citenamefont {White}}]{White2011}%
  \BibitemOpen
  \bibfield  {author} {\bibinfo {author} {\bibfnamefont {S.}~\bibnamefont
  {Yan}}, \bibinfo {author} {\bibfnamefont {D.~A.}\ \bibnamefont {Huse}},\ and\
  \bibinfo {author} {\bibfnamefont {S.~R.}\ \bibnamefont {White}},\ }\bibfield
  {title} {\bibinfo {title} {{Spin-Liquid Ground State of the $S=$1/2 Kagome
  Heisenberg Antiferromagnet}},\ }\href
  {https://doi.org/10.1126/science.1201080} {\bibfield  {journal} {\bibinfo
  {journal} {Science}\ }\textbf {\bibinfo {volume} {332}},\ \bibinfo {pages}
  {1173} (\bibinfo {year} {2011})}\BibitemShut {NoStop}%
\bibitem [{\citenamefont {Depenbrock}\ \emph {et~al.}(2012)\citenamefont
  {Depenbrock}, \citenamefont {McCulloch},\ and\ \citenamefont
  {Schollw\"ock}}]{Depenbrock2012}%
  \BibitemOpen
  \bibfield  {author} {\bibinfo {author} {\bibfnamefont {S.}~\bibnamefont
  {Depenbrock}}, \bibinfo {author} {\bibfnamefont {I.~P.}\ \bibnamefont
  {McCulloch}},\ and\ \bibinfo {author} {\bibfnamefont {U.}~\bibnamefont
  {Schollw\"ock}},\ }\bibfield  {title} {\bibinfo {title} {{Nature of the
  Spin-Liquid Ground State of the $S=1/2$ Heisenberg Model on the Kagome
  Lattice}},\ }\href {https://doi.org/10.1103/PhysRevLett.109.067201}
  {\bibfield  {journal} {\bibinfo  {journal} {Phys. Rev. Lett.}\ }\textbf
  {\bibinfo {volume} {109}},\ \bibinfo {pages} {067201} (\bibinfo {year}
  {2012})}\BibitemShut {NoStop}%
\bibitem [{\citenamefont {Tay}\ and\ \citenamefont
  {Motrunich}(2011)}]{Motrunich2011}%
  \BibitemOpen
  \bibfield  {author} {\bibinfo {author} {\bibfnamefont {T.}~\bibnamefont
  {Tay}}\ and\ \bibinfo {author} {\bibfnamefont {O.~I.}\ \bibnamefont
  {Motrunich}},\ }\bibfield  {title} {\bibinfo {title} {{Sign structures for
  short-range RVB states on small kagome clusters}},\ }\href
  {https://doi.org/10.1103/PhysRevB.84.193102} {\bibfield  {journal} {\bibinfo
  {journal} {Phys. Rev. B}\ }\textbf {\bibinfo {volume} {84}},\ \bibinfo
  {pages} {193102} (\bibinfo {year} {2011})}\BibitemShut {NoStop}%
\bibitem [{\citenamefont {Lu}\ \emph {et~al.}(2011)\citenamefont {Lu},
  \citenamefont {Ran},\ and\ \citenamefont {Lee}}]{Lu2011}%
  \BibitemOpen
  \bibfield  {author} {\bibinfo {author} {\bibfnamefont {Y.-M.}\ \bibnamefont
  {Lu}}, \bibinfo {author} {\bibfnamefont {Y.}~\bibnamefont {Ran}},\ and\
  \bibinfo {author} {\bibfnamefont {P.~A.}\ \bibnamefont {Lee}},\ }\bibfield
  {title} {\bibinfo {title} {{${\mathbb{Z}}_{2}$ spin liquids in the
  $S=\frac{1}{2}$ Heisenberg model on the kagome lattice: A projective
  symmetry-group study of Schwinger fermion mean-field states}},\ }\href
  {https://doi.org/10.1103/PhysRevB.83.224413} {\bibfield  {journal} {\bibinfo
  {journal} {Phys. Rev. B}\ }\textbf {\bibinfo {volume} {83}},\ \bibinfo
  {pages} {224413} (\bibinfo {year} {2011})}\BibitemShut {NoStop}%
\bibitem [{\citenamefont {Iqbal}\ \emph {et~al.}(2011)\citenamefont {Iqbal},
  \citenamefont {Becca},\ and\ \citenamefont {Poilblanc}}]{Iqbal2011}%
  \BibitemOpen
  \bibfield  {author} {\bibinfo {author} {\bibfnamefont {Y.}~\bibnamefont
  {Iqbal}}, \bibinfo {author} {\bibfnamefont {F.}~\bibnamefont {Becca}},\ and\
  \bibinfo {author} {\bibfnamefont {D.}~\bibnamefont {Poilblanc}},\ }\bibfield
  {title} {\bibinfo {title} {{Projected wave function study of
  ${\mathbb{Z}}_{2}$ spin liquids on the kagome lattice for the
  spin-$\frac{1}{2}$ quantum Heisenberg antiferromagnet}},\ }\href
  {https://doi.org/10.1103/PhysRevB.84.020407} {\bibfield  {journal} {\bibinfo
  {journal} {Phys. Rev. B}\ }\textbf {\bibinfo {volume} {84}},\ \bibinfo
  {pages} {020407} (\bibinfo {year} {2011})}\BibitemShut {NoStop}%
\bibitem [{\citenamefont {Zhang}\ \emph {et~al.}(2012)\citenamefont {Zhang},
  \citenamefont {Grover}, \citenamefont {Turner}, \citenamefont {Oshikawa},\
  and\ \citenamefont {Vishwanath}}]{Zhang2012}%
  \BibitemOpen
  \bibfield  {author} {\bibinfo {author} {\bibfnamefont {Y.}~\bibnamefont
  {Zhang}}, \bibinfo {author} {\bibfnamefont {T.}~\bibnamefont {Grover}},
  \bibinfo {author} {\bibfnamefont {A.}~\bibnamefont {Turner}}, \bibinfo
  {author} {\bibfnamefont {M.}~\bibnamefont {Oshikawa}},\ and\ \bibinfo
  {author} {\bibfnamefont {A.}~\bibnamefont {Vishwanath}},\ }\bibfield  {title}
  {\bibinfo {title} {{Quasiparticle statistics and braiding from ground-state
  entanglement}},\ }\href {https://doi.org/10.1103/PhysRevB.85.235151}
  {\bibfield  {journal} {\bibinfo  {journal} {Phys. Rev. B}\ }\textbf {\bibinfo
  {volume} {85}},\ \bibinfo {pages} {235151} (\bibinfo {year}
  {2012})}\BibitemShut {NoStop}%
\bibitem [{\citenamefont {Jiang}\ \emph {et~al.}(2012)\citenamefont {Jiang},
  \citenamefont {Wang},\ and\ \citenamefont {Balents}}]{Jiang2012}%
  \BibitemOpen
  \bibfield  {author} {\bibinfo {author} {\bibfnamefont {H.-C.}\ \bibnamefont
  {Jiang}}, \bibinfo {author} {\bibfnamefont {Z.}~\bibnamefont {Wang}},\ and\
  \bibinfo {author} {\bibfnamefont {L.}~\bibnamefont {Balents}},\ }\bibfield
  {title} {\bibinfo {title} {{Identifying topological order by entanglement
  entropy}},\ }\href {https://doi.org/10.1038/nphys2465} {\bibfield  {journal}
  {\bibinfo  {journal} {Nat. Phys.}\ }\textbf {\bibinfo {volume} {8}},\
  \bibinfo {pages} {902} (\bibinfo {year} {2012})}\BibitemShut {NoStop}%
\bibitem [{\citenamefont {Messio}\ \emph {et~al.}(2012)\citenamefont {Messio},
  \citenamefont {Bernu},\ and\ \citenamefont {Lhuillier}}]{Messio2012}%
  \BibitemOpen
  \bibfield  {author} {\bibinfo {author} {\bibfnamefont {L.}~\bibnamefont
  {Messio}}, \bibinfo {author} {\bibfnamefont {B.}~\bibnamefont {Bernu}},\ and\
  \bibinfo {author} {\bibfnamefont {C.}~\bibnamefont {Lhuillier}},\ }\bibfield
  {title} {\bibinfo {title} {{Kagome Antiferromagnet: A Chiral Topological Spin
  Liquid?}},\ }\href {https://doi.org/10.1103/PhysRevLett.108.207204}
  {\bibfield  {journal} {\bibinfo  {journal} {Phys. Rev. Lett.}\ }\textbf
  {\bibinfo {volume} {108}},\ \bibinfo {pages} {207204} (\bibinfo {year}
  {2012})}\BibitemShut {NoStop}%
\bibitem [{\citenamefont {Essin}\ and\ \citenamefont
  {Hermele}(2013)}]{Hermele2013}%
  \BibitemOpen
  \bibfield  {author} {\bibinfo {author} {\bibfnamefont {A.~M.}\ \bibnamefont
  {Essin}}\ and\ \bibinfo {author} {\bibfnamefont {M.}~\bibnamefont
  {Hermele}},\ }\bibfield  {title} {\bibinfo {title} {{Classifying
  fractionalization: Symmetry classification of gapped ${\mathbb{Z}}_{2}$ spin
  liquids in two dimensions}},\ }\href
  {https://doi.org/10.1103/PhysRevB.87.104406} {\bibfield  {journal} {\bibinfo
  {journal} {Phys. Rev. B}\ }\textbf {\bibinfo {volume} {87}},\ \bibinfo
  {pages} {104406} (\bibinfo {year} {2013})}\BibitemShut {NoStop}%
\bibitem [{\citenamefont {Huh}\ \emph {et~al.}(2011)\citenamefont {Huh},
  \citenamefont {Punk},\ and\ \citenamefont {Sachdev}}]{Huh2011}%
  \BibitemOpen
  \bibfield  {author} {\bibinfo {author} {\bibfnamefont {Y.}~\bibnamefont
  {Huh}}, \bibinfo {author} {\bibfnamefont {M.}~\bibnamefont {Punk}},\ and\
  \bibinfo {author} {\bibfnamefont {S.}~\bibnamefont {Sachdev}},\ }\bibfield
  {title} {\bibinfo {title} {{Vison states and confinement transitions of
  ${\mathbb{Z}}_{2}$ spin liquids on the kagome lattice}},\ }\href
  {https://doi.org/10.1103/PhysRevB.84.094419} {\bibfield  {journal} {\bibinfo
  {journal} {Phys. Rev. B}\ }\textbf {\bibinfo {volume} {84}},\ \bibinfo
  {pages} {094419} (\bibinfo {year} {2011})}\BibitemShut {NoStop}%
\bibitem [{\citenamefont {Wan}\ and\ \citenamefont
  {Tchernyshyov}(2013)}]{Wan2013}%
  \BibitemOpen
  \bibfield  {author} {\bibinfo {author} {\bibfnamefont {Y.}~\bibnamefont
  {Wan}}\ and\ \bibinfo {author} {\bibfnamefont {O.}~\bibnamefont
  {Tchernyshyov}},\ }\bibfield  {title} {\bibinfo {title} {{Phenomenological
  ${Z}_{2}$ lattice gauge theory of the spin-liquid state of the kagome
  Heisenberg antiferromagnet}},\ }\href
  {https://doi.org/10.1103/PhysRevB.87.104408} {\bibfield  {journal} {\bibinfo
  {journal} {Phys. Rev. B}\ }\textbf {\bibinfo {volume} {87}},\ \bibinfo
  {pages} {104408} (\bibinfo {year} {2013})}\BibitemShut {NoStop}%
\bibitem [{\citenamefont {Hwang}\ \emph {et~al.}(2015)\citenamefont {Hwang},
  \citenamefont {Huh},\ and\ \citenamefont {Kim}}]{Hwang2015}%
  \BibitemOpen
  \bibfield  {author} {\bibinfo {author} {\bibfnamefont {K.}~\bibnamefont
  {Hwang}}, \bibinfo {author} {\bibfnamefont {Y.}~\bibnamefont {Huh}},\ and\
  \bibinfo {author} {\bibfnamefont {Y.~B.}\ \bibnamefont {Kim}},\ }\bibfield
  {title} {\bibinfo {title} {{$\mathbb{Z}_2$ gauge theory for valence bond
  solids on the kagome lattice}},\ }\href
  {https://doi.org/10.1103/PhysRevB.92.205131} {\bibfield  {journal} {\bibinfo
  {journal} {Phys. Rev. B}\ }\textbf {\bibinfo {volume} {92}},\ \bibinfo
  {pages} {205131} (\bibinfo {year} {2015})}\BibitemShut {NoStop}%
\bibitem [{\citenamefont {Zaletel}\ and\ \citenamefont
  {Vishwanath}(2015)}]{Zaletel2015}%
  \BibitemOpen
  \bibfield  {author} {\bibinfo {author} {\bibfnamefont {M.~P.}\ \bibnamefont
  {Zaletel}}\ and\ \bibinfo {author} {\bibfnamefont {A.}~\bibnamefont
  {Vishwanath}},\ }\bibfield  {title} {\bibinfo {title} {{Constraints on
  Topological Order in Mott Insulators}},\ }\href
  {https://doi.org/10.1103/PhysRevLett.114.077201} {\bibfield  {journal}
  {\bibinfo  {journal} {Phys. Rev. Lett.}\ }\textbf {\bibinfo {volume} {114}},\
  \bibinfo {pages} {077201} (\bibinfo {year} {2015})}\BibitemShut {NoStop}%
\bibitem [{\citenamefont {Lu}\ \emph {et~al.}(2017)\citenamefont {Lu},
  \citenamefont {Cho},\ and\ \citenamefont {Vishwanath}}]{Lu2017}%
  \BibitemOpen
  \bibfield  {author} {\bibinfo {author} {\bibfnamefont {Y.-M.}\ \bibnamefont
  {Lu}}, \bibinfo {author} {\bibfnamefont {G.~Y.}\ \bibnamefont {Cho}},\ and\
  \bibinfo {author} {\bibfnamefont {A.}~\bibnamefont {Vishwanath}},\ }\bibfield
   {title} {\bibinfo {title} {{Unification of bosonic and fermionic theories of
  spin liquids on the kagome lattice}},\ }\href
  {https://doi.org/10.1103/PhysRevB.96.205150} {\bibfield  {journal} {\bibinfo
  {journal} {Phys. Rev. B}\ }\textbf {\bibinfo {volume} {96}},\ \bibinfo
  {pages} {205150} (\bibinfo {year} {2017})}\BibitemShut {NoStop}%
\bibitem [{\citenamefont {Schwandt}\ \emph {et~al.}(2010)\citenamefont
  {Schwandt}, \citenamefont {Mambrini},\ and\ \citenamefont
  {Poilblanc}}]{Poilblanc2010}%
  \BibitemOpen
  \bibfield  {author} {\bibinfo {author} {\bibfnamefont {D.}~\bibnamefont
  {Schwandt}}, \bibinfo {author} {\bibfnamefont {M.}~\bibnamefont {Mambrini}},\
  and\ \bibinfo {author} {\bibfnamefont {D.}~\bibnamefont {Poilblanc}},\
  }\bibfield  {title} {\bibinfo {title} {{Generalized hard-core dimer model
  approach to low-energy Heisenberg frustrated antiferromagnets: General
  properties and application to the kagome antiferromagnet}},\ }\href
  {https://doi.org/10.1103/PhysRevB.81.214413} {\bibfield  {journal} {\bibinfo
  {journal} {Phys. Rev. B}\ }\textbf {\bibinfo {volume} {81}},\ \bibinfo
  {pages} {214413} (\bibinfo {year} {2010})}\BibitemShut {NoStop}%
\bibitem [{\citenamefont {Hao}\ \emph {et~al.}(2014)\citenamefont {Hao},
  \citenamefont {Inglis},\ and\ \citenamefont {Melko}}]{Hao2014}%
  \BibitemOpen
  \bibfield  {author} {\bibinfo {author} {\bibfnamefont {Z.}~\bibnamefont
  {Hao}}, \bibinfo {author} {\bibfnamefont {S.}~\bibnamefont {Inglis}},\ and\
  \bibinfo {author} {\bibfnamefont {R.}~\bibnamefont {Melko}},\ }\bibfield
  {title} {\bibinfo {title} {Destroying a topological quantum bit by condensing
  ising vortices},\ }\href {https://doi.org/10.1038/ncomms6781} {\bibfield
  {journal} {\bibinfo  {journal} {Nature Communications}\ }\textbf {\bibinfo
  {volume} {5}},\ \bibinfo {pages} {5781} (\bibinfo {year} {2014})}\BibitemShut
  {NoStop}%
\bibitem [{\citenamefont {Rousochatzakis}\ \emph {et~al.}(2014)\citenamefont
  {Rousochatzakis}, \citenamefont {Wan}, \citenamefont {Tchernyshyov},\ and\
  \citenamefont {Mila}}]{Rousochatzakis2014}%
  \BibitemOpen
  \bibfield  {author} {\bibinfo {author} {\bibfnamefont {I.}~\bibnamefont
  {Rousochatzakis}}, \bibinfo {author} {\bibfnamefont {Y.}~\bibnamefont {Wan}},
  \bibinfo {author} {\bibfnamefont {O.}~\bibnamefont {Tchernyshyov}},\ and\
  \bibinfo {author} {\bibfnamefont {F.}~\bibnamefont {Mila}},\ }\bibfield
  {title} {\bibinfo {title} {Quantum dimer model for the spin-$\frac{1}{2}$
  kagome ${Z}_{2}$ spin liquid},\ }\href
  {https://doi.org/10.1103/PhysRevB.90.100406} {\bibfield  {journal} {\bibinfo
  {journal} {Phys. Rev. B}\ }\textbf {\bibinfo {volume} {90}},\ \bibinfo
  {pages} {100406} (\bibinfo {year} {2014})}\BibitemShut {NoStop}%
\bibitem [{\citenamefont {Ralko}\ \emph {et~al.}(2018)\citenamefont {Ralko},
  \citenamefont {Mila},\ and\ \citenamefont {Rousochatzakis}}]{Ralko2018}%
  \BibitemOpen
  \bibfield  {author} {\bibinfo {author} {\bibfnamefont {A.}~\bibnamefont
  {Ralko}}, \bibinfo {author} {\bibfnamefont {F.}~\bibnamefont {Mila}},\ and\
  \bibinfo {author} {\bibfnamefont {I.}~\bibnamefont {Rousochatzakis}},\
  }\bibfield  {title} {\bibinfo {title} {Microscopic theory of the
  nearest-neighbor valence bond sector of the spin-$\frac{1}{2}$ kagome
  antiferromagnet},\ }\href {https://doi.org/10.1103/PhysRevB.97.104401}
  {\bibfield  {journal} {\bibinfo  {journal} {Phys. Rev. B}\ }\textbf {\bibinfo
  {volume} {97}},\ \bibinfo {pages} {104401} (\bibinfo {year}
  {2018})}\BibitemShut {NoStop}%
\bibitem [{\citenamefont {Iqbal}\ \emph {et~al.}(2020)\citenamefont {Iqbal},
  \citenamefont {Casademunt},\ and\ \citenamefont {Schuch}}]{Iqbal2020}%
  \BibitemOpen
  \bibfield  {author} {\bibinfo {author} {\bibfnamefont {M.}~\bibnamefont
  {Iqbal}}, \bibinfo {author} {\bibfnamefont {H.}~\bibnamefont {Casademunt}},\
  and\ \bibinfo {author} {\bibfnamefont {N.}~\bibnamefont {Schuch}},\
  }\bibfield  {title} {\bibinfo {title} {{Topological spin liquids: Robustness
  under perturbations}},\ }\href {https://doi.org/10.1103/PhysRevB.101.115101}
  {\bibfield  {journal} {\bibinfo  {journal} {Phys. Rev. B}\ }\textbf {\bibinfo
  {volume} {101}},\ \bibinfo {pages} {115101} (\bibinfo {year}
  {2020})}\BibitemShut {NoStop}%
\bibitem [{\citenamefont {Huh}\ \emph {et~al.}(2010)\citenamefont {Huh},
  \citenamefont {Fritz},\ and\ \citenamefont {Sachdev}}]{Sachdev2010}%
  \BibitemOpen
  \bibfield  {author} {\bibinfo {author} {\bibfnamefont {Y.}~\bibnamefont
  {Huh}}, \bibinfo {author} {\bibfnamefont {L.}~\bibnamefont {Fritz}},\ and\
  \bibinfo {author} {\bibfnamefont {S.}~\bibnamefont {Sachdev}},\ }\bibfield
  {title} {\bibinfo {title} {{Quantum criticality of the kagome antiferromagnet
  with Dzyaloshinskii-Moriya interactions}},\ }\href
  {https://doi.org/10.1103/PhysRevB.81.144432} {\bibfield  {journal} {\bibinfo
  {journal} {Phys. Rev. B}\ }\textbf {\bibinfo {volume} {81}},\ \bibinfo
  {pages} {144432} (\bibinfo {year} {2010})}\BibitemShut {NoStop}%
\bibitem [{\citenamefont {Messio}\ \emph {et~al.}(2010)\citenamefont {Messio},
  \citenamefont {C\'epas},\ and\ \citenamefont {Lhuillier}}]{Messio2010}%
  \BibitemOpen
  \bibfield  {author} {\bibinfo {author} {\bibfnamefont {L.}~\bibnamefont
  {Messio}}, \bibinfo {author} {\bibfnamefont {O.}~\bibnamefont {C\'epas}},\
  and\ \bibinfo {author} {\bibfnamefont {C.}~\bibnamefont {Lhuillier}},\
  }\bibfield  {title} {\bibinfo {title} {{Schwinger-boson approach to the
  kagome antiferromagnet with Dzyaloshinskii-Moriya interactions: Phase diagram
  and dynamical structure factors}},\ }\href
  {https://doi.org/10.1103/PhysRevB.81.064428} {\bibfield  {journal} {\bibinfo
  {journal} {Phys. Rev. B}\ }\textbf {\bibinfo {volume} {81}},\ \bibinfo
  {pages} {064428} (\bibinfo {year} {2010})}\BibitemShut {NoStop}%
\bibitem [{\citenamefont {Dodds}\ \emph {et~al.}(2013)\citenamefont {Dodds},
  \citenamefont {Bhattacharjee},\ and\ \citenamefont {Kim}}]{Dodds2013}%
  \BibitemOpen
  \bibfield  {author} {\bibinfo {author} {\bibfnamefont {T.}~\bibnamefont
  {Dodds}}, \bibinfo {author} {\bibfnamefont {S.}~\bibnamefont
  {Bhattacharjee}},\ and\ \bibinfo {author} {\bibfnamefont {Y.~B.}\
  \bibnamefont {Kim}},\ }\bibfield  {title} {\bibinfo {title} {Quantum spin
  liquids in the absence of spin-rotation symmetry: Application to
  herbertsmithite},\ }\href {https://doi.org/10.1103/PhysRevB.88.224413}
  {\bibfield  {journal} {\bibinfo  {journal} {Phys. Rev. B}\ }\textbf {\bibinfo
  {volume} {88}},\ \bibinfo {pages} {224413} (\bibinfo {year}
  {2013})}\BibitemShut {NoStop}%
\bibitem [{\citenamefont {Punk}\ \emph {et~al.}(2014)\citenamefont {Punk},
  \citenamefont {Chowdhury},\ and\ \citenamefont {Sachdev}}]{Sachdev2014}%
  \BibitemOpen
  \bibfield  {author} {\bibinfo {author} {\bibfnamefont {M.}~\bibnamefont
  {Punk}}, \bibinfo {author} {\bibfnamefont {D.}~\bibnamefont {Chowdhury}},\
  and\ \bibinfo {author} {\bibfnamefont {S.}~\bibnamefont {Sachdev}},\
  }\bibfield  {title} {\bibinfo {title} {Topological excitations and the
  dynamic structure factor of spin liquids on the kagome lattice},\ }\href
  {https://doi.org/10.1038/nphys2887} {\bibfield  {journal} {\bibinfo
  {journal} {Nat. Phys.}\ }\textbf {\bibinfo {volume} {10}},\ \bibinfo {pages}
  {289} (\bibinfo {year} {2014})}\BibitemShut {NoStop}%
\bibitem [{\citenamefont {Messio}\ \emph {et~al.}(2017)\citenamefont {Messio},
  \citenamefont {Bieri}, \citenamefont {Lhuillier},\ and\ \citenamefont
  {Bernu}}]{Messio2017}%
  \BibitemOpen
  \bibfield  {author} {\bibinfo {author} {\bibfnamefont {L.}~\bibnamefont
  {Messio}}, \bibinfo {author} {\bibfnamefont {S.}~\bibnamefont {Bieri}},
  \bibinfo {author} {\bibfnamefont {C.}~\bibnamefont {Lhuillier}},\ and\
  \bibinfo {author} {\bibfnamefont {B.}~\bibnamefont {Bernu}},\ }\bibfield
  {title} {\bibinfo {title} {{Chiral Spin Liquid on a Kagome Antiferromagnet
  Induced by the Dzyaloshinskii-Moriya Interaction}},\ }\href
  {https://doi.org/10.1103/PhysRevLett.118.267201} {\bibfield  {journal}
  {\bibinfo  {journal} {Phys. Rev. Lett.}\ }\textbf {\bibinfo {volume} {118}},\
  \bibinfo {pages} {267201} (\bibinfo {year} {2017})}\BibitemShut {NoStop}%
\bibitem [{\citenamefont {Halimeh}\ and\ \citenamefont
  {Punk}(2016)}]{Halimeh2016}%
  \BibitemOpen
  \bibfield  {author} {\bibinfo {author} {\bibfnamefont {J.~C.}\ \bibnamefont
  {Halimeh}}\ and\ \bibinfo {author} {\bibfnamefont {M.}~\bibnamefont {Punk}},\
  }\bibfield  {title} {\bibinfo {title} {{Spin structure factors of chiral
  quantum spin liquids on the kagome lattice}},\ }\href
  {https://doi.org/10.1103/PhysRevB.94.104413} {\bibfield  {journal} {\bibinfo
  {journal} {Phys. Rev. B}\ }\textbf {\bibinfo {volume} {94}},\ \bibinfo
  {pages} {104413} (\bibinfo {year} {2016})}\BibitemShut {NoStop}%
\bibitem [{\citenamefont {Zhu}\ \emph {et~al.}(2019)\citenamefont {Zhu},
  \citenamefont {Gong},\ and\ \citenamefont {Sheng}}]{Sheng2019}%
  \BibitemOpen
  \bibfield  {author} {\bibinfo {author} {\bibfnamefont {W.}~\bibnamefont
  {Zhu}}, \bibinfo {author} {\bibfnamefont {S.-S.}\ \bibnamefont {Gong}},\ and\
  \bibinfo {author} {\bibfnamefont {D.~N.}\ \bibnamefont {Sheng}},\ }\bibfield
  {title} {\bibinfo {title} {{Identifying spinon excitations from dynamic
  structure factor of spin-1/2 Heisenberg antiferromagnet on the Kagome
  lattice}},\ }\href {https://doi.org/10.1073/pnas.1807840116} {\bibfield
  {journal} {\bibinfo  {journal} {Proc. Natl. Acad. Sci. U.S.A.}\ }\textbf {\bibinfo
  {volume} {116}},\ \bibinfo {pages} {5437} (\bibinfo {year}
  {2019})}\BibitemShut {NoStop}%
\bibitem [{\citenamefont {Halimeh}\ and\ \citenamefont
  {Singh}(2019)}]{Halimeh2019}%
  \BibitemOpen
  \bibfield  {author} {\bibinfo {author} {\bibfnamefont {J.~C.}\ \bibnamefont
  {Halimeh}}\ and\ \bibinfo {author} {\bibfnamefont {R.~R.~P.}\ \bibnamefont
  {Singh}},\ }\bibfield  {title} {\bibinfo {title} {{Rapid filling of the spin
  gap with temperature in the Schwinger-boson mean-field theory of the
  antiferromagnetic Heisenberg kagome model}},\ }\href
  {https://doi.org/10.1103/PhysRevB.99.155151} {\bibfield  {journal} {\bibinfo
  {journal} {Phys. Rev. B}\ }\textbf {\bibinfo {volume} {99}},\ \bibinfo
  {pages} {155151} (\bibinfo {year} {2019})}\BibitemShut {NoStop}%
\bibitem [{\citenamefont {Sun}\ \emph {et~al.}(2018)\citenamefont {Sun},
  \citenamefont {Wang}, \citenamefont {Fang}, \citenamefont {Qi}, \citenamefont
  {Cheng},\ and\ \citenamefont {Meng}}]{Meng2018BFG}%
  \BibitemOpen
  \bibfield  {author} {\bibinfo {author} {\bibfnamefont {G.-Y.}\ \bibnamefont
  {Sun}}, \bibinfo {author} {\bibfnamefont {Y.-C.}\ \bibnamefont {Wang}},
  \bibinfo {author} {\bibfnamefont {C.}~\bibnamefont {Fang}}, \bibinfo {author}
  {\bibfnamefont {Y.}~\bibnamefont {Qi}}, \bibinfo {author} {\bibfnamefont
  {M.}~\bibnamefont {Cheng}},\ and\ \bibinfo {author} {\bibfnamefont {Z.~Y.}\
  \bibnamefont {Meng}},\ }\bibfield  {title} {\bibinfo {title} {{Dynamical
  Signature of Symmetry Fractionalization in Frustrated Magnets}},\ }\href
  {https://doi.org/10.1103/PhysRevLett.121.077201} {\bibfield  {journal}
  {\bibinfo  {journal} {Phys. Rev. Lett.}\ }\textbf {\bibinfo {volume} {121}},\
  \bibinfo {pages} {077201} (\bibinfo {year} {2018})}\BibitemShut {NoStop}%
\bibitem [{\citenamefont {Becker}\ and\ \citenamefont
  {Wessel}(2018)}]{Wessel2018BFG}%
  \BibitemOpen
  \bibfield  {author} {\bibinfo {author} {\bibfnamefont {J.}~\bibnamefont
  {Becker}}\ and\ \bibinfo {author} {\bibfnamefont {S.}~\bibnamefont
  {Wessel}},\ }\bibfield  {title} {\bibinfo {title} {{Diagnosing
  Fractionalization from the Spin Dynamics of ${Z}_{2}$ Spin Liquids on the
  Kagome Lattice by Quantum Monte Carlo Simulations}},\ }\href
  {https://doi.org/10.1103/PhysRevLett.121.077202} {\bibfield  {journal}
  {\bibinfo  {journal} {Phys. Rev. Lett.}\ }\textbf {\bibinfo {volume} {121}},\
  \bibinfo {pages} {077202} (\bibinfo {year} {2018})}\BibitemShut {NoStop}%
\bibitem [{\citenamefont {Zhu}\ and\ \citenamefont {White}(2015)}]{White2015}%
  \BibitemOpen
  \bibfield  {author} {\bibinfo {author} {\bibfnamefont {Z.}~\bibnamefont
  {Zhu}}\ and\ \bibinfo {author} {\bibfnamefont {S.~R.}\ \bibnamefont
  {White}},\ }\bibfield  {title} {\bibinfo {title} {{Spin liquid phase of the
  $S=\frac{1}{2}\phantom{\rule{4.pt}{0ex}}{J}_{1}\ensuremath{-}{J}_{2}$
  Heisenberg model on the triangular lattice}},\ }\href
  {https://doi.org/10.1103/PhysRevB.92.041105} {\bibfield  {journal} {\bibinfo
  {journal} {Phys. Rev. B}\ }\textbf {\bibinfo {volume} {92}},\ \bibinfo
  {pages} {041105(R)} (\bibinfo {year} {2015})}\BibitemShut {NoStop}%
\bibitem [{\citenamefont {Hu}\ \emph {et~al.}(2015)\citenamefont {Hu},
  \citenamefont {Gong}, \citenamefont {Zhu},\ and\ \citenamefont
  {Sheng}}]{Sheng2015}%
  \BibitemOpen
  \bibfield  {author} {\bibinfo {author} {\bibfnamefont {W.-J.}\ \bibnamefont
  {Hu}}, \bibinfo {author} {\bibfnamefont {S.-S.}\ \bibnamefont {Gong}},
  \bibinfo {author} {\bibfnamefont {W.}~\bibnamefont {Zhu}},\ and\ \bibinfo
  {author} {\bibfnamefont {D.~N.}\ \bibnamefont {Sheng}},\ }\bibfield  {title}
  {\bibinfo {title} {{Competing spin-liquid states in the spin-$\frac{1}{2}$
  Heisenberg model on the triangular lattice}},\ }\href
  {https://doi.org/10.1103/PhysRevB.92.140403} {\bibfield  {journal} {\bibinfo
  {journal} {Phys. Rev. B}\ }\textbf {\bibinfo {volume} {92}},\ \bibinfo
  {pages} {140403(R)} (\bibinfo {year} {2015})}\BibitemShut {NoStop}%
\bibitem [{\citenamefont {Ralko}\ \emph {et~al.}(2007)\citenamefont {Ralko},
  \citenamefont {Ferrero}, \citenamefont {Becca}, \citenamefont {Ivanov},\ and\
  \citenamefont {Mila}}]{Mila2007}%
  \BibitemOpen
  \bibfield  {author} {\bibinfo {author} {\bibfnamefont {A.}~\bibnamefont
  {Ralko}}, \bibinfo {author} {\bibfnamefont {M.}~\bibnamefont {Ferrero}},
  \bibinfo {author} {\bibfnamefont {F.}~\bibnamefont {Becca}}, \bibinfo
  {author} {\bibfnamefont {D.}~\bibnamefont {Ivanov}},\ and\ \bibinfo {author}
  {\bibfnamefont {F.}~\bibnamefont {Mila}},\ }\bibfield  {title} {\bibinfo
  {title} {Crystallization of the resonating valence bond liquid as vortex
  condensation},\ }\href {https://doi.org/10.1103/PhysRevB.76.140404}
  {\bibfield  {journal} {\bibinfo  {journal} {Phys. Rev. B}\ }\textbf {\bibinfo
  {volume} {76}},\ \bibinfo {pages} {140404(R)} (\bibinfo {year}
  {2007})}\BibitemShut {NoStop}%
\bibitem [{\citenamefont {Misguich}\ and\ \citenamefont
  {Mila}(2008)}]{Misguich2008}%
  \BibitemOpen
  \bibfield  {author} {\bibinfo {author} {\bibfnamefont {G.}~\bibnamefont
  {Misguich}}\ and\ \bibinfo {author} {\bibfnamefont {F.}~\bibnamefont
  {Mila}},\ }\bibfield  {title} {\bibinfo {title} {{Quantum dimer model on the
  triangular lattice: Semiclassical and variational approaches to vison
  dispersion and condensation}},\ }\href
  {https://doi.org/10.1103/PhysRevB.77.134421} {\bibfield  {journal} {\bibinfo
  {journal} {Phys. Rev. B}\ }\textbf {\bibinfo {volume} {77}},\ \bibinfo
  {pages} {134421} (\bibinfo {year} {2008})}\BibitemShut {NoStop}%
\bibitem [{\citenamefont {Slagle}\ and\ \citenamefont {Xu}(2014)}]{Slagle2014}%
  \BibitemOpen
  \bibfield  {author} {\bibinfo {author} {\bibfnamefont {K.}~\bibnamefont
  {Slagle}}\ and\ \bibinfo {author} {\bibfnamefont {C.}~\bibnamefont {Xu}},\
  }\bibfield  {title} {\bibinfo {title} {Quantum phase transition between the
  ${Z}_{2}$ spin liquid and valence bond crystals on a triangular lattice},\
  }\href {https://doi.org/10.1103/PhysRevB.89.104418} {\bibfield  {journal}
  {\bibinfo  {journal} {Phys. Rev. B}\ }\textbf {\bibinfo {volume} {89}},\
  \bibinfo {pages} {104418} (\bibinfo {year} {2014})}\BibitemShut {NoStop}%
\bibitem [{\citenamefont {Yan}\ \emph {et~al.}(2021)\citenamefont {Yan},
  \citenamefont {Wang}, \citenamefont {Ma}, \citenamefont {Qi},\ and\
  \citenamefont {Meng}}]{Meng2021QDM}%
  \BibitemOpen
  \bibfield  {author} {\bibinfo {author} {\bibfnamefont {Z.}~\bibnamefont
  {Yan}}, \bibinfo {author} {\bibfnamefont {Y.-C.}\ \bibnamefont {Wang}},
  \bibinfo {author} {\bibfnamefont {N.}~\bibnamefont {Ma}}, \bibinfo {author}
  {\bibfnamefont {Y.}~\bibnamefont {Qi}},\ and\ \bibinfo {author}
  {\bibfnamefont {Z.~Y.}\ \bibnamefont {Meng}},\ }\bibfield  {title} {\bibinfo
  {title} {{Topological phase transition and single/multi anyon dynamics of Z2
  spin liquid}},\ }\href {https://doi.org/10.1038/s41535-021-00338-1}
  {\bibfield  {journal} {\bibinfo  {journal} {npj Quantum Mater.}\ }\textbf
  {\bibinfo {volume} {6}},\ \bibinfo {pages} {39} (\bibinfo {year}
  {2021})}\BibitemShut {NoStop}%
\bibitem [{\citenamefont {Yan}\ \emph {et~al.}(2022)\citenamefont {Yan},
  \citenamefont {Samajdar}, \citenamefont {Wang}, \citenamefont {Sachdev},\
  and\ \citenamefont {Meng}}]{Meng2022QDM}%
  \BibitemOpen
  \bibfield  {author} {\bibinfo {author} {\bibfnamefont {Z.}~\bibnamefont
  {Yan}}, \bibinfo {author} {\bibfnamefont {R.}~\bibnamefont {Samajdar}},
  \bibinfo {author} {\bibfnamefont {Y.-C.}\ \bibnamefont {Wang}}, \bibinfo
  {author} {\bibfnamefont {S.}~\bibnamefont {Sachdev}},\ and\ \bibinfo {author}
  {\bibfnamefont {Z.~Y.}\ \bibnamefont {Meng}},\ }\bibfield  {title} {\bibinfo
  {title} {Triangular lattice quantum dimer model with variable dimer
  density},\ }\href {https://doi.org/10.1038/s41467-022-33431-5} {\bibfield
  {journal} {\bibinfo  {journal} {Nat. Commun.}\ }\textbf {\bibinfo {volume}
  {13}},\ \bibinfo {pages} {5799} (\bibinfo {year} {2022})}\BibitemShut
  {NoStop}%
\bibitem [{\citenamefont {Satzinger}\ \emph {et~al.}(2021)\citenamefont
  {Satzinger}, \citenamefont {Liu}, \citenamefont {Smith}, \citenamefont
  {Knapp}, \citenamefont {Newman}, \citenamefont {Jones}, \citenamefont {Chen},
  \citenamefont {Quintana}, \citenamefont {Mi}, \citenamefont {Dunsworth} \emph
  {et~al.}}]{Satzinger2021}%
  \BibitemOpen
  \bibfield  {author} {\bibinfo {author} {\bibfnamefont {K.~J.}\ \bibnamefont
  {Satzinger}}, \bibinfo {author} {\bibfnamefont {Y.-J.}\ \bibnamefont {Liu}},
  \bibinfo {author} {\bibfnamefont {A.}~\bibnamefont {Smith}}, \bibinfo
  {author} {\bibfnamefont {C.}~\bibnamefont {Knapp}}, \bibinfo {author}
  {\bibfnamefont {M.}~\bibnamefont {Newman}}, \bibinfo {author} {\bibfnamefont
  {C.}~\bibnamefont {Jones}}, \bibinfo {author} {\bibfnamefont
  {Z.}~\bibnamefont {Chen}}, \bibinfo {author} {\bibfnamefont {C.}~\bibnamefont
  {Quintana}}, \bibinfo {author} {\bibfnamefont {X.}~\bibnamefont {Mi}},
  \bibinfo {author} {\bibfnamefont {A.}~\bibnamefont {Dunsworth}}, \emph
  {et~al.},\ }\bibfield  {title} {\bibinfo {title} {{Realizing topologically
  ordered states on a quantum processor}},\ }\href
  {https://www.science.org/doi/10.1126/science.abi8378} {\bibfield  {journal}
  {\bibinfo  {journal} {Science}\ }\textbf {\bibinfo {volume} {374}},\ \bibinfo
  {pages} {1237} (\bibinfo {year} {2021})}\BibitemShut {NoStop}%
\bibitem [{\citenamefont {Semeghini}\ \emph {et~al.}(2021)\citenamefont
  {Semeghini}, \citenamefont {Levine}, \citenamefont {Keesling}, \citenamefont
  {Ebadi}, \citenamefont {Wang}, \citenamefont {Bluvstein}, \citenamefont
  {Verresen}, \citenamefont {Pichler}, \citenamefont {Kalinowski},
  \citenamefont {Samajdar} \emph {et~al.}}]{Semeghini2021}%
  \BibitemOpen
  \bibfield  {author} {\bibinfo {author} {\bibfnamefont {G.}~\bibnamefont
  {Semeghini}}, \bibinfo {author} {\bibfnamefont {H.}~\bibnamefont {Levine}},
  \bibinfo {author} {\bibfnamefont {A.}~\bibnamefont {Keesling}}, \bibinfo
  {author} {\bibfnamefont {S.}~\bibnamefont {Ebadi}}, \bibinfo {author}
  {\bibfnamefont {T.~T.}\ \bibnamefont {Wang}}, \bibinfo {author}
  {\bibfnamefont {D.}~\bibnamefont {Bluvstein}}, \bibinfo {author}
  {\bibfnamefont {R.}~\bibnamefont {Verresen}}, \bibinfo {author}
  {\bibfnamefont {H.}~\bibnamefont {Pichler}}, \bibinfo {author} {\bibfnamefont
  {M.}~\bibnamefont {Kalinowski}}, \bibinfo {author} {\bibfnamefont
  {R.}~\bibnamefont {Samajdar}}, \emph {et~al.},\ }\bibfield  {title} {\bibinfo
  {title} {{Probing topological spin liquids on a programmable quantum
  simulator}},\ }\href
  {https://www.science.org/doi/full/10.1126/science.abi8794} {\bibfield
  {journal} {\bibinfo  {journal} {Science}\ }\textbf {\bibinfo {volume}
  {374}},\ \bibinfo {pages} {1242} (\bibinfo {year} {2021})}\BibitemShut
  {NoStop}%
\bibitem [{\citenamefont {Samajdar}\ \emph {et~al.}(2021)\citenamefont
  {Samajdar}, \citenamefont {Ho}, \citenamefont {Pichler}, \citenamefont
  {Lukin},\ and\ \citenamefont {Sachdev}}]{Samajdar2020}%
  \BibitemOpen
  \bibfield  {author} {\bibinfo {author} {\bibfnamefont {R.}~\bibnamefont
  {Samajdar}}, \bibinfo {author} {\bibfnamefont {W.~W.}\ \bibnamefont {Ho}},
  \bibinfo {author} {\bibfnamefont {H.}~\bibnamefont {Pichler}}, \bibinfo
  {author} {\bibfnamefont {M.~D.}\ \bibnamefont {Lukin}},\ and\ \bibinfo
  {author} {\bibfnamefont {S.}~\bibnamefont {Sachdev}},\ }\bibfield  {title}
  {\bibinfo {title} {{Quantum phases of Rydberg atoms on a kagome lattice}},\
  }\href {https://doi.org/10.1073/pnas.2015785118} {\bibfield  {journal}
  {\bibinfo  {journal} {Proc. Natl. Acad. Sci. U.S.A.}\ }\textbf {\bibinfo
  {volume} {118}},\ \bibinfo {pages} {e2015785118} (\bibinfo {year}
  {2021})}\BibitemShut {NoStop}%
\bibitem [{\citenamefont {Verresen}\ \emph {et~al.}(2021)\citenamefont
  {Verresen}, \citenamefont {Lukin},\ and\ \citenamefont
  {Vishwanath}}]{Verresen2021}%
  \BibitemOpen
  \bibfield  {author} {\bibinfo {author} {\bibfnamefont {R.}~\bibnamefont
  {Verresen}}, \bibinfo {author} {\bibfnamefont {M.~D.}\ \bibnamefont
  {Lukin}},\ and\ \bibinfo {author} {\bibfnamefont {A.}~\bibnamefont
  {Vishwanath}},\ }\bibfield  {title} {\bibinfo {title} {{Prediction of Toric
  Code Topological Order from Rydberg Blockade}},\ }\href
  {https://doi.org/10.1103/PhysRevX.11.031005} {\bibfield  {journal} {\bibinfo
  {journal} {Phys. Rev. X}\ }\textbf {\bibinfo {volume} {11}},\ \bibinfo
  {pages} {031005} (\bibinfo {year} {2021})}\BibitemShut {NoStop}%
\bibitem [{\citenamefont {Verresen}\ and\ \citenamefont
  {Vishwanath}(2022)}]{Verresen2022}%
  \BibitemOpen
  \bibfield  {author} {\bibinfo {author} {\bibfnamefont {R.}~\bibnamefont
  {Verresen}}\ and\ \bibinfo {author} {\bibfnamefont {A.}~\bibnamefont
  {Vishwanath}},\ }\bibfield  {title} {\bibinfo {title} {{Unifying Kitaev
  Magnets, Kagom\'e Dimer Models, and Ruby Rydberg Spin Liquids}},\ }\href
  {https://doi.org/10.1103/PhysRevX.12.041029} {\bibfield  {journal} {\bibinfo
  {journal} {Phys. Rev. X}\ }\textbf {\bibinfo {volume} {12}},\ \bibinfo
  {pages} {041029} (\bibinfo {year} {2022})}\BibitemShut {NoStop}%
\bibitem [{\citenamefont {Samajdar}\ \emph {et~al.}(2023)\citenamefont
  {Samajdar}, \citenamefont {Joshi}, \citenamefont {Teng},\ and\ \citenamefont
  {Sachdev}}]{Samajdar2023}%
  \BibitemOpen
  \bibfield  {author} {\bibinfo {author} {\bibfnamefont {R.}~\bibnamefont
  {Samajdar}}, \bibinfo {author} {\bibfnamefont {D.~G.}\ \bibnamefont {Joshi}},
  \bibinfo {author} {\bibfnamefont {Y.}~\bibnamefont {Teng}},\ and\ \bibinfo
  {author} {\bibfnamefont {S.}~\bibnamefont {Sachdev}},\ }\bibfield  {title}
  {\bibinfo {title} {{Emergent ${\mathbb{Z}}_{2}$ Gauge Theories and
  Topological Excitations in Rydberg Atom Arrays}},\ }\href
  {https://doi.org/10.1103/PhysRevLett.130.043601} {\bibfield  {journal}
  {\bibinfo  {journal} {Phys. Rev. Lett.}\ }\textbf {\bibinfo {volume} {130}},\
  \bibinfo {pages} {043601} (\bibinfo {year} {2023})}\BibitemShut {NoStop}%
\bibitem [{\citenamefont {Bais}\ and\ \citenamefont
  {Slingerland}(2009)}]{Bais2009}%
  \BibitemOpen
  \bibfield  {author} {\bibinfo {author} {\bibfnamefont {F.~A.}\ \bibnamefont
  {Bais}}\ and\ \bibinfo {author} {\bibfnamefont {J.~K.}\ \bibnamefont
  {Slingerland}},\ }\bibfield  {title} {\bibinfo {title} {{Condensate-induced
  transitions between topologically ordered phases}},\ }\href
  {https://doi.org/10.1103/PhysRevB.79.045316} {\bibfield  {journal} {\bibinfo
  {journal} {Phys. Rev. B}\ }\textbf {\bibinfo {volume} {79}},\ \bibinfo
  {pages} {045316} (\bibinfo {year} {2009})}\BibitemShut {NoStop}%
\bibitem [{\citenamefont {Burnell}(2018)}]{Burnell2018}%
  \BibitemOpen
  \bibfield  {author} {\bibinfo {author} {\bibfnamefont {F.~J.}\ \bibnamefont
  {Burnell}},\ }\bibfield  {title} {\bibinfo {title} {{Anyon Condensation and
  Its Applications}},\ }\href
  {https://doi.org/10.1146/annurev-conmatphys-033117-054154} {\bibfield
  {journal} {\bibinfo  {journal} {Annu. Rev. Condens. Matter Phys.}\ }\textbf
  {\bibinfo {volume} {9}},\ \bibinfo {pages} {307} (\bibinfo {year}
  {2018})}\BibitemShut {NoStop}%
\bibitem [{\citenamefont {Hwang}(2024)}]{Hwang2024}%
  \BibitemOpen
  \bibfield  {author} {\bibinfo {author} {\bibfnamefont {K.}~\bibnamefont
  {Hwang}},\ }\bibfield  {title} {\bibinfo {title} {Anyon condensation and
  confinement transition in a kitaev spin liquid bilayer},\ }\href
  {https://doi.org/10.1103/PhysRevB.109.134412} {\bibfield  {journal} {\bibinfo
   {journal} {Phys. Rev. B}\ }\textbf {\bibinfo {volume} {109}},\ \bibinfo
  {pages} {134412} (\bibinfo {year} {2024})}\BibitemShut {NoStop}%
\bibitem [{\citenamefont {Wilson}(1974)}]{Wilson1974}%
  \BibitemOpen
  \bibfield  {author} {\bibinfo {author} {\bibfnamefont {K.~G.}\ \bibnamefont
  {Wilson}},\ }\bibfield  {title} {\bibinfo {title} {Confinement of quarks},\
  }\href {https://doi.org/10.1103/PhysRevD.10.2445} {\bibfield  {journal}
  {\bibinfo  {journal} {Phys. Rev. D}\ }\textbf {\bibinfo {volume} {10}},\
  \bibinfo {pages} {2445} (\bibinfo {year} {1974})}\BibitemShut {NoStop}%
\bibitem [{\citenamefont {Kogut}(1979)}]{Kogut1979}%
  \BibitemOpen
  \bibfield  {author} {\bibinfo {author} {\bibfnamefont {J.~B.}\ \bibnamefont
  {Kogut}},\ }\bibfield  {title} {\bibinfo {title} {{An introduction to lattice
  gauge theory and spin systems}},\ }\href
  {https://doi.org/10.1103/RevModPhys.51.659} {\bibfield  {journal} {\bibinfo
  {journal} {Rev. Mod. Phys.}\ }\textbf {\bibinfo {volume} {51}},\ \bibinfo
  {pages} {659} (\bibinfo {year} {1979})}\BibitemShut {NoStop}%
\bibitem [{Note1()}]{Note1}%
  \BibitemOpen
  \bibinfo {note} {In our notation, $X$ and $Z$ symbols represent the usual
  $\sigma _x$ and $\sigma _z$ Pauli operators/matrices. When we mention $X=+1$
  or $X=-1$, it means the eigenvalue/eigenstate of the Pauli-$X$ operator
  (similarly for the Pauli-$Z$ operator). Relations such as Eqs.~(\ref
  {eq:dimer-parity}) and (\ref {eq:vison-string-op}) are easily understood in
  the eigenbasis of the Pauli-$X$ operator.}\BibitemShut {Stop}%
\bibitem [{Note2()}]{Note2}%
  \BibitemOpen
  \bibinfo {note} {In Fig.~\ref {fig:DSF}(d), the DSF shows peak structures in
  a symmetric fashion because the ED wave function is a mixture of the 12-site
  pinwheel VBS states with different dimer orientations. Also, the absence of
  the peak at ${\protect \bf k}={\protect \bf 0}$ is due to the cancellation by
  $\langle d_{ij} \rangle \langle d_{kl} \rangle $ in Eq.~(\ref {eq:DSF}). In
  Fig.~\ref {fig:VBS}, the DSF is calculated only with the dimer-dimer
  correlator $\langle d_{ij} d_{kl} \rangle $ since the other term $\langle
  d_{ij} \rangle \langle d_{kl} \rangle $ makes a complete cancellation in the
  DSF for pure dimer states. Taking these points into account, we see that both
  DSFs are consistent with each other.}\BibitemShut {Stop}%
\end{thebibliography}

%

\end{document}